\documentclass{amsart}

\usepackage{caption}

\usepackage{amsmath,amssymb}

\usepackage{graphicx}

\usepackage{epstopdf}

\usepackage{color}
\usepackage{subfigure}
\usepackage{ulem}
\usepackage{tikz}
\usepackage{amsthm}

\usetikzlibrary{decorations.pathmorphing, patterns,shapes}

\newtheorem{theorem}{Theorem}[section]

\theoremstyle{definition}
\newtheorem{definition}[theorem]{Definition}

\theoremstyle{remark}

\numberwithin{equation}{section}

\begin{document}
\title[IST for the nonlocal RST Sine/Sinh-Gordon and NLS with NZBCs] {Inverse scattering transform for the 
nonlocal reverse space-time sine-gordon, sinh-gordon and nonlinear Schr\"{o}dinger equations with nonzero boundary conditions}

\author{Mark J. Ablowitz}
\address{Department of Applied Mathematics, University of Colorado, Campus Box 526, Boulder, Colorado 80309-0526}
\email{mark.ablowitz@colorado.edu}
\thanks{}

\author{Bao-Feng Feng}
\address{School of Mathematical and Statistical Sciences, University of Texas Rio Grande Valley}
\email{baofeng.feng@utgv.edu}

\author{Xu-Dan Luo}
\address{Department of Applied Mathematics, University of Colorado at Boulder, Boulder, Colorado 80309}
\email{lxdmathematics@gmail.com}

\author{Ziad H. Musslimani}
\address{Department of Mathematics, Florida State University, Tallahassee, FL 32306-4510}
\email{muslimani@math.fsu.edu}

\begin{abstract}
The reverse space-time (RST) Sine-Gordon, Sinh-Gordon and nonlinear Schr\"odinger equations were recently introduced and shown to be integrable infinite-dimensional dynamical systems. The inverse scattering transform (IST) for rapidly decaying data was also
constructed. In this paper, IST  for these equations with nonzero boundary conditions (NZBCs) at infinity is presented. The NZBC problem is more complicated due to the associated branching structure of the associated linear eigenfunctions.
With constant amplitude at infinity, four cases are analyzed; they correspond to two different signs of nonlinearity and two different values of the phase at infinity. Special soliton solutions are discussed and explicit 1-soliton and 2-soliton solutions are found. In terms of IST, the difference between the RST Sine-Gordon/Sinh-Gordon equations and the RST NLS equation is the time dependence of the scattering data. Spatially dependent boundary conditions are also briefly considered.
\end{abstract}
\maketitle
\section{Introduction}
In 1974, AKNS \cite{AKNS} published ``Inverse scattering transform -- Fourier analysis for nonlinear problems" where a general framework to generate integrable systems solvable by the method of inverse scattering transform was developed. The idea is to consider the scattering problem
\begin{equation}
\label{E:AKNS}
v_{x}=Xv=
\left(\begin{array}{cc}
-ik& q(x,t)\\
r(x,t)& ik
\end{array}\right)v,
\end{equation}
where $v(x,t)=(v_{1}(x,t), v_{2}(x,t))^{T}$, $k$ is a time-independent spectral parameter and $q(x,t)$, $r(x,t)$ are complex-valued functions of the real variables $x, t$. Furthermore,
associated with the AKNS scattering problem (\ref{E:AKNS}) is the time evolution equation
\begin{equation}
\label{time dependence}
v_{t}=Tv=
\left(\begin{array}{cc}
A& B\\
C& -A
\end{array}\right)v,
\end{equation}
where the quantities $A$, $B$ and $C$ are scalar functions of $q(x,t)$, $r(x,t)$ and spectral parameter $k$. Depending on the choices of $A$, $B$ and $C$ the compatibility condition
$v_{xt}=v_{tx}$ gives rise to a coupled partial differential equations for $q(x,t)$ and $r(x,t).$ Under certain relation between $q(x,t)$ and $r(x,t)$ (also referred to as symmetry reduction) the resulting system is compatible and leads to a single integrable evolution equation for
$q(x,t)$ or $r(x,t)$. Examples include, the nonlinear Schr\"{o}dinger (NLS), complex modified Korteweg-deVries (KdV) and complex Sine/Sinh-Gordon equations.
All these equations originate from the same scattering problem (\ref{E:AKNS}),
and are found using the symmetry reduction \cite{AKNS}
\begin{equation}
\label{AKNS_sym-1}
r(x,t)=\sigma q^{*}(x,t), \;\;\;\sigma=\mp 1\;.
\end{equation}
The only differences are the coefficients $A,B,C$ in the time evolution equation (\ref{time dependence}). For decaying data, this means that once the scattering problem is solved, one can obtain the solutions of all these equations by taking different time evolutions for scattering data which are determined by the value of $A$ at infinity.

If $r(x,t)=\sigma q(x,t)$ and $q(x,t)$ are real, then the eigenfunctions $v(x,t)$ possess two symmetries. Associated with the proper time evolution equations, we are able to recover and linearize
the real Sine-Gordon equation with $r(x,t)=-q(x,t)=\frac{1}{2}u_{x}$ and the real Sinh-Gordon equation with $r(x,t)=q(x,t)=\frac{1}{2}u_{x}$.

Most studies of IST deal with  initial-value problems with rapidly decaying data; e.g. $q(x,t)$, $r(x,t) \rightarrow 0$ rapidly as $x\rightarrow\pm \infty$. However there has been keen interest in other NLS type problems with non-zero boundary conditions (NZBCs). Pioneering studies investigating NZBCs were developed for the NLS equation \cite{ZS2}. The original method to carry out the inverse problem for NZBCs employed  two Riemann surfaces associated with square root branch points in the eigenfunctions/scattering data. A significant improvement was made  with the introduction of a uniformization variable \cite{FT87}. This transforms the inverse problem to the more standard inverse problem in the upper lower/half planes in the new variable.  Subsequently a number of researchers have also studied NLS problems in this manner,  cf. \cite{Prinari06,Demontis, Demontis2, BK14},  hence substantially enhancing the applicability of the IST technique.\\

The solutions of the complex Sine/Sinh-Gordon equations with non-decaying boundary conditions can be deduced from the same scattering problem as the NLS equation; the only difference is the associated time evolution equation (see section 9).\\

Recently, four new nonlocal symmetry reductions for the AKNS scattering problem have been identified \cite{AblowitzMusslimani4}. These are (here $\sigma=\mp 1$)
\begin{itemize}

\item $r(x,t)=\sigma q^*(x,-t)\;,$

\item $r(x,t)=\sigma q^*(-x,t)\;,$
\item $r(x,t)=\sigma q(-x,-t)\;,$
\item $r(x,t)=\sigma q^*(-x,-t)\;.$
\end{itemize}
Each of these symmetry reduction leads to a new types of inverse problems and new classes of nonlocal nonlinear integrable equations \cite{AblowitzMusslimani4}; furthermore, the IST with decaying data for many associated equations was constructed. But the IST with nonzero boundary conditions (NZBCs) was only considered for the $PT$ symmetric case: $r(x,t)=\sigma q^{*}(-x,t)$ \cite{AblowitzLuoMusslimani}. Recall that an evolution equation is said to be $PT$ symmetric if it is invariant under the combined action of parity operator $P (x\rightarrow -x)$ and time-reversal symmetry and complex conjugation $T$.\\

In addition to these nonlocal NLS equations, the integrability of the nonlocal `reverse space-time' (RST) Sine/Sinh-Gordon equations was also established
in \cite{AblowitzMusslimani4}. They correspond to the symmetry reduction $r(x,t)=\sigma q(-x,-t)$ with $\sigma=-1$ for the nonlocal Sin-Gordon equation and $\sigma=1$ for the nonlocal Sinh-Gordon equation. Although the IST under the symmetry reduction $r(x,t)=\sigma q(-x,-t)$ was analyzed for decaying data, the IST for the nonlocal RST NLS equation and nonlocal RST Sine/Sinh-Gordon equations with NZBCs is still new and open. There are significant differences between these cases and the $PT$ symmetric case considered in \cite{AblowitzLuoMusslimani}.
The real Sine-Gordon equation
\begin{equation}
\label{SineG}
u_{xt}=\sin u
\end{equation}
arises in many branches of mathematical physics. In mathematics, it has arisen classically in the study of differential geometry and in physics it has arisen in the study of Josephson junctions, models of particle physics, stability of fluid motions etc. The nonlocal RST
Sine-Gordon and Sinh-Gordon equations \cite{AblowitzMusslimani3}, \cite{AblowitzMusslimani4} are given by
\begin{equation}
\label{E:nonlocal Sine/Sinh}
q_{xt}(x,t)+2s(x,t)q(x,t)=0,
\end{equation}
with
\begin{equation}
\label{seqSine/Sinh}
s_{x}(x,t)=-\sigma \partial_t (q(x,t)q(-x,-t)), ~~\sigma=\mp 1.
\end{equation}
We will define $s(x,t)$ as
\begin{equation}
\label{defin_s}
s(x,t)= \sigma \int_x^{\infty} \partial_t (q(x',t)q(-x',-t) )dx'
\end{equation}
and require $s(x,t) \rightarrow 0 ~\text{as} ~ |x| \rightarrow \infty$;  alternatively, (\ref{defin_s}) implies
\[   \int_{-\infty}^{\infty} \partial_t (q(x',t)q(-x',-t) )dx'=0 \]
It follows that  $s(-x,-t)=s(x,t)$. When $\sigma=-1$, (\ref{E:nonlocal Sine/Sinh})-(\ref{defin_s}) is referred to as the nonlocal Sine-Gordon equation; when $\sigma=1$, (\ref{E:nonlocal Sine/Sinh})-(\ref{defin_s}) is referred to as the nonlocal Sinh-Gordon equation; both are integrable. We note that for the classical real Sine-Gordon/Sinh-Gordon equations $s_{x}(x,t)=-\sigma(q^2(x,t))_{t}$ and there is a further reduction; i.e. we can reduce to to (\ref{SineG})  when $q= -u_x/2$. For the complex sine/Sinh-Gordon equation we have $s_{x}(x,t)=-\sigma(|q|^2(x,t))_{t}$, but there is no known reduction beyond that. The same is true for the nonlocal RST Sine/Sinh-Gordon equations.\\
In this paper, we study the direct and inverse scattering transform associated with the
RST nonlocal Sine/Sinh-Gordon equation with the following nonzero boundary conditions
\begin{equation}
\label{E:NZBC}
q(x,t)\rightarrow q_{\pm}(t)=q_{0}e^{i(\alpha t+\theta_{\pm})}, ~\text{and}~s(x,t) \rightarrow 0 \ \ as\  x\rightarrow\pm\infty,
\end{equation}
where $q_{0}>0$, $0\leq\theta_{\pm}<2\pi$ and $\alpha$ is real.

\medskip

Taking into account results from the scattering problem, 
the following cases are considered. For both $\sigma=1$ and $\sigma=-1$: $\theta_{+}+\theta_{-}=0$ or $\theta_{+}+\theta_{-}=\pi$, so that the branch points of eigenvalues in the scattering problem either lie in the real axis ($\pm q_{0}$) or imaginary axis ($\pm i q_{0}$).

First, we consider the case of $\sigma=1$ (the nonlocal Sinh-Gordon equation), $\theta_{+}+\theta_{-}=0$. We find the following nonsingular `dark' 1-soliton solution which in terms of magnitude has a dip from its asymptotic values at infinity:
\begin{equation}
q(x,t)=q_{0}e^{i\alpha t}\cos\left(\theta_{+}-iq_{0}x\sin\theta_{+}-\frac{i}{2}\alpha t\tan\theta_{+}\right) \text{sech}\left(q_{0}x\sin\theta_{+}+\frac{1}{2}\alpha t\tan\theta_{+}\right),
\end{equation}
~~$\theta_{+} \neq \pi/2$. This is discussed in section 3 below.

Second, we consider the case of $\sigma=1$ (the nonlocal Sinh-Gordon equation), $\theta_{+}+\theta_{-}=\pi$. In this case a single eigenvalue is found to be in the continuous spectrum; there is no `proper exponentially decaying' one soliton solution.  The simplest decaying pure reflectionless potential generates a 2-soliton solution. There are nonsingular 2-soliton solutions. See section 4 below.

%

Third, we consider the case of $\sigma=-1$ (the nonlocal Sine-Gordon equation), $\theta_{+}+\theta_{-}=\pi$. We find the following nonsingular dark 1-soliton solution
\begin{equation}
q(x,t)=q_{0}e^{i\alpha t}\left[i\sin\theta_{+}+\cos\theta_{+}\tanh\left(q_{0}x\cos\theta_{+}-\frac{\alpha t}{2}\cot\theta_{+}\right)\right];
\end{equation}
see section 6.
%
%

%
Finally we analyze the case of $\sigma=-1$ (the nonlocal Sine-Gordon equation), $\theta_{+}+\theta_{-}=0$. We have a similar conclusion as that  discussed in the first case above  when $\sigma=1$ and $\theta_{+}+\theta_{-}=\pi$; i.e there are no exponentially decaying 1-soltion solutions, but there are nonsingular 2-soliton solutions; this is discussed in section 7.


In addition the IST is developed for the nonlocal reverse space-time NLS (RST-NLS) equation
\begin{equation}
\label{RSTNLS}
iq_{t}(x,t)=q_{xx}(x,t)-2\sigma q^{2}(x,t)q(-x,-t),
\end{equation}
with the nonzero boundary conditions:
\begin{equation}
q(x,t)\rightarrow q_{0}e^{i(2\sigma q_{0}^{2}t+\theta_{\pm})}
\end{equation}
as $x\rightarrow\pm \infty$, where $\theta_{+}+\theta_{-}=0$, and the case
\begin{equation}
q(x,t)\rightarrow q_{0}e^{i(-2\sigma q_{0}^{2}t+\theta_{\pm})}
\end{equation}
as $x\rightarrow\pm \infty$, where $\theta_{+}+\theta_{-}=\pi$. 
Since the RST-NLS equation has the same scattering problem as the nonlocal RST Sine/Sinh-Gordon equations, we obtain the corresponding soliton solutions for all four cases by modifying the time evolution of the scattering data.

We note that 
some properties of the nonzero boundary conditions associated with RST-NLS equation (\ref{RSTNLS}) can be obtained directly. We assume that as $q \rightarrow q_{\pm}(t)$ as $x \rightarrow \pm \infty$ , then equation (\ref{RSTNLS}) yields

\begin{equation}
\label{a1RSTNLS}
iq_{\pm,t}(t)=-2\sigma q_{\pm}^{2}(t)q_{\mp}(-t).
\end{equation}

This implies that
\begin{equation}
\label{a2RSTNLS}
q_{+}(t)q_{-}(-t)=C_0=\text{const.}
\end{equation}
hence equation (\ref{a1RSTNLS}) simplifies to
\begin{equation}
\label{a3RSTNLS}
iq_{\pm,t}(t)=-2\sigma C_0q_{\pm}(t).
\end{equation}
This equation has the solution
\begin{equation}
\label{a4RSTNLS}
q_{\pm}(t)= |q_{\pm}(0)|e^{2 i \sigma C_0  t}e^{i \theta_{\pm}}
\end{equation}
where $\theta_{\pm}$ are constant. Since $C_{0}=|q_{+}(0)|\cdot|q_{-}(0)|\cdot e^{i(\theta_{+}+\theta_{-})}$ , thus, if $\theta_{+}+\theta_{-}=0$ or $\theta_{+}+\theta_{-}=\pi$, then $C_{0}$ is real. Otherwise it is complex and the background either grows or decays exponentially as $|t|\rightarrow \infty$. Without loss of generality we take $q_{\pm}(0)=q_0=$const.

For the RST-NLS equation in section 8 we then consider the IST for the same four cases as for the Sine/Sinh-Gordon equations: $\sigma=\pm 1$ with $\theta_{+}+\theta_{-} = 0, \pi$.

First, when $\sigma=1$, $\theta_{+}+\theta_{-}=0$ , we find the following nonsingular dark 1-soliton solution
\begin{equation}
\begin{split}
q(x,t)=\frac{q_{0}e^{2iq_{0}^{2} t}\left[e^{i\theta_{+}}\cdot e^{2q_{0}x\sin\theta_{+}}+e^{-i\theta_{+}}\cdot e^{2q_{0}^{2}t\sin(2\theta_{+})}\right]}{e^{2q_{0}x\sin\theta_{+}}+e^{2q_{0}^{2}t\sin(2\theta_{+})}}.
\end{split}
\end{equation}
Second, when $\sigma=1$, $\theta_{+}+\theta_{-}=\pi$ , there is no `proper exponentially decaying'
1-soliton solution as a single eigenvalue is found to be in the continuous spectrum.  The simplest decaying pure reflectionless potential generates a 2-soliton solution. There are nonsingular 2-soliton solutions.

Third, when $\sigma=-1$, $\theta_{+}+\theta_{-}=\pi$ , we find the following nonsingular dark 1-soliton solution
\begin{equation}
q(x,t)=q_{0}e^{2iq_{0}^{2}t}[i\sin\theta_{+}+\cos\theta_{+}\tanh(q_{0}\cos\theta_{+}\cdot(x+2q_{0}t\sin\theta_{+}))].
\end{equation}
Lastly, when $\sigma=-1$, $\theta_{+}+\theta_{-}=0$ , we have a similar conclusion as that  discussed in the  case  when $\sigma=1$ and $\theta_{+}+\theta_{-}=\pi$.

In section 9 we show that there also exist solutions for the nonlocal Sine/Sinh-Gordon equations, complex Sine/Sinh-Gordon equations and the nonlinear Schr\"{o}dinger equation, which satisfy the following spatially dependent boundary conditions
\begin{equation}
q(x,t)\rightarrow q_{0}e^{i(\alpha t+\beta x+\theta_{\pm})},
\end{equation}
as $x\rightarrow \pm\infty$, where both $\alpha$ and $\beta$ are real.
For the NLS equation Galilean invariance is verified.

In section 10, we also present, some novel types of solutions to the above equations which are singular along space-time lines.

%
%
%

\section{Nonlocal Sinh-Gordon equation}
The nonlocal Sinh-Gordon equation
\begin{equation}
q_{xt}(x,t)+2s(x,t)q(x,t)=0,
\end{equation}
where $s_{x}(x,t)=-(q(x,t)q(-x,-t))_{t}, s(x,t) \rightarrow 0$ as  $|x| \rightarrow  \infty$ and $s(-x,-t)=s(x,t)$ 
is associated with the following $2\times2$
compatible systems:
\begin{equation}
\label{E:nonlocal scattering}
v_{x}=Xv=
\left(\begin{array}{cc}
-ik& q(x,t)\\
q(-x,-t)& ik
\end{array}\right)v,
\end{equation}
\begin{equation}
v_{t}=Tv=
\left(\begin{array}{cc}
-\frac{is(x,t)}{2k}& -\frac{iq_{t}(x,t)}{2k}\\
\frac{iq_{t}(-x,-t)}{2k}& \frac{is(x,t)}{2k}
\end{array}\right)v,
\end{equation}
where $q(x,t)$ is a complex-valued function of the real variables $x$ and $t$ with the corresponding linear operators given by
\begin{equation}
X=
\left(\begin{array}{cc}
-ik& q(x,t)\\
q(-x,-t)& ik
\end{array}\right),\ \
T=
\left(\begin{array}{cc}
-\frac{is(x,t)}{2k}& -\frac{iq_{t}(x,t)}{2k}\\
\frac{iq_{t}(-x,-t)}{2k}& \frac{is(x,t)}{2k}
\end{array}\right).
\end{equation}
Alternatively the space part of the compatible system may be written in the form
\begin{equation}
v_{x}=(ikJ+Q)v, \ \ \ x\in \mathbb{R},
\end{equation}
where
\begin{equation}J=
\left(\begin{matrix}
-1& 0\\
0& 1
\end{matrix}\right),\ \ \
Q=
\left(\begin{matrix}
0& q(x,t)\\
q(-x,-t)& 0
\end{matrix}\right).
\end{equation}
Here, $q(x,t)$ is called the potential and $k$ is a complex spectral parameter.

As $x\rightarrow\pm\infty$, the eigenfunctions of the scattering problem asymptotically satisfy
\begin{equation}
\label{E:eigenvalue asymptotic}
\left(\begin{array}{cc}
v_{1}\\
v_{2}
\end{array}\right)_{x}
=
\left(\begin{array}{cc}
-ik& q_{0}e^{i(\alpha t+\theta_{\pm})}\\
q_{0}e^{i(-\alpha t+\theta_{\mp})}& ik
\end{array}\right)
\left(\begin{array}{cc}
v_{1}\\
v_{2}
\end{array}\right),
\end{equation}
i.e.,
\begin{equation}
v_{x}=(ikJ+Q_{\pm}(t))v, \ \ \ Q_{\pm}(t)=\left(\begin{array}{cc}
0& q_{0}e^{i(\alpha t+\theta_{\pm})}\\
q_{0}e^{i(-\alpha t+\theta_{\mp})}& 0
\end{array}\right).
\end{equation}

\section{Nonlocal Sinh-Gordon equation with $\theta_{+}+\theta_{-}=0$}
\subsection{Direct scattering}
In this section we consider the nonzero boundary conditions (NZBCs) given above in (\ref{E:NZBC}) and $\theta_{+}+\theta_{-}=0$. With this condition, equation (\ref{E:eigenvalue asymptotic}) conveniently reduces to
\begin{equation}
\frac{\partial^{2}v_{j}}{\partial x^{2}}=-(k^{2}-q_{0}^{2})v_{j}, \ \ j=1,2.
\end{equation}
Each of the two equations has two linearly independent solutions $e^{i\lambda x}$ and $e^{-i\lambda x}$ as $|x|\rightarrow\infty$, where
$\lambda=\sqrt{k^{2}-q_{0}^{2}}$.
The variable $k$ is then considered to belong to a Riemann surface $\mathbb{K}$ consisting of two sheets $\mathbb{C}_{1}$ and $\mathbb{C}_{2}$ 
with the complex plane cut along $(-\infty, -q_{0}]\cup [q_{0}, +\infty)$
with its edges glued in such a way that $\lambda(k)$ is continuous through the cut. We introduce the local polar coordinates
\begin{equation}
k-q_{0}=r_{1}e^{i\theta_{1}}, \ \ \ 0\leq\theta_{1}<2\pi,
\end{equation}
\begin{equation}
k+q_{0}=r_{2}e^{i\theta_{2}}, \ \ \ -\pi\leq\theta_{2}<\pi,
\end{equation}
where $r_{1}=|k-q_{0}|$ and $r_{2}=|k+q_{0}|$. Then the function $\lambda(k)$ becomes single-valued on $\mathbb{K}$, i.e.,
\begin{equation}
\label{E:lambda}
\lambda(k)=\left\{\begin{array}{ll}
\lambda_{1}(k)=(r_{1}r_{2})^{\frac{1}{2}}\cdot e^{i\frac{\theta_{1}+\theta_{2}}{2}},\ \ k\in\mathbb{C}_{1},\\
\lambda_{2}(k)=-(r_{1}r_{2})^{\frac{1}{2}}\cdot e^{i\frac{\theta_{1}+\theta_{2}}{2}}, \ \ k\in\mathbb{C}_{2}.\\
\end{array}\right.
\end{equation}
Moreover, if $k\in\mathbb{C}_{1}$, then $\Im \lambda\geq 0$; and if $k\in\mathbb{C}_{2}$, then $\Im \lambda\leq 0$.
Hence, the variable $\lambda$ is thought of as belonging to the complex plane consisting of the upper half plane $U_{+}$ :$\Im \lambda\geq 0$, and lower half plane $U_{-}$ :$\Im \lambda\leq 0$,  glued together along the real axis; the transition occurs at  $\Im \lambda= 0$. The transformation $k\rightarrow\lambda$ maps $\mathbb{C}_{1}$ onto $U_{+}$, $\mathbb{C}_{2}$ onto $U_{-}$, the cut $(-\infty, -q_{0}]\cup [q_{0}, +\infty)$ onto the real axis, and the points $\pm q_{0}$ to $0$ (See \cite{AblowitzLuoMusslimani}, Page 6, Fig. 1 and Fig. 2).


\subsection{Eigenfunctions}
It is natural to introduce the eigenfunctions defined by the following boundary conditions
\begin{equation}
\label{E:asymptotic 1}
\phi(x,k)\sim w e^{-i\lambda x}, \ \ \ \overline{\phi}(x,k)\sim \overline{w}e^{i\lambda x}
\end{equation}
as $x\rightarrow-\infty$,
\begin{equation}
\label{E:asymptotic 2}
\psi(x,k)\sim v e^{i\lambda x}, \ \ \ \overline{\psi}(x,k)\sim \overline{v}e^{-i\lambda x}
\end{equation}
as $x\rightarrow +\infty$. We substitute the above into (\ref{E:eigenvalue asymptotic}), obtaining
\begin{equation}
\label{E:boundary conditions 1}
w=\left(\begin{array}{cc}
\lambda+k\\
iq_{+}
\end{array}\right), \ \ \
\overline{w}=\left(\begin{array}{cc}
-iq_{-}\\
\lambda+k
\end{array}\right),
\end{equation}
\begin{equation}
\label{E:boundary conditions 2}
v=\left(\begin{array}{cc}
-iq_{+}\\
\lambda+k
\end{array}\right), \ \ \
\overline{v}=
\left(\begin{array}{cc}
\lambda+k\\
iq_{-}
\end{array}\right),
\end{equation}
which satisfy the boundary conditions, but they are not unique.

In the following analysis, it is convenient to consider functions with constant boundary conditions. We define the bounded eigenfunctions as follows:
\begin{equation}
\label{E:definition 1}
M(x,k)=e^{i\lambda x}\phi(x,k), \ \ \ \overline{M}(x,k)=e^{-i\lambda x}\overline{\phi}(x,k),
\end{equation}
\begin{equation}
\label{E:definition 2}
N(x,k)=e^{-i\lambda x}\psi(x,k), \ \ \ \overline{N}(x,k)=e^{i\lambda x}\overline{\psi}(x,k).
\end{equation}
The eigenfunctions can be represented by means of the following integral equations
\begin{equation}
M(x,k)=
\left(\begin{array}{cc}
\lambda+k\\
iq_{+}
\end{array}\right)
+\int_{-\infty}^{+\infty}G_{-}(x-x',k)((Q-Q_{-})M)(x',k)dx',
\end{equation}
\begin{equation}
\overline{M}(x,k)=
\left(\begin{array}{cc}
-iq_{-}\\
\lambda+k
\end{array}\right)
+\int_{-\infty}^{+\infty}\overline{G}_{-}(x-x',k)((Q-Q_{-})M)(x',k)dx',
\end{equation}
\begin{equation}
N(x,k)=
\left(\begin{array}{cc}
-iq_{+}\\
\lambda+k
\end{array}\right)
+\int_{-\infty}^{+\infty}G_{+}(x-x',k)((Q-Q_{+})M)(x',k)dx',
\end{equation}
\begin{equation}
\overline{N}(x,k)=\left(\begin{array}{cc}
\lambda+k\\
iq_{-}
\end{array}\right)
+\int_{-\infty}^{+\infty}\overline{G}_{+}(x-x',k)((Q-Q_{+})M)(x',k)dx'.
\end{equation}
Using the Fourier transform method, we get
\begin{equation}
G_{-}(x,k)=\frac{\theta(x)}{2\lambda}[(1+e^{2i\lambda x})\lambda I-i(e^{2i\lambda x}-1)(ikJ+Q_{-})],
\end{equation}
\begin{equation}
\overline{G}_{-}(x,k)=\frac{\theta(x)}{2\lambda}[(1+e^{-2i\lambda x})\lambda I+i(e^{-2i\lambda x}-1)(ikJ+Q_{-})],
\end{equation}
\begin{equation}
G_{+}(x,k)=-\frac{\theta(-x)}{2\lambda}[(1+e^{-2i\lambda x})\lambda I+i(e^{-2i\lambda x}-1)(ikJ+Q_{+})],
\end{equation}
\begin{equation}
\overline{G}_{+}(x,k)=-\frac{\theta(-x)}{2\lambda}[(1+e^{2i\lambda x})\lambda I-i(e^{2i\lambda x}-1)(ikJ+Q_{+})],
\end{equation}
where $\theta(x)$ is the Heaviside function, i.e., $\theta(x)=1$ if $x>0$ and $\theta(x)=0$ if $x<0$.
\begin{definition}
We say $f\in L^{1}(\mathbb{R})$  if $\int_{-\infty}^{+\infty}|f(x)|dx<\infty$, and $f\in L^{1,2}(\mathbb{R})$ if $\int_{-\infty}^{+\infty}|f(x)|\cdot(1+|x|)^{2}dx<\infty$.
\end{definition}
Then we have the following result.
\begin{theorem}
\label{T:1}
Suppose the entries of $Q-Q_{\pm}$ belong to $L^{1}(\mathbb{R})$, then for each $x\in\mathbb{R}$, the eigenfunctions $M(x,k)$ and $N(x,k)$ are continuous for $k\in \overline{\mathbb{C}}_{1}\setminus\{\pm q_{0}\}$ and analytic for $k\in \mathbb{C}_{1}$, $\overline{M}(x,k)$ and $\overline{N}(x,k)$ are continuous for $k\in \overline{\mathbb{C}}_{2}\setminus\{\pm q_{0}\}$ and analytic for $k\in \mathbb{C}_{2}$. In addition, if the entries of $Q-Q_{\pm}$ belong to $L^{1,2}(\mathbb{R})$, then for each $x\in\mathbb{R}$, the eigenfunctions $M(x,k)$ and $N(x,k)$ are continuous for $k\in \overline{\mathbb{C}}_{1}$ and analytic for $k\in \mathbb{C}_{1}$, $\overline{M}(x,k)$ and $\overline{N}(x,k)$ are continuous for $k\in \overline{\mathbb{C}}_{2}$ and analytic for $k\in \mathbb{C}_{2}$.

The proof makes use of Neumann series; it is similar to \cite{AblowitzLuoMusslimani}.

\end{theorem}
\subsection{Scattering data}
The two eigenfunctions with non-fixed or fixed boundary conditions as $x\rightarrow-\infty$ are linearly independent, so as are the two eigenfunctions with fixed boundary conditions as $x\rightarrow+\infty$. Indeed, if $u(x,k)=(u_{1}(x,k), u_{2}(x,k))^{T}$ and $v(x,k)=(v_{1}(x,k), v_{2}(x,k))^{T}$ are any two solutions of (\ref{E:nonlocal scattering}), we have
\begin{equation}
\frac{d}{dx}W(u,v)=0,
\end{equation}
where the Wronskian of $u$ and $v$, $W(u,v)$ is given by
\begin{equation}
W(u,v)=u_{1}v_{2}-u_{2}v_{1}.
\end{equation}
From the asymptotics (\ref{E:asymptotic 1}) and (\ref{E:asymptotic 2}), it follows that
\begin{equation}
\label{E:Wronskian 1}
W(\phi,\overline{\phi})=\lim_{x\rightarrow-\infty}W(\phi(x,k),\overline{\phi}(x,k))=2\lambda(\lambda+k)
\end{equation}
and
\begin{equation}
\label{E:Wronskian 2}
W(\psi,\overline{\psi})=\lim_{x\rightarrow+\infty}W(\psi(x,k),\overline{\psi}(x,k))=-2\lambda(\lambda+k),
\end{equation}
which proves that the functions $\phi(x,k)$ and $\overline{\phi}(x,k)$ are linearly independent, as are $\psi$ and $\overline{\psi}$,
with only the branch points $\pm q_{0}$ being excluded. Hence, we can write $\phi(x,k)$ and $\overline{\phi}(x,k)$ as linear combinations of
$\psi(x,k)$ and $\overline{\psi}(x,k)$, or vice versa. Thus, the relations
\begin{equation}
\label{E:linear combination 1}
\phi(x,k)=b(k)\psi(x,k)+a(k)\overline{\psi}(x,k)
\end{equation}
and
\begin{equation}
\label{E:linear combination 2}
\overline{\phi}(x,k)=\overline{a}(k)\psi(x,k)+\overline{b}(k)\overline{\psi}(x,k)
\end{equation}
hold for any $k$ such that all four eigenfunctions exist. Combining (\ref{E:Wronskian 1}) and (\ref{E:Wronskian 2}), we can deduce
that the scattering data satisfy the following characterization equation
\begin{equation}
a(k)\overline{a}(k)-b(k)\overline{b}(k)=1.
\end{equation}
The scattering data can be represented in terms of  Wronskians of the eigenfunctions, i.e.,
\begin{equation}
a(k)=\frac{W(\phi(x,k),\psi(x,k))}{W(\overline{\psi}(x,k),\psi(x,k))}=\frac{W(\phi(x,k),\psi(x,k))}{2\lambda(\lambda+k)},
\end{equation}
\begin{equation}
\overline{a}(k)=-\frac{W(\overline{\phi}(x,k),\overline{\psi}(x,k))}{W(\overline{\psi}(x,k),\psi(x,k))}
=-\frac{W(\overline{\phi}(x,k),\overline{\psi}(x,k))}{2\lambda(\lambda+k)},
\end{equation}
\begin{equation}
b(k)=-\frac{W(\phi(x,k),\overline{\psi}(x,k))}{W(\overline{\psi}(x,k),\psi(x,k))}=-\frac{W(\phi(x,k),\overline{\psi}(x,k))}{2\lambda(\lambda+k)},
\end{equation}
\begin{equation}
\overline{b}(k)=\frac{W(\overline{\phi}(x,k),\psi(x,k))}{W(\overline{\psi}(x,k),\psi(x,k))}
=\frac{W(\overline{\phi}(x,k),\psi(x,k))}{2\lambda(\lambda+k)}.
\end{equation}
Then from the analytic behavior of the eigenfunctions we have the following theorem.
\begin{theorem}
\label{T:2}
Suppose the entries of $Q-Q_{\pm}$ belong to $L^{1}(\mathbb{R})$, then $a(k)$ is continuous for $k\in \overline{\mathbb{C}}_{1}\setminus\{\pm q_{0}\}$ and analytic for $k\in \mathbb{C}_{1}$, and $\overline{a}(k)$ is continuous for $k\in \overline{\mathbb{C}}_{2}\setminus\{\pm q_{0}\}$ and analytic for $k\in \mathbb{C}_{2}$. Moreover, $b(k)$ and $\overline{b}(k)$ are continuous in $k\in(-\infty,-q_{0})\cup (q_{0},+\infty)$. In addition, if the entries of $Q-Q_{\pm}$ belong to $L^{1,2}(\mathbb{R})$,
then $a(k)\lambda(k)$ is continuous for $k\in \overline{\mathbb{C}}_{1}$ and analytic for $k\in \mathbb{C}_{1}$, and $\overline{a}(k)\lambda(k)$ is continuous for $k\in \overline{\mathbb{C}}_{2}$ and analytic for $k\in \mathbb{C}_{2}$. Moreover, $b(k)\lambda(k)$ and $\overline{b}(k)\lambda(k)$ are continuous for $k\in \mathbb{R}$. If the entries of $Q-Q_{\pm}$ do not grow faster than $e^{-ax^{2}}$, where $a$ is a positive real number, then $a(k)\lambda(k)$, $\overline{a}(k)\lambda(k)$, $b(k)\lambda(k)$ and $\overline{b}(k)\lambda(k)$ are analytic for $k\in\mathbb{K}$.
\end{theorem}
The proof follows from the Wronskian relations; see also \cite{AblowitzLuoMusslimani}.

Note that (\ref{E:linear combination 1}) and (\ref{E:linear combination 2}) can be written as
\begin{equation}
\mu(x,k)=\rho(k)e^{2i\lambda x}N(x,k)+\overline{N}(x,k)
\end{equation}
and
\begin{equation}
\overline{\mu}(x,k)=N(x,k)+\overline{\rho}(k)e^{-2i\lambda x}\overline{N}(x,k),
\end{equation}
where $\mu(x,k)=M(x,k)a^{-1}(k)$, $\overline{\mu}(x,k)=\overline{M}(x,k)\overline{a}^{-1}(k)$, $\rho(k)=b(k)a^{-1}(k)$ and $\overline{\rho}(k)=\overline{b}(k)\overline{a}^{-1}(k)$. Introducing the $2\times2$ matrices
\begin{equation}
m_{+}(x,k)=(\mu(x,k), N(x,k)), \ \ \ m_{-}(x,k)=(\overline{N}(x,k), \overline{\mu}(x,k)),
\end{equation}
which are meromorphic in $\mathbb{C}_{1}$ and $\mathbb{C}_{2}$ respectively. Hence, we can write the Riemann-Hilbert problem or `jump' conditions in the $k$-plane as
\begin{equation}
m_{+}(x,k)-m_{-}(x,k)=m_{-}(x,k)\left(\begin{array}{cc}
-\rho(k)\overline{\rho}(k)& -\overline{\rho}(k)e^{-2i\lambda x}\\
\rho(k)e^{2i\lambda x}& 0
\end{array}\right)
\end{equation}
on the contour $\Sigma: k \in(-\infty,-q_{0}]\cup [q_{0},+\infty)$.
\subsection{Symmetry reductions}
The symmetry in the potential induces a symmetry between the eigenfunctions. Indeed, if $v(x,k)=(v_{1}(x,k), v_{2}(x,k))^{T}$ solves
(\ref{E:nonlocal scattering}), then $(v_{2}(-x,k), -v_{1}(-x,k))^{T}$ also solves (\ref{E:nonlocal scattering}). Taking into account boundary conditions (\ref{E:boundary conditions 1}) and (\ref{E:boundary conditions 2}), we can obtain
\begin{equation}
\psi(x,k)=\left(\begin{array}{cc}
0& -1\\
1& 0
\end{array}\right)\phi(-x,k)
\end{equation}
and
\begin{equation}
\overline{\psi}(x,k)=\left(\begin{array}{cc}
0& 1\\
-1& 0
\end{array}\right)\overline{\phi}(-x,k).
\end{equation}
By (\ref{E:definition 1}) and (\ref{E:definition 2}), we can get the symmetry relations of the eigenfunctions, i.e.,
\begin{equation}
N(x,k)=\left(\begin{array}{cc}
0& -1\\
1& 0
\end{array}\right)M(-x,k)
\end{equation}
and
\begin{equation}
\overline{N}(x,k)=\left(\begin{array}{cc}
0& 1\\
-1& 0
\end{array}\right)\overline{M}(-x,k).
\end{equation}
From the Wronskian representations for the scattering data and the above symmetry relations, we have
\begin{equation}
\overline{b}(k)=b(k).
\end{equation}
\subsection{Uniformization coordinates}
Before discussing the properties of scattering data and solving the inverse problem, we introduce a uniformization variable $z$, defined by the conformal mapping:
\begin{equation}
z=z(k)=k+\lambda(k),
\end{equation}
where $\lambda= \sqrt{k^2-q_0^2}$ and the inverse mapping is given by
\begin{equation}
k=k(z)=\frac{1}{2}\left(z+\frac{q_{0}^{2}}{z}\right).
\end{equation}
Then
\begin{equation}
\lambda(z)=\frac{1}{2}\left(z-\frac{q_{0}^{2}}{z}\right).
\end{equation}
We observe that

(1) the upper sheet $\mathbb{C}_{1}$ and lower sheet $\mathbb{C}_{2}$ of the Riemann surface $\mathbb{K}$ are mapped onto the upper half plane $\mathbb{C}^{+}$ and lower half plane $\mathbb{C}^{-}$ of the complex $z-$plane respectively;

(2) the cut $(-\infty, -q_{0}]\cup [q_{0}, +\infty)$ on the Riemann surface is mapped onto the real $z$ axis;

(3) the segments $[-q_{0}, q_{0}]$ on $\mathbb{C}_{1}$ and $\mathbb{C}_{2}$ are mapped onto the upper and lower semicircles of radius $q_{0}$ and centered at the origin of the complex $z-$plane respectively.

From Theorem \ref{T:1}, we have the eigenfunctions $M(x,z)$ and $N(x,z)$ are analytic in the upper half $z-$ plane: i.e  $ z \in \mathbb{C}^{+}$, and $\overline{M}(x,z)$ and $\overline{N}(x,z)$ are analytic in the lower half $z-$ plane: i.e. $z \in \mathbb{C}^{-}$. Moreover, by Theorem \ref{T:2},
we find that $a(z)$ is analytic in the upper half $z-$ plane:  $z \in \mathbb{C}^{+}$ and $\overline{a}(z)$ is analytic in the lower half plane: $z \in \mathbb{C}^{-}$.
\subsection{Symmetries via uniformization coordinates}
From the eigenfunction symmetries above we have

\begin{equation}
\psi(x,z)=\left(\begin{array}{cc}
0& -1\\
1& 0
\end{array}\right)\phi(-x,z),
\end{equation}
\begin{equation}
\overline{\psi}(x,z)=\left(\begin{array}{cc}
0& 1\\
-1& 0
\end{array}\right)\overline{\phi}(-x,z).
\end{equation}
Further, when $z\rightarrow \frac{q_{0}^{2}}{z}$, then $(k, \lambda)\rightarrow (k, -\lambda)$. Hence,

\begin{equation}
\phi\left(x,\frac{q_{0}^{2}}{z}\right)=\frac{\frac{q_{0}^{2}}{z}}{-iq_{-}}\cdot\overline{\phi}(x,z), \ \ \ \psi\left(x,\frac{q_{0}^{2}}{z}\right)=\frac{-iq_{+}}{z}\cdot\overline{\psi}(x,z), \ \ \ \Im z<0.
\end{equation}
Similarly, we can get
\begin{equation}
N(x,z)=\left(\begin{array}{cc}
0& -1\\
1& 0
\end{array}\right)M(-x,z),
\end{equation}
\begin{equation}
\overline{N}(x,z)=\left(\begin{array}{cc}
0& 1\\
-1& 0
\end{array}\right)\overline{M}(-x,z),
\end{equation}
\begin{equation}
\overline{b}(z)=b(z),
\end{equation}
\begin{equation}
a\left(\frac{q_{0}^{2}}{z}\right)=e^{2i\theta_{+}}\overline{a}(z),\ \ \ \Im z<0; \ \ \ b\left(\frac{q_{0}^{2}}{z}\right)=-\overline{b}(z).
\end{equation}
\subsection{Asymptotic behavior of eigenfunctions and scattering data}
In order to solve the inverse problem, one has to determine the asymptotic behavior of eigenfunctions and scattering data both as $z\rightarrow\infty$ and as $z\rightarrow 0$. From the integral equations (in terms of Green's functions), we have
\begin{equation}
M(x,z)\sim\left\{\begin{array}{ll}
\left(\begin{array}{cc}
z\\
iq(-x)
\end{array}\right), \ \ \ z\rightarrow\infty\\
\left(\begin{array}{cc}
z\cdot\frac{q(x)}{q_{-}}\\
iq_{+}
\end{array}\right), \ \ \ z\rightarrow 0,\\
\end{array}\right.
\end{equation}

\begin{equation}
N(x,z)\sim\left\{\begin{array}{ll}
\left(\begin{array}{cc}
-iq(x)\\
z
\end{array}\right), \ \ \ z\rightarrow\infty\\
\left(\begin{array}{cc}
-iq_{+}\\
z\cdot \frac{q(-x)}{q_{-}}
\end{array}\right), \ \ \ z\rightarrow 0,\\
\end{array}\right.
\end{equation}

\begin{equation}
\overline{M}(x,z)\sim\left\{\begin{array}{ll}
\left(\begin{array}{cc}
-iq(x)\\
z
\end{array}\right), \ \ \ z\rightarrow\infty\\
\left(\begin{array}{cc}
-iq_{-}\\
z\cdot \frac{q(-x)}{q_{+}}
\end{array}\right), \ \ \ z\rightarrow 0,\\
\end{array}\right.
\end{equation}

\begin{equation}
\label{E:asymptotic 3}
\overline{N}(x,z)\sim\left\{\begin{array}{ll}
\left(\begin{array}{cc}
z\\
iq(-x)
\end{array}\right), \ \ \ z\rightarrow\infty\\
\left(\begin{array}{cc}
z\cdot\frac{q(x)}{q_{+}}\\
iq_{-}
\end{array}\right), \ \ \ z\rightarrow 0,\\
\end{array}\right.
\end{equation}

\begin{equation}
a(z)=
\left\{\begin{array}{ll}
1,\ \ \ z\rightarrow\infty,\\
e^{2i\theta_{+}}, \ \ \ z\rightarrow 0,\\
\end{array}\right.
\end{equation}

\begin{equation}
\overline{a}(z)=
\left\{\begin{array}{ll}
1,\ \ \ z\rightarrow\infty,\\
e^{-2i\theta_{+}}, \ \ \ z\rightarrow 0,\\
\end{array}\right.
\end{equation}
\begin{equation}
\lim_{z\rightarrow\infty}zb(z)=0, \ \ \ \lim_{z\rightarrow0}\frac{b(z)}{z^{2}}=0.
\end{equation}
\subsection{Riemann-Hilbert problem via uniformization coordinates}
\subsubsection{Left scattering problem}
In order to take into account the behavior of the eigenfunctions, the `jump' conditions at the real $z-$ axis can be written from the left end as
\begin{equation}
\frac{M(x,z)}{za(z)}-\frac{\overline{N}(x,z)}{z}=\rho(z)e^{i\big(z-\frac{q_{0}^{2}}{z}\big)x}\cdot \frac{N(x,z)}{z}
\end{equation}
and
\begin{equation}
\frac{\overline{M}(x,z)}{z\overline{a}(z)}-\frac{N(x,z)}{z}=\overline{\rho}(z)e^{-i\big(z-\frac{q_{0}^{2}}{z}\big)x}\cdot \frac{\overline{N}(x,z)}{z},
\end{equation}
so that the functions will be bounded at infinity, though having an additional pole at $z=0$.
Note that $M(x,z)/a(z)$, as a function of
$z$, is defined in the upper half plane $\mathbb{C}^{+}$, where it has (by assumption) simple poles $z_{j}$, i.e., $a(z_{j})=0$, and $\overline{M}(x,z)/\overline{a}(z)$,
is defined in the lower half plane $\mathbb{C}^{-}$, where it has simple poles $\overline{z}_{j}$, i.e., $\overline{a}(\overline{z}_{j})=0$. At zeros of $a,\overline{a}$, we have
\begin{equation}
\label{E:M}
M(x,z_{j})=b(z_{j})e^{i\big(z_{j}-\frac{q_{0}^{2}}{z_{j}}\big)x}\cdot N(x,z_{j})
\end{equation}
and
\begin{equation}
\overline{M}(x,\overline{z}_{j})=\overline{b}(\overline{z}_{j})e^{-i\big(\overline{z}_{j}-\frac{q_{0}^{2}}{\overline{z}_{j}}\big)x}\cdot \overline{N}(x,\overline{z}_{j}).
\end{equation}
Then subtracting the values at infinity, the induced pole at the origin and the poles, assumed simple, in the upper/lower half planes respectively, at $a(z_j)=0, j=1,2...J$ and  $\bar{a}(\overline{z}_{j}), j=1,2...\bar{J} $ (later we will see that $J=\bar{J}$) gives
\begin{equation}
\label{E:jump 1}
\begin{split}
&\left[\frac{M(x,z)}{za(z)}-\left(\begin{array}{cc}
1\\
0
\end{array}\right)-
\frac{1}{z}\left(\begin{array}{cc}
0\\
iq_{-}
\end{array}\right)
-\sum_{j=1}^{J}\frac{M(x,z_{j})}{(z-z_{j})z_{j}a'(z_{j})}\right]\\
&-\left[\frac{\overline{N}(x,z)}{z}-\left(\begin{array}{cc}
1\\
0
\end{array}\right)-
\frac{1}{z}\left(\begin{array}{cc}
0\\
iq_{-}
\end{array}\right)
-\sum_{j=1}^{J}\frac{b(z_{j})e^{i\big(z_{j}-\frac{q_{0}^{2}}{z_{j}}\big)x}\cdot N(x,z_{j})}{(z-z_{j})z_{j}a'(z_{j})}\right]\\
&=\rho(z)e^{i\big(z-\frac{q_{0}^{2}}{z}\big)x}\cdot \frac{N(x,z)}{z}
\end{split}
\end{equation}
and
\begin{equation}
\label{E:jump 2}
\begin{split}
&\left[\frac{\overline{M}(x,z)}{z\overline{a}(z)}-\left(\begin{array}{cc}
0\\
1
\end{array}\right)-
\frac{1}{z}\left(\begin{array}{cc}
-iq_{+}\\
0
\end{array}\right)
-\sum_{j=1}^{\overline{J}}\frac{\overline{M}(x,\overline{z}_{j})}{(z-\overline{z}_{j})\overline{z}_{j}a'(\overline{z}_{j})}\right]\\
&-\left[\frac{N(x,z)}{z}-\left(\begin{array}{cc}
0\\
1
\end{array}\right)-
\frac{1}{z}\left(\begin{array}{cc}
-iq_{+}\\
0
\end{array}\right)
-\sum_{j=1}^{\overline{J}}\frac{\overline{b}(\overline{z}_{j})e^{-i\big(\overline{z}_{j}-\frac{q_{0}^{2}}{\overline{z}_{j}}\big)x}\cdot \overline{N}(x,\overline{z}_{j})}{(z-\overline{z}_{j})\overline{z}_{j}\overline{a}'(\overline{z}_{j})}\right]\\
&=\overline{\rho}(z)e^{-i\big(z-\frac{q_{0}^{2}}{z}\big)x}\cdot \frac{\overline{N}(x,z)}{z}.
\end{split}
\end{equation}
We now introduce the projection operators
\begin{equation}
P_{\pm}(f)(z)=\frac{1}{2\pi i}\int_{-\infty}^{+\infty}\frac{f(\xi)}{\xi-(z\pm i0)}d\xi,
\end{equation}
which are well-defined for any function $f(\xi)$ that is integrable on the real axis. If $f_{\pm}(\xi)$ is analytic in the upper/lower $z-$ plane and $f_{\pm}(\xi)$ is decaying at large $\xi$, then
\begin{equation}
P_{\pm}(f_{\pm})(z)=\pm f_{\pm}(z), \ \ \ P_{\mp}(f_{\pm})(z)=0.
\end{equation}
Applying $P_{-}$ to (\ref{E:jump 1}) and $P_{+}$ to (\ref{E:jump 2}), we can obtain
\begin{equation}
\begin{split}
\frac{\overline{N}(x,z)}{z}&=\left(\begin{array}{cc}
1\\
0
\end{array}\right)+
\frac{1}{z}\left(\begin{array}{cc}
0\\
iq_{-}
\end{array}\right)
+\sum_{j=1}^{J}\frac{b(z_{j})e^{i\big(z_{j}-\frac{q_{0}^{2}}{z_{j}}\big)x}\cdot N(x,z_{j})}{(z-z_{j})z_{j}a'(z_{j})}\\
&+\frac{1}{2\pi i}\int_{-\infty}^{+\infty}\frac{\rho(\xi)}{\xi(\xi-z)}\cdot e^{i\big(\xi-\frac{q_{0}^{2}}{\xi}\big)x}\cdot N(x,\xi)d\xi
\end{split}
\end{equation}
and
\begin{equation}
\begin{split}
\frac{N(x,z)}{z}&=\left(\begin{array}{cc}
0\\
1
\end{array}\right)+
\frac{1}{z}\left(\begin{array}{cc}
-iq_{+}\\
0
\end{array}\right)
+\sum_{j=1}^{\overline{J}}\frac{\overline{b}(\overline{z}_{j})e^{-i\big(\overline{z}_{j}-\frac{q_{0}^{2}}{\overline{z}_{j}}\big)x}\cdot \overline{N}(x,\overline{z}_{j})}{(z-\overline{z}_{j})\overline{z}_{j}\overline{a}'(\overline{z}_{j})}\\
&-\frac{1}{2\pi i}\int_{-\infty}^{+\infty}\frac{\overline{\rho}(\xi)}{\xi(\xi-z)}\cdot e^{-i\big(\xi-\frac{q_{0}^{2}}{\xi}\big)x}\cdot \overline{N}(x,\xi)d\xi,
\end{split}
\end{equation}
i.e.,
\begin{equation}
\label{E:eigenfunction 1}
\begin{split}
\overline{N}(x,z)&=\left(\begin{array}{cc}
z\\
iq_{-}
\end{array}\right)
+\sum_{j=1}^{J}\frac{z\cdot b(z_{j})e^{i\big(z_{j}-\frac{q_{0}^{2}}{z_{j}}\big)x}\cdot N(x,z_{j})}{(z-z_{j})z_{j}a'(z_{j})}\\
&+\frac{z}{2\pi i}\int_{-\infty}^{+\infty}\frac{\rho(\xi)}{\xi(\xi-z)}\cdot e^{i\big(\xi-\frac{q_{0}^{2}}{\xi}\big)x}\cdot N(x,\xi)d\xi
\end{split}
\end{equation}
and
\begin{equation}
\label{E:eigenfunction 2}
\begin{split}
N(x,z)&=\left(\begin{array}{cc}
-iq_{+}\\
z
\end{array}\right)
+\sum_{j=1}^{\overline{J}}\frac{z\cdot\overline{b}(\overline{z}_{j})e^{-i\big(\overline{z}_{j}-\frac{q_{0}^{2}}{\overline{z}_{j}}\big)x}\cdot \overline{N}(x,\overline{z}_{j})}{(z-\overline{z}_{j})\overline{z}_{j}\overline{a}'(\overline{z}_{j})}\\
&-\frac{z}{2\pi i}\int_{-\infty}^{+\infty}\frac{\overline{\rho}(\xi)}{\xi(\xi-z)}\cdot e^{-i\big(\xi-\frac{q_{0}^{2}}{\xi}\big)x}\cdot \overline{N}(x,\xi)d\xi.
\end{split}
\end{equation}

Since the symmetries are between eigenfunctions defined at both $\pm \infty$ we proceed to obtain the inverse scattering integral equations defined from the right end.

\subsubsection{Right scattering problem}
The right scattering problem can be written as
\begin{equation}
\psi(x,z)=\alpha(z)\overline{\phi}(x,z)+\beta(z)\phi(x,z)
\end{equation}
and
\begin{equation}
\overline{\psi}(x,z)=\overline{\alpha}(z)\phi(x,z)+\overline{\beta}(z)\overline{\phi}(x,z),
\end{equation}
where $\alpha(z)$, $\overline{\alpha}(z)$, $\beta(z)$ and $\overline{\beta}(z)$ are the right scattering data. Moreover, we can get the right scattering data and left scattering data satisfy the following relations
\begin{equation}
\overline{\alpha}(z)=\overline{a}(z), \ \ \ \alpha(z)=a(z), \ \ \ \overline{\beta}(z)=-b(z), \ \ \ \beta(z)=-\overline{b}(z).
\end{equation}
Thus,
\begin{equation}
\begin{split}
N(x,z)&=\alpha(z)\overline{M}(x,z)+\beta(z)M(x,z)e^{-i\big(z-\frac{q_{0}^{2}}{z}\big)x}\\
&=a(z)\overline{M}(x,z)-\overline{b}(z)M(x,z)e^{-i\big(z-\frac{q_{0}^{2}}{z}\big)x}
\end{split}
\end{equation}
and
\begin{equation}
\begin{split}
\overline{N}(x,z)&=\overline{\alpha}(z)M(x,z)+\overline{\beta}(z)\overline{M}(x,z)e^{i\big(z-\frac{q_{0}^{2}}{z}\big)x}\\
&=\overline{a}(z)M(x,z)-b(z)\overline{M}(x,z)e^{i\big(z-\frac{q_{0}^{2}}{z}\big)x}.
\end{split}
\end{equation}
The above two equations can be rewritten as
\begin{equation}
\frac{N(x,z)}{za(z)}-\frac{\overline{M}(x,z)}{z}=-\frac{\overline{b}(z)}{a(z)}\cdot e^{-i\big(z-\frac{q_{0}^{2}}{z}\big)x}\cdot \frac{M(x,z)}{z}
\end{equation}
and
\begin{equation}
\frac{\overline{N}(x,z)}{z\overline{a}(z)}-\frac{M(x,z)}{z}=-\frac{b(z)}{\overline{a}(z)}\cdot e^{i\big(z-\frac{q_{0}^{2}}{z}\big)x}\cdot \frac{\overline{M}(x,z)}{z}.
\end{equation}
By the symmetry relations of scattering data, we have
\begin{equation}
\frac{N(x,z)}{za(z)}-\frac{\overline{M}(x,z)}{z}=\rho^{*}(-z^{*})\cdot e^{-i\big(z-\frac{q_{0}^{2}}{z}\big)x}\cdot \frac{M(x,z)}{z}
\end{equation}
and
\begin{equation}
\frac{\overline{N}(x,z)}{z\overline{a}(z)}-\frac{M(x,z)}{z}=\overline{\rho}^{*}(-z^{*})\cdot e^{i\big(z-\frac{q_{0}^{2}}{z}\big)x}\cdot \frac{\overline{M}(x,z)}{z},
\end{equation}
so that the functions will be bounded at infinity, though having an additional pole at $z=0$. Note that $N(x,z)/a(z)$, as a function of
$z$, is defined in the upper half plane $\mathbb{C}^{+}$, where it has simple poles $z_{j}$, i.e., $a(z_{j})=0$, and $\overline{N}(x,z)/\overline{a}(z)$,
is defined in the lower half plane $\mathbb{C}^{-}$, where it has simple poles $\overline{z}_{j}$, i.e., $\overline{a}(\overline{z}_{j})=0$. At the zeros of $a,\overline{a}$,
\begin{equation}
N(x,z_{j})=-\overline{b}(z_{j})M(x,z_{j})e^{-i\big(z_{j}-\frac{q_{0}^{2}}{z_{j}}\big)x}
\end{equation}
and
\begin{equation}
\overline{N}(x,\overline{z}_{j})=-b(\overline{z}_{j})\overline{M}(x,\overline{z}_{j})e^{i\big(\overline{z}_{j}-\frac{q_{0}^{2}}{\overline{z}_{j}}\big)x}.
\end{equation}
Then, as before, subtracting the values at infinity, the induced pole at the origin and the poles, assumed simple, in the upper/lower half planes respectively, at $a(z_j)=0, j=1,2...J$ and  $\bar{a}(\overline{z}_{j}), j=1,2...\bar{J} $, gives
\begin{equation}
\label{E:jump 3}
\begin{split}
&\left[\frac{N(x,z)}{za(z)}-\left(\begin{array}{cc}
0\\
1
\end{array}\right)
-\frac{1}{z}\left(\begin{array}{cc}
-iq_{-}\\
0
\end{array}\right)
-\sum_{j=1}^{J}\frac{N(x,z_{j})}{(z-z_{j})z_{j}a'(z_{j})}\right]\\
&-\left[\frac{\overline{M}(x,z)}{z}-\left(\begin{array}{cc}
0\\
1
\end{array}\right)
-\frac{1}{z}\left(\begin{array}{cc}
-iq_{-}\\
0
\end{array}\right)-\sum_{j=1}^{J}\frac{-\overline{b}(z_{j})M(x,z_{j})e^{-i\big(z_{j}-\frac{q_{0}^{2}}{z_{j}}\big)x}}{(z-z_{j})z_{j}a'(z_{j})}\right]\\
&=\rho^{*}(-z^{*})\cdot e^{-i\big(z-\frac{q_{0}^{2}}{z}\big)x}\cdot \frac{M(x,z)}{z}
\end{split}
\end{equation}
and
\begin{equation}
\label{E:jump 4}
\begin{split}
&\left[\frac{\overline{N}(x,z)}{z\overline{a}(z)}-\left(\begin{array}{cc}
1\\
0
\end{array}\right)
-\frac{1}{z}\left(\begin{array}{cc}
0\\
iq_{+}
\end{array}\right)
-\sum_{j=1}^{\overline{J}}\frac{\overline{N}(x,\overline{z}_{j})}{(z-\overline{z}_{j})\overline{z}_{j}\overline{a}'(\overline{z}_{j})}\right]\\
&-\left[\frac{M(x,z)}{z}-\left(\begin{array}{cc}
1\\
0
\end{array}\right)
-\frac{1}{z}\left(\begin{array}{cc}
0\\
iq_{+}
\end{array}\right)-\sum_{j=1}^{\overline{J}}\frac{-b(\overline{z}_{j})\overline{M}(x,\overline{z}_{j})e^{i\big(\overline{z}_{j}-\frac{q_{0}^{2}}{\overline{z}_{j}}\big)x}}
{(z-\overline{z}_{j})\overline{z}_{j}\overline{a}'(\overline{z}_{j})}\right]\\
&=\overline{\rho}^{*}(-z^{*})\cdot e^{i\big(z-\frac{q_{0}^{2}}{z}\big)x}\cdot \frac{\overline{M}(x,z)}{z}.
\end{split}
\end{equation}
Applying $P_{-}$ to (\ref{E:jump 3}) and $P_{+}$ to (\ref{E:jump 4}), we can obtain
\begin{equation}
\label{E:eigenfunction 3}
\begin{split}
\overline{M}(x,z)&=\left(\begin{array}{cc}
-iq_{-}\\
z
\end{array}\right)+
\sum_{j=1}^{J}\frac{-z\cdot\overline{b}(z_{j})M(x,z_{j})e^{-i\big(z_{j}-\frac{q_{0}^{2}}{z_{j}}\big)x}}{(z-z_{j})z_{j}a'(z_{j})}\\
&+\frac{z}{2\pi i}\int_{-\infty}^{+\infty}\frac{\rho^{*}(-\xi)}{\xi(\xi-z)}\cdot e^{-i\big(\xi-\frac{q_{0}^{2}}{\xi}\big)x}\cdot M(x,\xi)d\xi
\end{split}
\end{equation}
and
\begin{equation}
\label{E:eigenfunction 4}
\begin{split}
M(x,z)&=\left(\begin{array}{cc}
z\\
iq_{+}
\end{array}\right)+
\sum_{j=1}^{\overline{J}}\frac{-z\cdot b(\overline{z}_{j})\overline{M}(x,\overline{z}_{j})e^{i\big(\overline{z}_{j}-\frac{q_{0}^{2}}{\overline{z}_{j}}\big)x}}
{(z-\overline{z}_{j})\overline{z}_{j}\overline{a}'(\overline{z}_{j})}\\
&-\frac{z}{2\pi i}\int_{-\infty}^{+\infty}\frac{\overline{\rho}^{*}(-\xi)}{\xi(\xi-z)}\cdot e^{i\big(\xi-\frac{q_{0}^{2}}{\xi}\big)x}\cdot \overline{M}(x,\xi)d\xi.
\end{split}
\end{equation}
\subsection{Recovery of the potentials}
In order to reconstruct the potential, we use asymptotics . For example from equation (\ref{E:asymptotic 3}), we have
\begin{equation}
\frac{\overline{N}_{1}(x,z)}{z}\sim \frac{q(x)}{q_{+}}
\end{equation}
as $z\rightarrow 0$. By (\ref{E:eigenfunction 1}), we can get
\begin{equation}
\frac{\overline{N}_{1}(x,z)}{z}\sim 1+\sum_{j=1}^{J}\frac{b(z_{j})e^{i\big(z_{j}-\frac{q_{0}^{2}}{z_{j}}\big)x}}{-z_{j}^{2}a'(z_{j})}\cdot N_{1}(x,z_{j})+\frac{1}{2\pi i}\int_{-\infty}^{+\infty}\frac{\rho(\xi)}{\xi^{2}}\cdot e^{i\big(\xi-\frac{q_{0}^{2}}{\xi}\big)x}\cdot N_{1}(x,\xi)d\xi
\end{equation}
as $z\rightarrow 0$. Hence,
\begin{equation}
\label{asympN1c}
q(x)=q_{+}\cdot\left[1+\sum_{j=1}^{J}\frac{b(z_{j})e^{i\big(z_{j}-\frac{q_{0}^{2}}{z_{j}}\big)x}}{-z_{j}^{2}a'(z_{j})}\cdot N_{1}(x,z_{j})+\frac{1}{2\pi i}\int_{-\infty}^{+\infty}\frac{\rho(\xi)}{\xi^{2}}\cdot e^{i\big(\xi-\frac{q_{0}^{2}}{\xi}\big)x}\cdot N_{1}(x,\xi)d\xi\right].
\end{equation}
\subsection{Closing the system}
We can find $J=\overline{J}$ from $a\left(\frac{q_{0}^{2}}{z}\right)=e^{2i\theta_{+}}\overline{a}(z)$. By combining the prior integral equations we find
\begin{equation}
\label{E:closing system 1}
\begin{split}
&\left(\begin{array}{cc}
N_{1}(x,z)\\
N_{2}(x,z)
\end{array}\right)=\left(\begin{array}{cc}
-iq_{+}\\
z
\end{array}\right)
+\sum_{j=1}^{J}\frac{z\cdot\overline{b}(\overline{z}_{j})e^{-i\big(\overline{z}_{j}-\frac{q_{0}^{2}}{\overline{z}_{j}}\big)x} }{(z-\overline{z}_{j})\overline{z}_{j}\overline{a}'(\overline{z}_{j})}\cdot\\
&\left(\begin{array}{cc}
\overline{z}_{j}+\sum_{l=1}^{J}\frac{\overline{z}_{j}\cdot b(z_{l})e^{i\big(z_{l}-\frac{q_{0}^{2}}{z_{l}}\big)x}}{(\overline{z}_{j}-z_{l})z_{l}a'(z_{l})}\cdot N_{1}(x, z_{l})+\frac{\overline{z}_{j}}{2\pi i}\int_{-\infty}^{+\infty}\frac{\rho(\xi)}{\xi(\xi-\overline{z}_{j})}\cdot e^{i\big(\xi-\frac{q_{0}^{2}}{\xi}\big)x}\cdot N_{1}(x, \xi)d\xi\\
iq_{-}+\sum_{l=1}^{J}\frac{\overline{z}_{j}\cdot b(z_{l})e^{i\big(z_{l}-\frac{q_{0}^{2}}{z_{l}}\big)x}}{(\overline{z}_{j}-z_{l})z_{l}a'(z_{l})}\cdot N_{2}(x, z_{l})+\frac{\overline{z}_{j}}{2\pi i}\int_{-\infty}^{+\infty}\frac{\rho(\xi)}{\xi(\xi-\overline{z}_{j})}\cdot e^{i\big(\xi-\frac{q_{0}^{2}}{\xi}\big)x}\cdot N_{2}(x, \xi)d\xi
\end{array}\right)\\
&-\frac{z}{2\pi i}\int_{-\infty}^{+\infty}\frac{\overline{\rho}(\xi)}{\xi(\xi-z)}\cdot e^{-i\big(\xi-\frac{q_{0}^{2}}{\xi}\big)x}\cdot\\
&\left(\begin{array}{cc}
\xi+\sum_{l=1}^{J}\frac{\xi\cdot b(z_{l})e^{i\big(z_{l}-\frac{q_{0}^{2}}{z_{l}}\big)x}}{(\xi-z_{l})z_{l}a'(z_{l})}\cdot N_{1}(x, z_{l})+\frac{\xi}{2\pi i}\int_{-\infty}^{+\infty}\frac{\rho(\eta)}{\eta(\eta-\xi)}\cdot e^{i\big(\eta-\frac{q_{0}^{2}}{\eta}\big)x}\cdot N_{1}(x, \eta)d\eta\\
iq_{-}+\sum_{l=1}^{J}\frac{\xi\cdot b(z_{l})e^{i\big(z_{l}-\frac{q_{0}^{2}}{z_{l}}\big)x}}{(\xi-z_{l})z_{l}a'(z_{l})}\cdot N_{2}(x, z_{l})+\frac{\xi}{2\pi i}\int_{-\infty}^{+\infty}\frac{\rho(\eta)}{\eta(\eta-\xi)}\cdot e^{i\big(\eta-\frac{q_{0}^{2}}{\eta}\big)x}\cdot N_{2}(x, \eta)d\eta
\end{array}\right)d\xi,
\end{split}
\end{equation}
\begin{equation}
\label{E:closing system 2}
\begin{split}
&\left(\begin{array}{cc}
\overline{M}_{1}(x,z)\\
\overline{M}_{2}(x,z)
\end{array}\right)=\left(\begin{array}{cc}
-iq_{-}\\
z
\end{array}\right)+
\sum_{j=1}^{J}\frac{-z\cdot\overline{b}(z_{j})e^{-i\big(z_{j}-\frac{q_{0}^{2}}{z_{j}}\big)x}}{(z-z_{j})z_{j}a'(z_{j})}\cdot\\
& \left(\begin{array}{cc}
z_{j}+
\sum_{l=1}^{J}\frac{-z_{j}\cdot b(\overline{z}_{l})e^{i\big(\overline{z}_{l}-\frac{q_{0}^{2}}{\overline{z}_{l}}\big)x}}
{(z_{j}-\overline{z}_{l})\overline{z}_{l}\overline{a}'(\overline{z}_{l})}\cdot \overline{M}_{1}(x, \overline{z}_{l})-\frac{z_{j}}{2\pi i}\int_{-\infty}^{+\infty}\frac{\overline{\rho}^{*}(-\xi)}{\xi(\xi-z_{j})}\cdot e^{i\big(\xi-\frac{q_{0}^{2}}{\xi}\big)x}\cdot\overline{M}_{1}(x,\xi)d\xi\\
iq_{+}+
\sum_{l=1}^{J}\frac{-z_{j}\cdot b(\overline{z}_{l})e^{i\big(\overline{z}_{l}-\frac{q_{0}^{2}}{\overline{z}_{l}}\big)x}}
{(z_{j}-\overline{z}_{l})\overline{z}_{l}\overline{a}'(\overline{z}_{l})}\cdot \overline{M}_{2}(x, \overline{z}_{l})-\frac{z_{j}}{2\pi i}\int_{-\infty}^{+\infty}\frac{\overline{\rho}^{*}(-\xi)}{\xi(\xi-z_{j})}\cdot e^{i\big(\xi-\frac{q_{0}^{2}}{\xi}\big)x}\cdot\overline{M}_{2}(x,\xi)d\xi
\end{array}\right)\\
&+\frac{z}{2\pi i}\int_{-\infty}^{+\infty}\frac{\rho^{*}(-\xi)}{\xi(\xi-z)}\cdot e^{-i\big(\xi-\frac{q_{0}^{2}}{\xi}\big)x}\cdot\\
&\left(\begin{array}{cc}
\xi+
\sum_{l=1}^{J}\frac{-\xi\cdot b(\overline{z}_{l})e^{i\big(\overline{z}_{l}-\frac{q_{0}^{2}}{\overline{z}_{l}}\big)x}}
{(\xi-\overline{z}_{l})\overline{z}_{l}\overline{a}'(\overline{z}_{l})}\cdot\overline{M}_{1}(x,\overline{z}_{l})-\frac{\xi}{2\pi i}\int_{-\infty}^{+\infty}\frac{\overline{\rho}^{*}(-\eta)}{\eta(\eta-\xi)}\cdot e^{i\big(\eta-\frac{q_{0}^{2}}{\eta}\big)x}\cdot \overline{M}_{1}(x,\eta)d\eta\\
iq_{+}+
\sum_{l=1}^{J}\frac{-\xi\cdot b(\overline{z}_{l})e^{i\big(\overline{z}_{l}-\frac{q_{0}^{2}}{\overline{z}_{l}}\big)x}}
{(\xi-\overline{z}_{l})\overline{z}_{l}\overline{a}'(\overline{z}_{l})}\cdot\overline{M}_{2}(x,\overline{z}_{l})-\frac{\xi}{2\pi i}\int_{-\infty}^{+\infty}\frac{\overline{\rho}^{*}(-\eta)}{\eta(\eta-\xi)}\cdot e^{i\big(\eta-\frac{q_{0}^{2}}{\eta}\big)x}\cdot \overline{M}_{2}(x,\eta)d\eta
\end{array}\right)d\xi.
\end{split}
\end{equation}
The potential (\ref{asympN1c}) can be reconstructed from the solution of the integral equation (\ref{E:closing system 1}).
\subsection{Trace formula}
We have shown that $a(z)$ and $\overline{a}(z)$ are analytic in the upper and lower $z-$plane respectively. As mentioned above, we assume that $a(z)$ has simple zeros, which we call
$z_{j}$. By the symmetry relation $a\left(\frac{q_{0}^{2}}{z}\right)=e^{2i\theta_{+}}\overline{a}(z)$, we can deduce that $\overline{a}(z)$ has simple zeros $\frac{q_{0}^{2}}{z_{j}}$.

We define
\begin{equation}
\gamma(z)=a(z)\cdot \prod_{j=1}^{J} \frac{z-\frac{q_{0}^{2}}{z_{j}}}{z-z_{j}},
\end{equation}
\begin{equation}
\overline{\gamma}(z)=\overline{a}(z)\cdot
\prod_{j=1}^{J}\frac{z-z_{j}}{z-\frac{q_{0}^{2}}{z_{j}}}.
\end{equation}

Then $\gamma(z)$ and $\overline{\gamma}(z)$ are analytic in the upper and lower $z-$plane respectively and have no zeros in their respective half planes. We can get
\begin{equation}
\log \gamma(z)=\frac{1}{2\pi i}\int_{-\infty}^{+\infty}\frac{\log \gamma(\xi)}{\xi-z}d\xi, \ \ \
\frac{1}{2\pi i}\int_{-\infty}^{+\infty}\frac{\log \overline{\gamma}(\xi)}{\xi-z}d\xi=0, \ \ \ \Im z>0,
\end{equation}

\begin{equation}
\log \overline{\gamma}(z)=-\frac{1}{2\pi i}\int_{-\infty}^{+\infty}\frac{\log \overline{\gamma}(\xi)}{\xi-z}d\xi, \ \ \
\frac{1}{2\pi i}\int_{-\infty}^{+\infty}\frac{\log \gamma(\xi)}{\xi-z}d\xi=0, \ \ \ \Im z<0.
\end{equation}
Adding or subtracting the above equations in each half plane respectively, we obtain
\begin{equation}
\log \gamma(z)=\frac{1}{2\pi i}\int_{-\infty}^{+\infty}\frac{\log \gamma(\xi)\overline{\gamma}(\xi)}{\xi-z}d\xi, \ \ \ \Im z>0,
\end{equation}
\begin{equation}
\log \overline{\gamma}(z)=-\frac{1}{2\pi i}\int_{-\infty}^{+\infty}\frac{\log \gamma(\xi)\overline{\gamma}(\xi)}{\xi-z}d\xi, \ \ \ \Im z<0.
\end{equation}
Hence,
\begin{equation}
\log a(z)=\log\left(\prod_{j=1}^{J}\frac{z-z_{j}}{z-\frac{q_{0}^{2}}{z_{j}}}\right)+\frac{1}{2\pi i}\int_{-\infty}^{+\infty}\frac{\log \gamma(\xi)\overline{\gamma}(\xi)}{\xi-z}d\xi, \ \ \ \Im z>0,
\end{equation}

\begin{equation}
\log \overline{a}(z)=\log\left(\prod_{j=1}^{J} \frac{z-\frac{q_{0}^{2}}{z_{j}}}{z-z_{j}}\right)-\frac{1}{2\pi i}\int_{-\infty}^{+\infty}\frac{\log \gamma(\xi)\overline{\gamma}(\xi)}{\xi-z}d\xi, \ \ \ \Im z<0.
\end{equation}
Note that
\begin{equation}
\gamma(z)\overline{\gamma}(z)=a(z)\overline{a}(z),
\end{equation}
from the unitarity condition
\begin{equation}
a(z)\overline{a}(z)-b(z)\overline{b}(z)=1,
\end{equation}
it follows
\begin{equation}
\log a(z)=\log\left(\prod_{j=1}^{J}\frac{z-z_{j}}{z-\frac{q_{0}^{2}}{z_{j}}}\right)+\frac{1}{2\pi i}\int_{-\infty}^{+\infty}\frac{\log (1+b(\xi)\overline{b}(\xi))}{\xi-z}d\xi, \ \ \ \Im z>0,
\end{equation}

\begin{equation}
\log \overline{a}(z)=\log\left(\prod_{j=1}^{J} \frac{z-\frac{q_{0}^{2}}{z_{j}}}{z-z_{j}}\right)-\frac{1}{2\pi i}\int_{-\infty}^{+\infty}\frac{\log (1+b(\xi)\overline{b}(\xi))}{\xi-z}d\xi, \ \ \ \Im z<0.
\end{equation}
By the symmetry $b(z)=\overline{b}(z)$, we can obtain
\begin{equation}
\log a(z)=\log\left(\prod_{j=1}^{J}\frac{z-z_{j}}{z-\frac{q_{0}^{2}}{z_{j}}}\right)+\frac{1}{2\pi i}\int_{-\infty}^{+\infty}\frac{\log (1+b^{2}(\xi))}{\xi-z}d\xi, \ \ \ \Im z>0,
\end{equation}

\begin{equation}
\log \overline{a}(z)=\log\left(\prod_{j=1}^{J} \frac{z-\frac{q_{0}^{2}}{z_{j}}}{z-z_{j}}\right)-\frac{1}{2\pi i}\int_{-\infty}^{+\infty}\frac{\log (1+b^{2}(\xi))}{\xi-z}d\xi, \ \ \ \Im z<0.
\end{equation}
Thus we can reconstruct $a(k), \bar{a}(k)$ in terms of the eigenvalues (zero's)  and only {\bf one function $b(k)$}.

Since $a(z)\sim e^{2i\theta_{+}}$ as $z\rightarrow 0$, from the trace formula when $b(\xi)=0$ in the real axis, we have the following constraint for the reflectionless potentials

\begin{equation}
\prod_{j=1}^{J}z_{j}^2=q_{0}^{2J}e^{2i\theta_{+}}
\end{equation}


\subsection{Discrete scattering data and their symmetries}

In order to find reflectionless potentials/solitons, we need to calculate the relevant scattering data:

\[ b(z_j) ~\mbox{and} ~~~\bar{b}(\bar{z}_j) ,~~j=1,2,...J.\]
and

\[a'(z_j), ~~\bar{a}'(z_j) \]
The latter functions can be calculated via the trace formulae. So we concentrate on the former.

Since
\begin{equation}
N_{1}(x,z)=-M_{2}(-x,z), \ \ \ N_{2}(x,z)=M_{1}(-x,z),
\end{equation}
\begin{equation}
M_{1}(x,z_{j})=b(z_{j})e^{i\big(z_{j}-\frac{q_{0}^{2}}{z_{j}}\big)x}\cdot N_{1}(x,z_{j})
\end{equation}
and
\begin{equation}
M_{2}(x,z_{j})=b(z_{j})e^{i\big(z_{j}-\frac{q_{0}^{2}}{z_{j}}\big)x}\cdot N_{2}(x,z_{j}),
\end{equation}
we have
\begin{equation}
\label{E:N1}
N_{1}(x,z_{j})=-b(z_{j})\cdot e^{-i\left(z_{j}-\frac{q_{0}^{2}}{z_{j}}\right)x}\cdot N_{2}(-x,z_{j}),
\end{equation}
\begin{equation}
\label{E:N2}
N_{2}(x,z_{j})=b(z_{j})\cdot e^{-\left(z_{j}-\frac{q_{0}^{2}}{z_{j}}\right)x}\cdot N_{1}(-x,z_{j}).
\end{equation}
By rewriting (\ref{E:N2}), we obtain
\begin{equation}
N_{2}(-x,z_{j})=b(z_{j})\cdot e^{\left(z_{j}-\frac{q_{0}^{2}}{z_{j}}\right)x}\cdot N_{1}(x,z_{j}).
\end{equation}
Combining (\ref{E:N1}), we can deduce the following symmetry condition on the discrete data $b(z_j)$
\begin{equation}
\label{E:b}
-b^{2}(z_{j})=1.
\end{equation}

Similar analysis shows that $\bar{b}(\bar{z}_j) $ satisfies an analogous equation

\[ -\bar{b}^{2}(\bar{z}_j)  =1,\]
i.e.,
\begin{equation}
b(z_{j})=\pm i, \ \ \ \overline{b}(\overline{z}_{j})=\pm i.
\end{equation}
By the symmetry relation
\begin{equation}
\overline{b}(z)=b(z),
\end{equation}
we have
\begin{equation}
\overline{b}(z_{j})=b(z_{j}), \ \ \ b(\overline{z}_{j})=\overline{b}(\overline{z}_{j}).
\end{equation}

When $J=1$,  assuming that $0<\theta_{+}<\pi$, then $z_{1}=q_{0}e^{i\theta_{+}}$. By the trace formula when $b(\xi)=0$ in the real axis, we can get
\begin{equation}
a'(z_{1})=a'(q_{0}e^{i\theta_{+}})=\frac{1}{q_{0}(e^{i\theta_{+}}-e^{-i\theta_{+}})}, \ \ \
\overline{a}'(\overline{z}_{1})=a'(q_{0}e^{-i\theta_{+}})=\frac{1}{q_{0}(e^{-i\theta_{+}}-e^{i\theta_{+}})}.
\end{equation}
Moreover, from the symmetry relation
\begin{equation}
b\left(\frac{q_{0}^{2}}{z}\right)=-\overline{b}(z),
\end{equation}
we have
\begin{equation}
\overline{b}(q_{0}e^{-i\theta_{+}})=-b(q_{0}e^{i\theta_{+}}).
\end{equation}
For convenience, we write $b(q_{0}e^{i\theta_{+}})=\delta i$, then $\overline{b}(q_{0}e^{-i\theta_{+}})=-\delta i$, where $\delta=\pm 1$.
\subsection{Time evolution}
Since
\begin{equation}
\label{E:time evolution}
v_{t}=Tv=
\left(\begin{array}{cc}
-\frac{is(x,t)}{2k}& -\frac{iq_{t}(x,t)}{2k}\\
\frac{iq_{t}(-x,-t)}{2k}& \frac{is(x,t)}{2k}
\end{array}\right)v:=\left(\begin{array}{cc}
A& B\\
C& -A
\end{array}\right)v,
\end{equation}
we have
\begin{equation}
\frac{\partial v_{1}}{\partial t}=-\frac{is(x,t)}{2k}v_{1}-\frac{iq_{t}(x,t)}{2k}v_{2}
\end{equation}
and
\begin{equation}
\frac{\partial v_{2}}{\partial t}=\frac{iq_{t}(-x,-t)}{2k}v_{1}+\frac{is(x,t)}{2k}v_{2}.
\end{equation}
Note that
\begin{equation}
q_{xt}(x,t)+2s(x,t)q(x,t)=0,
\end{equation}
where
$s(x,t)= -\int_x^{\infty} \partial_t (q(x,t)q(-x,-t))$. As mentioned in the introduction we require $s(x,t) \rightarrow 0 ~\text{as}~|x| \rightarrow \infty$ and

%

\begin{equation}
q(x,t)\rightarrow q_{\pm}(t)=q_{0}e^{i(\alpha t+\theta_{\pm})}, \ \ ~\text{as}~\  x\rightarrow\pm\infty,
\end{equation}
where $q_{0}>0$, $0\leq\theta_{\pm}<2\pi$, $\theta_{+}+\theta_{-}=0$ and $\alpha$ is real.
%

Thus, we have
\begin{equation}
\frac{\partial v_{1}}{\partial t}\sim \frac{\alpha q_{0}e^{i(\alpha t+\theta_{+})}}{2k}v_{2}
\end{equation}
\begin{equation}
\frac{\partial v_{2}}{\partial t}\sim \frac{\alpha q_{0}e^{i(-\alpha t+\theta_{-})}}{2k}v_{1}
\end{equation}
as $x\rightarrow +\infty$; and
\begin{equation}
\frac{\partial v_{1}}{\partial t}\sim \frac{\alpha q_{0}e^{i(\alpha t+\theta_{-})}}{2k}v_{2}
\end{equation}
\begin{equation}
\frac{\partial v_{2}}{\partial t}\sim \frac{\alpha q_{0}e^{i(-\alpha t+\theta_{+})}}{2k}v_{1}
\end{equation}
as $x\rightarrow -\infty$. As $x\rightarrow\pm\infty$, the eigenfunctions of the spatial portion of the scattering problem asymptotically satisfy
\begin{equation}
\left(\begin{array}{cc}
v_{1}\\
v_{2}
\end{array}\right)_{x}
=
\left(\begin{array}{cc}
-ik& q_{0}e^{i(\alpha t+\theta_{\pm})}\\
q_{0}e^{i(-\alpha t+\theta_{\mp})}& ik
\end{array}\right)
\left(\begin{array}{cc}
v_{1}\\
v_{2}
\end{array}\right),
\end{equation}
we can get
\begin{equation}
q_{0}e^{i(\alpha t+\theta_{\pm})} v_{2}\sim \frac{\partial v_{1}}{\partial x}+ikv_{1}
\end{equation}
and
\begin{equation}
q_{0}e^{i(-\alpha t+\theta_{\mp})} v_{1}\sim \frac{\partial v_{2}}{\partial x}-ikv_{2}
\end{equation}
as $x\rightarrow\pm\infty$. Hence,
\begin{equation}
\frac{\partial v_{1}}{\partial t}\sim \frac{i\alpha}{2}v_{1}+\frac{\alpha}{2k}\frac{\partial v_{1}}{\partial x}
\end{equation}
and
\begin{equation}
\frac{\partial v_{2}}{\partial t}\sim -\frac{i\alpha}{2}v_{2}+\frac{\alpha}{2k}\frac{\partial v_{2}}{\partial x}
\end{equation}
as $x\rightarrow\pm\infty$.

Note that the eigenfunctions themselves, whose boundary values at space infinities, are not compatible with this time evolution. Therefore, one introduces time-dependent eigenfunctions
\begin{equation}
\Phi(x,t)=e^{i A_{\infty}t}\cdot \phi(x,t), \ \ \ \overline{\Phi}(x,t)=e^{i B_{\infty}t}\cdot \overline{\phi}(x,t),
\end{equation}
\begin{equation}
\Psi(x,t)=e^{i C_{\infty}t}\cdot \psi(x,t), \ \ \ \overline{\Psi}(x,t)=e^{i D_{\infty}t}\cdot \overline{\psi}(x,t)
\end{equation}
to be solutions of (\ref{E:time evolution}). We have
\begin{equation}
\frac{\partial \Phi_{1}(x,t)}{\partial t}=i A_{\infty} \Phi_{1}(x,t)+e^{iA_{\infty}t} \frac{\partial \phi_{1}(x,t)}{\partial t},
\ \ \ \frac{\partial \Phi_{2}(x,t)}{\partial t}=i A_{\infty} \Phi_{2}(x,t)+e^{iA_{\infty}t} \frac{\partial \phi_{2}(x,t)}{\partial t},
\end{equation}
\begin{equation}
\frac{\partial \overline{\Phi}_{1}(x,t)}{\partial t}=i B_{\infty} \overline{\Phi}_{1}(x,t)+e^{iB_{\infty}t} \frac{\partial \overline{\phi}_{1}(x,t)}{\partial t},
\ \ \ \frac{\partial \overline{\Phi}_{2}(x,t)}{\partial t}=i B_{\infty} \overline{\Phi}_{2}(x,t)+e^{iB_{\infty}t} \frac{\partial \overline{\phi}_{2}(x,t)}{\partial t},
\end{equation}
\begin{equation}
\frac{\partial \Psi_{1}(x,t)}{\partial t}=i C_{\infty} \Psi_{1}(x,t)+e^{iC_{\infty}t} \frac{\partial \psi_{1}(x,t)}{\partial t},
\ \ \ \frac{\partial \Psi_{2}(x,t)}{\partial t}=i C_{\infty} \Psi_{2}(x,t)+e^{iC_{\infty}t} \frac{\partial \psi_{2}(x,t)}{\partial t},
\end{equation}
\begin{equation}
\frac{\partial \overline{\Psi}_{1}(x,t)}{\partial t}=i D_{\infty} \overline{\Psi}_{1}(x,t)+e^{iD_{\infty}t} \frac{\partial \overline{\psi}_{1}(x,t)}{\partial t},
\ \ \ \frac{\partial \overline{\Psi}_{2}(x,t)}{\partial t}=i D_{\infty} \overline{\Psi}_{2}(x,t)+e^{iD_{\infty}t} \frac{\partial \overline{\psi}_{2}(x,t)}{\partial t}.
\end{equation}
We recall that
\begin{equation}
\phi(x,t)\sim\left(\begin{array}{cc}
\lambda+k\\
iq_{0}e^{i(\alpha t+\theta_{-})}
\end{array}\right)e^{-i\lambda x}, ~~\text{so that}\ \ \
\frac{\partial \phi(x,t)}{\partial t}\sim \left(\begin{array}{cc}
0\\
-\alpha q_{0}e^{i(\alpha t+\theta_{-})}
\end{array}\right)e^{-i\lambda x}
\end{equation}
as $x\rightarrow -\infty$.


From
\begin{equation}
\frac{\partial \Phi_{1}}{\partial t}\sim \frac{i\alpha}{2}\Phi_{1}(x,t)+\frac{\alpha}{2k}\frac{\partial \Phi_{1}(x,t)}{\partial x}=i A_{\infty} \Phi_{1}(x,t)+e^{iA_{\infty}t} \frac{\partial \phi_{1}(x,t)}{\partial t}
\end{equation}
as $x\rightarrow -\infty$, we can deduce
\begin{equation}
A_{\infty}=\frac{\alpha}{2}-\frac{\alpha \lambda}{2k}.
\end{equation}
Similarly, we have
\begin{equation}
B_{\infty}=-\frac{\alpha}{2}+\frac{\alpha \lambda}{2k},
\end{equation}
\begin{equation}
C_{\infty}=-\frac{\alpha}{2}+\frac{\alpha \lambda}{2k},
\end{equation}
\begin{equation}
D_{\infty}=\frac{\alpha}{2}-\frac{\alpha \lambda}{2k}.
\end{equation}
Then
\begin{equation}
\frac{\partial\phi(x,k)}{\partial t}=\left(\begin{array}{cc}
A-iA_{\infty}& B\\
C& -A-iA_{\infty}
\end{array}\right)\phi(x,k),
\end{equation}
\begin{equation}
\frac{\partial\overline{\phi}(x,k)}{\partial t}=\left(\begin{array}{cc}
A+iA_{\infty}& B\\
C& -A+iA_{\infty}
\end{array}\right)\overline{\phi}(x,k),
\end{equation}
\begin{equation}
\frac{\partial\psi(x,k)}{\partial t}=\left(\begin{array}{cc}
A+iA_{\infty}& B\\
C& -A+iA_{\infty}
\end{array}\right)\psi(x,k),
\end{equation}
\begin{equation}
\frac{\partial\overline{\psi}(x,k)}{\partial t}=\left(\begin{array}{cc}
A-iA_{\infty}& B\\
C& -A-iA_{\infty}
\end{array}\right)\overline{\psi}(x,k).
\end{equation}
Recall that
\begin{equation}
\phi(x,t)=b(t)\psi(x,t)+a(t)\overline{\psi}(x,t)
\end{equation}
and
\begin{equation}
\overline{\phi}(x,t)=\overline{a}(t)\psi(x,t)+\overline{b}(t)\overline{\psi}(x,t),
\end{equation}
we have
\begin{equation}
\begin{split}
&\left(\begin{array}{cc}
A-iA_{\infty}& B\\
C& -A-iA_{\infty}
\end{array}\right)[b\psi+a\overline{\psi}]\\
&=b_{t}\psi+b\left(\begin{array}{cc}
A+iA_{\infty}& B\\
C& -A+iA_{\infty}
\end{array}\right)\psi+a_{t}\overline{\psi}+a\left(\begin{array}{cc}
A-iA_{\infty}& B\\
C& -A-iA_{\infty}
\end{array}\right)\overline{\psi},
\end{split}
\end{equation}
which implies that
\begin{equation}
\frac{\partial a(t)}{\partial t}=0, \ \ \ \frac{\partial b(t)}{\partial t}=-2iA_{\infty}b(t).
\end{equation}
Similarly, we can obtain
\begin{equation}
\frac{\partial \overline{a}(t)}{\partial t}=0, \ \ \ \frac{\partial \overline{b}(t)}{\partial t}=2iA_{\infty}\overline{b}(t).
\end{equation}
Therefore, both $a(t)$ and $\overline{a}(t)$ are time independent, and
\begin{equation}
b(z;t)=b(z;0)e^{-i\left(\alpha-\frac{\alpha \lambda}{k}\right)t},
\end{equation}
and
\begin{equation}
\overline{b}(z;t)=\overline{b}(z;0)e^{i\left(\alpha-\frac{\alpha \lambda}{k}\right)t}.
\end{equation}
Then we have
\begin{equation}
b(q_{0}e^{i\theta_{+}},t)= \delta i e^{-\frac{i\alpha t }{\cos \theta_{+}}\cdot e^{-i\theta_{+}}}
\end{equation}
and
\begin{equation}
\overline{b}(q_{0}e^{-i\theta_{+}},t)=- \delta i e^{\frac{i\alpha t }{\cos \theta_{+}}\cdot e^{i\theta_{+}}}.
\end{equation}

\subsection{Pure 1-Soliton Solution}
\begin{figure}[h]
\begin{tabular}{cc}
\includegraphics[width=0.5\textwidth]{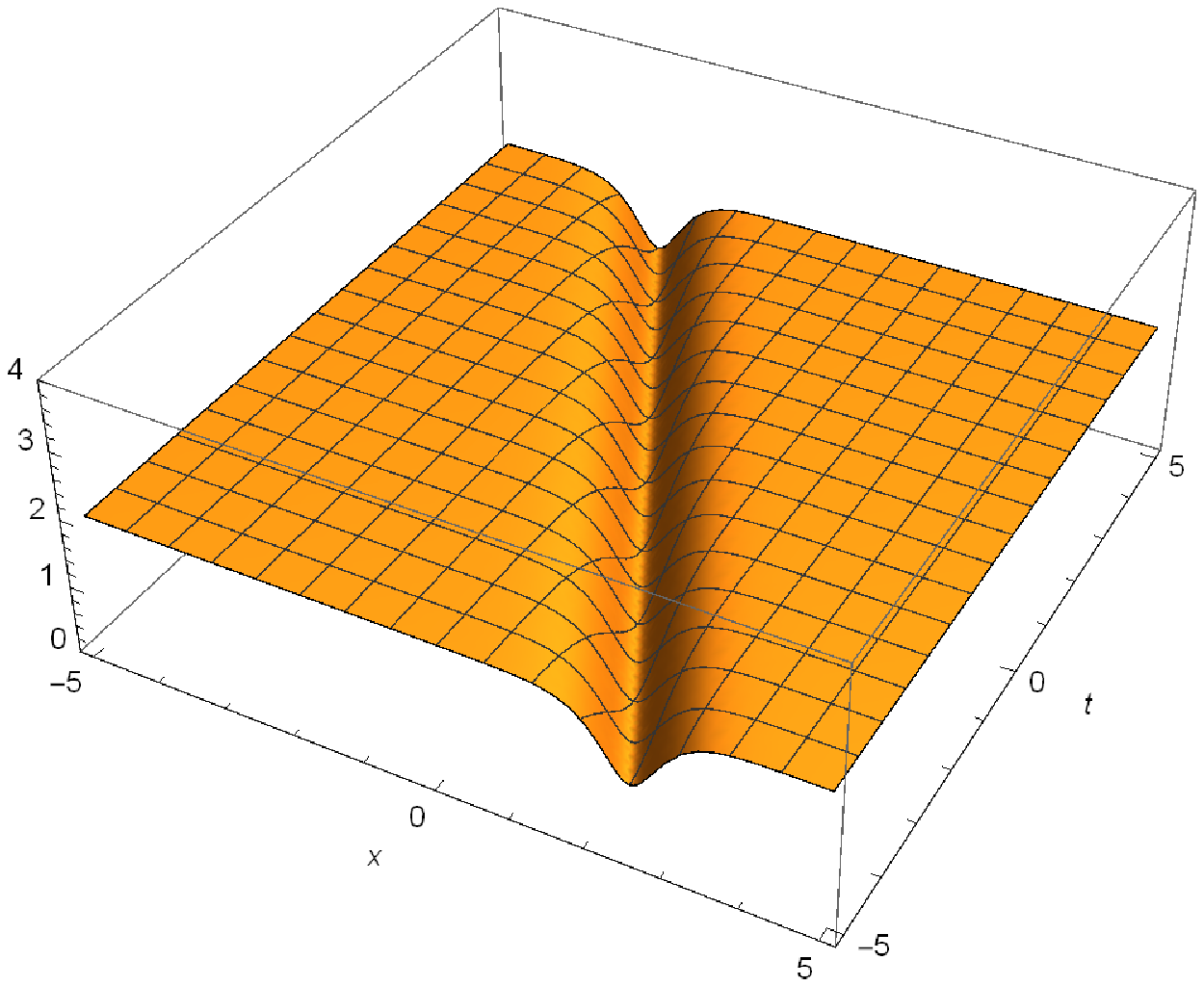}&
\includegraphics[width=0.5\textwidth]{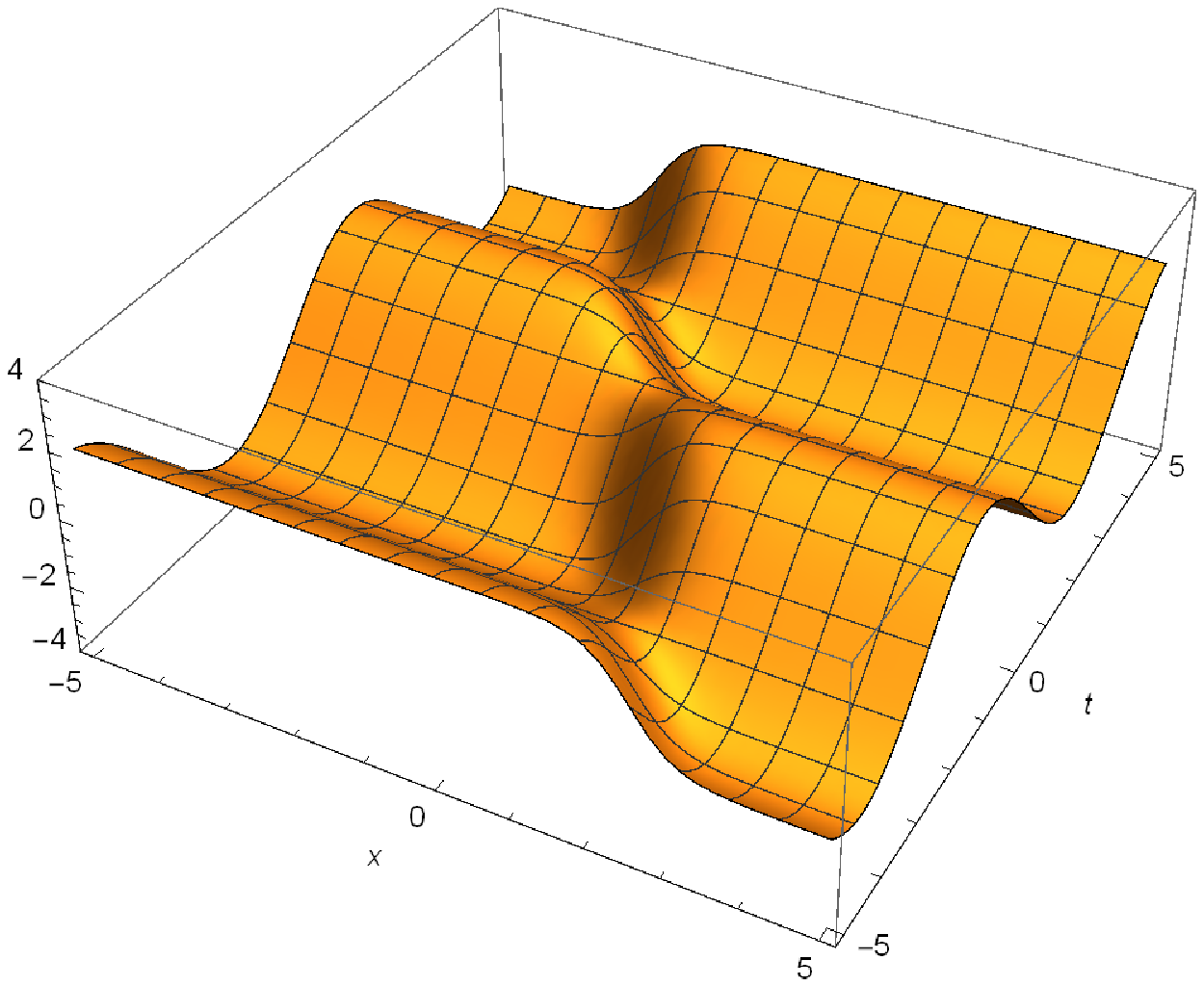}\\
(a) & (b)
\end{tabular}
\caption{(a) The amplitude of $q(x,t)$ with $\delta=1$, $\theta_{+}=\frac{\pi}{3}$, $\alpha=1$ and $q_{0}=2$.  (b) The real part of $q(x,t)$ with $\delta=1$, $\theta_{+}=\frac{\pi}{3}$, $\alpha=1$ and $q_{0}=2$. *}
\label{Sinh 1}
\end{figure}

With $J=1$, and  $b(\xi,t)=0$ (reflectionless potential) we have from  (\ref{asympN1c})
\begin{equation}
\label{E:soliton 1}
q(x,t)=q_{0}e^{i(\alpha t+\theta_{+})}\left[1-\frac{2i\sin\theta_{+}\cdot b(q_{0}e^{i\theta_{+}})e^{-2q_{0}x\sin\theta_{+}}}{q_{0}e^{2i\theta_{+}}}\cdot N_{1}(x, q_{0}e^{i\theta_{+}}, t)\right].
\end{equation}
Solving the corresponding discrete system (\ref{E:closing system 1}) with $J=1$
\begin{equation}
N_{1}(x, q_{0}e^{i\theta_{+}}, t)=\frac{-iq_{0}e^{i(\alpha t+\theta_{+})}-\overline{b}(q_{0}e^{-i\theta_{+}})q_{0}e^{i\theta_{+}}e^{-2q_{0}x\sin\theta_{+}}}{1-b(q_{0}e^{i\theta_{+}})\overline{b}(q_{0}e^{-i\theta_{+}})e^{-4q_{0}x\sin\theta_{+}}}.
\end{equation}
Note that $\delta$ can be chosen as either $1$ or $-1$, when $\delta=1$, there is a nonsingular pure "dark" 1-soliton solution (see the following); when $\delta=-1$, we find a pure "bright" 1-soliton solution, which is singular along a space-time  line (see **Section 10).
When $\delta=1$, it yields
\begin{equation}
\begin{split}
&q(x,t)=q_{0}e^{i(\alpha t+\theta_{+})}\Big[1+\frac{2\sin\theta_{+}\cdot e^{-\frac{i\alpha t }{\cos \theta_{+}}\cdot e^{-i\theta_{+}}} \cdot e^{-2q_{0}x\sin\theta_{+}}}{q_{0}e^{2i\theta_{+}}}\\
&\cdot \frac{-iq_{0}e^{i(\alpha t+\theta_{+})}+ie^{\frac{i\alpha t }{\cos \theta_{+}}\cdot e^{i\theta_{+}}}q_{0}e^{i\theta_{+}}\cdot e^{-2q_{0}x\sin\theta_{+}}}{1-e^{-2\alpha t\cdot \tan \theta_{+}}\cdot e^{-4q_{0}x\sin\theta_{+}}}\Big]\\
&=q_{0}e^{i\alpha t}\cos\left(\theta_{+}-iq_{0}x\sin\theta_{+}-\frac{i}{2}\alpha t\tan\theta_{+}\right) \text{sech}\left(q_{0}x\sin\theta_{+}+\frac{1}{2}\alpha t\tan\theta_{+}\right)
\end{split}
\end{equation}
and
\begin{equation}
s(x,t)=\frac{1}{2}q_{0}\alpha \cdot \text{sech}^{2}\left(q_{0}x\sin\theta_{+}+\frac{1}{2}\alpha t \tan\theta_{+}\right)\sin\theta_{+}\tan\theta_{+}.
\end{equation}
We see that $s \rightarrow 0$ as $|x| \rightarrow \infty$. This solution is not singular because \\
$e^{q_{0}x\sin\theta_{+}+\frac{1}{2}\alpha t \tan\theta_{+}}>0$.
In Fig. \ref{Sinh 1} below we give a typical dark 1-soliton solution. Note also, when $\theta_+=\pi/2$ then the solution becomes elementary.

\section{Nonlocal Sinh-Gordon equation with $\theta_{+}+\theta_{-}=\pi$}
In this section we consider the nonzero boundary conditions (NZBCs) given above in (\ref{E:NZBC}) and $\theta_{+}+\theta_{-}=\pi$. With this condition, equation (\ref{E:eigenvalue asymptotic}) conveniently reduces to
\begin{equation}
\frac{\partial^{2}v_{j}}{\partial x^{2}}=-(k^{2}+q_{0}^{2})v_{j}, \ \ j=1,2.
\end{equation}
Each of the two equations has two linearly independent solutions $e^{i\lambda x}$ and $e^{-i\lambda x}$ as $|x|\rightarrow\infty$, where
$\lambda=\sqrt{k^{2}+q_{0}^{2}}$. We introduce the local polar coordinates
\begin{equation}
k-iq_{0}=r_{1}e^{i\theta_{1}}, \ \ \ -\frac{\pi}{2}\leq\theta_{1}<\frac{3\pi}{2},
\end{equation}
\begin{equation}
k+iq_{0}=r_{2}e^{i\theta_{2}}, \ \ \ -\frac{\pi}{2}\leq\theta_{2}<\frac{3\pi}{2},
\end{equation}
where $r_{1}=|k-iq_{0}|$ and $r_{2}=|k+iq_{0}|$. One can write $\lambda(k)=(r_{1}r_{2})^{\frac{1}{2}}\cdot e^{i\cdot \frac{\theta_{1}+\theta_{2}}{2}+im\pi}$, and $m=0,1$ respectively on sheets I ($\mathbb{K}_{1}$) and II ($\mathbb{K}_{2}$). The variable $k$ is then thought of as belonging to a Riemann surface $\mathbb{K}$ consisting of sheets I and II with both coinciding with the complex plane cut along $\Sigma:=[-iq_{0}, iq_{0}]$
with its edges glued in such a way that $\lambda(k)$ is continuous through the cut. Along the real $k$ axis, we have $\lambda(k)=\pm sign(k) \sqrt{k^{2}+q_{0}^{2}}$, where the plus/minus signs apply, respectively, on sheet I and sheet II of the Riemann surface, and where the square root sign denotes the principal branch of the real-valued square root function. We denote $\mathbb{C}^{\pm}$ the open upper/lower complex half planes, and $\mathbb{K}^{\pm}$ the open upper/lower complex half planes cut along $\Sigma$. Then $\lambda$ provides one-to-one correspondences between the following sets:
~\\1. $k\in\mathbb{K}^{+}=\mathbb{C}^{+}\setminus (0, iq_{0}]$ and $\lambda\in\mathbb{C}^{+}$,
~\\2. $k\in \partial\mathbb{K}^{+}=\mathbb{R}\cup\{is-0^{+}:0<s<q_{0}\}\cup\{iq_{0}\}\cup\{is+0^{+}:0<s<q_{0}\}$ and $\lambda\in\mathbb{R}$,
~\\3. $k\in\mathbb{K}^{-}=\mathbb{C}^{-}\setminus [-iq_{0}, 0)$ and $\lambda\in\mathbb{C}^{-}$,
~\\4. $k\in \partial\mathbb{K}^{-}=\mathbb{R}\cup\{is-0^{+}:-q_{0}<s<0\}\cup\{-iq_{0}\}\cup\{is+0^{+}:-q_{0}<s<0\}$ and $\lambda\in\mathbb{R}$.

Moreover, $\lambda^{\pm}(k)$ will denote the boundary values taken by $\lambda(k)$ for $k\in\Sigma$ from the right/left edge of the cut, with $\lambda^{\pm}(k)=\pm\sqrt{q_{0}^{2}-|k|^{2}}$, $k=is\pm 0^{+}$, $|s|<q_{0}$ on the right/left edge of the cut (See \cite{AblowitzLuoMusslimani}, Page 30, Fig. 5 and Fig. 6).
\subsection{Eigenfunctions}
It is natural to introduce the eigenfunctions defined by the following boundary conditions
\begin{equation}
\phi(x,k)\sim w e^{-i\lambda x}, \ \ \ \overline{\phi}(x,k)\sim \overline{w}e^{i\lambda x}
\end{equation}
as $x\rightarrow-\infty$,
\begin{equation}
\psi(x,k)\sim v e^{i\lambda x}, \ \ \ \overline{\psi}(x,k)\sim \overline{v}e^{-i\lambda x}
\end{equation}
as $x\rightarrow +\infty$. We substitute the above into (\ref{E:eigenvalue asymptotic}), obtaining
\begin{equation}
w=\left(\begin{array}{cc}
\lambda+k\\
iq_{+}
\end{array}\right), \ \ \
\overline{w}=\left(\begin{array}{cc}
-iq_{-}\\
\lambda+k
\end{array}\right),
\end{equation}
\begin{equation}
v=\left(\begin{array}{cc}
-iq_{+}\\
\lambda+k
\end{array}\right), \ \ \
\overline{v}=
\left(\begin{array}{cc}
\lambda+k\\
iq_{-}
\end{array}\right),
\end{equation}
which satisfy the boundary conditions, which are not unique.

In the following analysis, it is convenient to consider functions with constant boundary conditions. We define the bounded eigenfunctions as follows:
\begin{equation}
M(x,k)=e^{i\lambda x}\phi(x,k), \ \ \ \overline{M}(x,k)=e^{-i\lambda x}\overline{\phi}(x,k),
\end{equation}
\begin{equation}
N(x,k)=e^{-i\lambda x}\psi(x,k), \ \ \ \overline{N}(x,k)=e^{i\lambda x}\overline{\psi}(x,k).
\end{equation}
The eigenfunctions can be represented by means of the following integral equations
\begin{equation}
M(x,k)=
\left(\begin{array}{cc}
\lambda+k\\
iq_{+}
\end{array}\right)
+\int_{-\infty}^{+\infty}G_{-}(x-x',k)((Q-Q_{-})M)(x',k)dx',
\end{equation}
\begin{equation}
\overline{M}(x,k)=
\left(\begin{array}{cc}
-iq_{-}\\
\lambda+k
\end{array}\right)
+\int_{-\infty}^{+\infty}\overline{G}_{-}(x-x',k)((Q-Q_{-})M)(x',k)dx',
\end{equation}
\begin{equation}
N(x,k)=
\left(\begin{array}{cc}
-iq_{+}\\
\lambda+k
\end{array}\right)
+\int_{-\infty}^{+\infty}G_{+}(x-x',k)((Q-Q_{+})M)(x',k)dx',
\end{equation}
\begin{equation}
\overline{N}(x,k)=\left(\begin{array}{cc}
\lambda+k\\
iq_{-}
\end{array}\right)
+\int_{-\infty}^{+\infty}\overline{G}_{+}(x-x',k)((Q-Q_{+})M)(x',k)dx'.
\end{equation}
Using the Fourier transform method, we get
\begin{equation}
G_{-}(x,k)=\frac{\theta(x)}{2\lambda}[(1+e^{2i\lambda x})\lambda I-i(e^{2i\lambda x}-1)(ikJ+Q_{-})],
\end{equation}
\begin{equation}
\overline{G}_{-}(x,k)=\frac{\theta(x)}{2\lambda}[(1+e^{-2i\lambda x})\lambda I+i(e^{-2i\lambda x}-1)(ikJ+Q_{-})],
\end{equation}
\begin{equation}
G_{+}(x,k)=-\frac{\theta(-x)}{2\lambda}[(1+e^{-2i\lambda x})\lambda I+i(e^{-2i\lambda x}-1)(ikJ+Q_{+})],
\end{equation}
\begin{equation}
\overline{G}_{+}(x,k)=-\frac{\theta(-x)}{2\lambda}[(1+e^{2i\lambda x})\lambda I-i(e^{2i\lambda x}-1)(ikJ+Q_{+})],
\end{equation}
where $\theta(x)$ is the Heaviside function, i.e., $\theta(x)=1$ if $x>0$ and $\theta(x)=0$ if $x<0$.
\begin{definition}
We say $f\in L^{1,1}(\mathbb{R})$ if $\int_{-\infty}^{+\infty}|f(x)|\cdot(1+|x|)dx<\infty$.
\end{definition}
Then, using similar methods as in the prior case ($\theta_{+}+\theta_{-} = 0$), we find the following result.
\begin{theorem}
Suppose the entries of $Q-Q_{\pm}$ belong to $L^{1,1}(\mathbb{R})$, then for each $x\in\mathbb{R}$, the eigenfunctions $M(x,k)$ and $N(x,k)$ are continuous for $k\in \overline{\mathbb{K}^{+}}\cup \partial \overline{\mathbb{K}^{-}}$ and analytic for $k\in \mathbb{K}^{+}$, $\overline{M}(x,k)$ and $\overline{N}(x,k)$ are continuous for $k\in \overline{\mathbb{K}^{-}}\cup \partial \overline{\mathbb{K}^{+}}$ and analytic for $k\in \mathbb{K}^{-}$.
\end{theorem}
The proof makes use of Neumann series; it is similar to \cite{AblowitzLuoMusslimani}.
\subsubsection{Scattering data}
We have
\begin{equation}
\phi(x,k)=b(k)\psi(x,k)+a(k)\overline{\psi}(x,k)
\end{equation}
and
\begin{equation}
\overline{\phi}(x,k)=\overline{a}(k)\psi(x,k)+\overline{b}(k)\overline{\psi}(x,k)
\end{equation}
hold for any $k$ such that all four eigenfunctions exist. Moreover,
\begin{equation}
a(k)\overline{a}(k)-b(k)\overline{b}(k)=1,
\end{equation}
where
\begin{equation}
a(k)=\frac{W(\phi(x,k),\psi(x,k))}{2\lambda(\lambda+k)}, \ \ \ \overline{a}(k)=-\frac{W(\overline{\phi}(x,k),\overline{\psi}(x,k))}{2\lambda(\lambda+k)},
\end{equation}
\begin{equation}
b(k)=-\frac{W(\phi(x,k),\overline{\psi}(x,k))}{2\lambda(\lambda+k)}, \ \ \ \overline{b}(k)=\frac{W(\overline{\phi}(x,k),\psi(x,k))}{2\lambda(\lambda+k)}.
\end{equation}
When $k\in (-iq_{0}, iq_{0})$, the above scattering data and eigenfunctions are defined by means of the corresponding values on the right/left edge of the cut, are labeled with superscripts $\pm$ as clarified below. Explicitly, one has
\begin{equation}
a^{\pm}(k)=\frac{W(\phi^{\pm}(x,k),\psi^{\pm}(x,k))}{2\lambda^{\pm}(\lambda^{\pm}+k)}, \ \ \ k\in (-iq_{0}, iq_{0}),
\end{equation}
\begin{equation}
\overline{a}^{\pm}(k)=-\frac{W(\overline{\phi}^{\pm}(x,k),\overline{\psi}^{\pm}(x,k))}{2\lambda^{\pm}(\lambda^{\pm}+k)}, \ \ \ k\in (-iq_{0}, iq_{0}),
\end{equation}
\begin{equation}
b^{\pm}(k)=-\frac{W(\phi^{\pm}(x,k),\overline{\psi}^{\pm}(x,k))}{2\lambda^{\pm}(\lambda^{\pm}+k)}, \ \ \ k\in (-iq_{0}, iq_{0}),
\end{equation}
\begin{equation}
\overline{b}^{\pm}(k)=\frac{W(\overline{\phi}^{\pm}(x,k),\psi^{\pm}(x,k))}{2\lambda^{\pm}(\lambda^{\pm}+k)},  \ \ \ k\in (-iq_{0}, iq_{0}).
\end{equation}
Then from the analytic behavior of the eigenfunctions we have the following theorem.
\begin{theorem}
Suppose the entries of $Q-Q_{\pm}$ belong to $L^{1,1}(\mathbb{R})$, then $a(k)$ is continuous for $k\in \overline{\mathbb{K}^{+}}\cup \partial \overline{\mathbb{K}^{-}}\setminus\{\pm iq_{0}\}$ and analytic for $k\in \mathbb{K}^{+}$, and $\overline{a}(k)$ is continuous for $k\in \overline{\mathbb{K}^{-}}\cup \partial \overline{\mathbb{K}^{+}}\setminus\{\pm iq_{0}\}$ and analytic for $k\in \mathbb{K}^{-}$. Moreover, $b(k)$ and $\overline{b}(k)$ are continuous in $k\in \mathbb{R}\cup (-iq_{0}, iq_{0})$. In addition, if the entries of $Q-Q_{\pm}$ do not grow faster than $e^{-ax^{2}}$, where $a$ is a positive real number, then $a(k)\lambda(k)$, $\overline{a}(k)\lambda(k)$, $b(k)\lambda(k)$ and $\overline{b}(k)\lambda(k)$ are analytic for $k\in\mathbb{K}$.
\end{theorem}
The proof makes use of Neumann series; it is similar to \cite{AblowitzLuoMusslimani}.
\subsection{Symmetry reductions}
The symmetry in the potential induces a symmetry between the eigenfunctions. Indeed, if $v(x,k)=(v_{1}(x,k), v_{2}(x,k))^{T}$ solves
(\ref{E:nonlocal scattering}), then $(v_{2}(-x,k), -v_{1}(-x,k))^{T}$ also solves (\ref{E:nonlocal scattering}). Taking into account boundary conditions, we can obtain
\begin{equation}
\psi(x,k)=\left(\begin{array}{cc}
0& -1\\
1& 0
\end{array}\right)\phi(-x,k)
\end{equation}
and
\begin{equation}
\overline{\psi}(x,k)=\left(\begin{array}{cc}
0& 1\\
-1& 0
\end{array}\right)\overline{\phi}(-x,k).
\end{equation}
Similarly, we can get the symmetry relations of the eigenfunctions, i.e.,
\begin{equation}
N(x,k)=\left(\begin{array}{cc}
0& -1\\
1& 0
\end{array}\right)M(-x,k)
\end{equation}
and
\begin{equation}
\overline{N}(x,k)=\left(\begin{array}{cc}
0& 1\\
-1& 0
\end{array}\right)\overline{M}(-x,k).
\end{equation}
From the Wronskian representations for the scattering data and the above symmetry relations, we have
\begin{equation}
\overline{b}(k)=b(k).
\end{equation}
\subsection{Uniformization coordinates}
Similarly, we introduce a uniformization variable $z$ , defined by the conformal mapping:
\begin{equation}
\label{zsect4}
z=z(k)=k+\lambda(k),
\end{equation}
where $ \lambda= \sqrt{k^2+q_0^2}$ and the inverse mapping is given by
\begin{equation}
k=k(z)=\frac{1}{2}\left(z-\frac{q_{0}^{2}}{z}\right).
\end{equation}
Then
\begin{equation}
\lambda(z)=\frac{1}{2}\left(z+\frac{q_{0}^{2}}{z}\right).
\end{equation}
We let $C_{0}$ be the circle of radius $q_{0}$ centered at the origin in $z-$ plane. We observe that

(1) The branch cut on either sheet is mapped onto $C_{0}$. In particular, $z(\pm iq_{0})=\pm iq_{0}$ from either sheet, $z(0_{I}^{\pm})=\pm q_{0}$ and $z(0_{II}^{\pm})=\mp q_{0}$.

(2) $\mathbb{K}_{1}$ is mapped onto the exterior of $C_{0}$, $\mathbb{K}_{2}$ is mapped onto the interior of $C_{0}$. In particular, $z(\infty_{I})=\infty$ and $z(\infty_{II})=0$; the first/second quadrants of $\mathbb{K}_{1}$ are mapped into the first/second quadrants outside $C_{0}$ respectively; the first/second quadrants of $\mathbb{K}_{2}$ are mapped into the second/first quadrants inside $C_{0}$ respectively; $z_{I}z_{II}=q_{0}^{2}$.

(3) The regions in $k-$ plane such that $\Im \lambda>0$ and $\Im \lambda<0$ are mapped onto $D^{+}=\{z\in\mathbb{C}:(|z|^{2}-q_{0}^{2})\cdot\Im z>0\}$ and $D^{-}=\{z\in\mathbb{C}:(|z|^{2}-q_{0}^{2})\cdot\Im z<0\}$ respectively (See \cite{AblowitzLuoMusslimani}, Page 36, Fig. 11).

Then we have that the eigenfunctions $M,N$ are analytic for $z \in D^{+}$ and
the eigenfunctions $\bar{M},\bar{N}$  are analytic for in $z \in D^{-}$.

\subsection{Symmetries via uniformization coordinates}
From the above eigenfunction symmetry relations, we have
\begin{equation}
\psi(x,z)=\left(\begin{array}{cc}
0& -1\\
1& 0
\end{array}\right)\phi(-x,z),
\end{equation}
\begin{equation}
\overline{\psi}(x,z)=\left(\begin{array}{cc}
0& 1\\
-1& 0
\end{array}\right)\overline{\phi}(-x,z).
\end{equation}
Further, from $z\rightarrow -\frac{q_{0}^{2}}{z}$, then $(k, \lambda)\rightarrow (k, -\lambda)$. Hence,
\begin{equation}
\phi\left(x,-\frac{q_{0}^{2}}{z}\right)=\frac{\frac{q_{0}^{2}}{z}}{iq_{-}}\overline{\phi}(x,z), \ \ \ \psi\left(x,-\frac{q_{0}^{2}}{z}\right)=\frac{-iq_{+}}{z}\overline{\psi}(x,z), \ \ \ z\in D^{-}.
\end{equation}
Similarly, we can get
\begin{equation}
N(x,z)=\left(\begin{array}{cc}
0& -1\\
1& 0
\end{array}\right)M(-x,z),
\end{equation}
\begin{equation}
\overline{N}(x,z)=\left(\begin{array}{cc}
0& 1\\
-1& 0
\end{array}\right)\overline{M}(-x,z),
\end{equation}
\begin{equation}
\overline{b}(z)=b(z),
\end{equation}
\begin{equation}
a\left(-\frac{q_{0}^{2}}{z}\right)=-e^{2i\theta_{+}}\overline{a}(z),\ \ \ z\in D^{-}; \ \ \ b\left(-\frac{q_{0}^{2}}{z}\right)=-\overline{b}(z).
\end{equation}
\subsection{Asymptotic behavior of eigenfunctions and scattering data}
In order to solve the inverse problem, one has to determine the asymptotic behavior of eigenfunctions and scattering data both as $z\rightarrow\infty$ in $\mathbb{K}_{1}$ and as $z\rightarrow 0$ in $\mathbb{K}_{2}$. We have
\begin{equation}
M(x,z)\sim\left\{\begin{array}{ll}
\left(\begin{array}{cc}
z\\
iq(-x)
\end{array}\right), \ \ \ z\rightarrow\infty\\
\left(\begin{array}{cc}
z\cdot\frac{q(x)}{q_{-}}\\
iq_{+}
\end{array}\right), \ \ \ z\rightarrow 0,\\
\end{array}\right.
\end{equation}

\begin{equation}
N(x,z)\sim\left\{\begin{array}{ll}
\left(\begin{array}{cc}
-iq(x)\\
z
\end{array}\right), \ \ \ z\rightarrow\infty\\
\left(\begin{array}{cc}
-iq_{+}\\
z\cdot \frac{q(-x)}{q_{-}}
\end{array}\right), \ \ \ z\rightarrow 0,\\
\end{array}\right.
\end{equation}

\begin{equation}
\overline{M}(x,z)\sim\left\{\begin{array}{ll}
\left(\begin{array}{cc}
-iq(x)\\
z
\end{array}\right), \ \ \ z\rightarrow\infty\\
\left(\begin{array}{cc}
-iq_{-}\\
z\cdot \frac{q(-x)}{q_{+}}
\end{array}\right), \ \ \ z\rightarrow 0,\\
\end{array}\right.
\end{equation}

\begin{equation}
\overline{N}(x,z)\sim\left\{\begin{array}{ll}
\left(\begin{array}{cc}
z\\
iq(-x)
\end{array}\right), \ \ \ z\rightarrow\infty\\
\left(\begin{array}{cc}
z\cdot\frac{q(x)}{q_{+}}\\
iq_{-}
\end{array}\right), \ \ \ z\rightarrow 0,\\
\end{array}\right.
\end{equation}

\begin{equation}
a(z)=
\left\{\begin{array}{ll}
1,\ \ \ z\rightarrow\infty,\\
-e^{2i\theta_{+}}, \ \ \ z\rightarrow 0,\\
\end{array}\right.
\end{equation}

\begin{equation}
\overline{a}(z)=
\left\{\begin{array}{ll}
1,\ \ \ z\rightarrow\infty,\\
-e^{-2i\theta_{+}}, \ \ \ z\rightarrow 0,\\
\end{array}\right.
\end{equation}
\begin{equation}
\lim_{z\rightarrow\infty}zb(z)=0, \ \ \ \lim_{z\rightarrow0}\frac{b(z)}{z^{2}}=0.
\end{equation}
\subsection{Riemann-Hilbert problem via uniformization coordinates}
\subsubsection{Left scattering problem}
In order to take into account the behavior of the eigenfunctions,  the `jump' conditions at $\Sigma$, where $\Sigma:=(-\infty,-q_{0})\cup (q_{0}, +\infty)\cup \overrightarrow{(q_{0}, -q_{0})}\cup \{q_{0}e^{i\theta}, \pi\leq \theta\leq 2\pi\}_{clockwise, upper \ circle}\cup \{q_{0}e^{i\theta}, -\pi\leq \theta\leq 0\}_{anticlockwise, lower \ circle}$, can be written as
\begin{equation}
\frac{M(x,z)}{za(z)}-\frac{\overline{N}(x,z)}{z}=\rho(z)e^{i\big(z+\frac{q_{0}^{2}}{z}\big)x}\cdot \frac{N(x,z)}{z}
\end{equation}
and
\begin{equation}
\frac{\overline{M}(x,z)}{z\overline{a}(z)}-\frac{N(x,z)}{z}=\overline{\rho}(z)e^{-i\big(z+\frac{q_{0}^{2}}{z}\big)x}\cdot \frac{\overline{N}(x,z)}{z},
\end{equation}
so that the functions will be bounded at infinity, though having an additional pole at $z=0$. Note that $M(x,z)/a(z)$, as a function of
$z$, is defined in $D^{+}$, where it has simple poles $z_{j}$, i.e., $a(z_{j})=0$, and $\overline{M}(x,z)/\overline{a}(z)$,
is defined in $D^{-}$, where it has simple poles $\overline{z}_{j}$, i.e., $\overline{a}(\overline{z}_{j})=0$. It follows that
\begin{equation}
M(x,z_{j})=b(z_{j})e^{i\big(z_{j}+\frac{q_{0}^{2}}{z_{j}}\big)x}\cdot N(x,z_{j})
\end{equation}
and
\begin{equation}
\overline{M}(x,\overline{z}_{j})=\overline{b}(\overline{z}_{j})e^{-i\big(\overline{z}_{j}+\frac{q_{0}^{2}}{\overline{z}_{j}}\big)x}\cdot \overline{N}(x,\overline{z}_{j}).
\end{equation}
Then subtracting the values at infinity, the induced pole at the origin and the poles, assumed simple in $D^{+}$/$D^{-}$ respectively, at $a(z_j)=0, j=1,2...J$ and  $\bar{a}(\overline{z}_{j}), j=1,2...\bar{J} $ gives
\begin{equation}
\label{E:jump 1'}
\begin{split}
&\left[\frac{M(x,z)}{za(z)}-\left(\begin{array}{cc}
1\\
0
\end{array}\right)-
\frac{1}{z}\left(\begin{array}{cc}
0\\
iq_{-}
\end{array}\right)
-\sum_{j=1}^{J}\frac{M(x,z_{j})}{(z-z_{j})z_{j}a'(z_{j})}\right]\\
&-\left[\frac{\overline{N}(x,z)}{z}-\left(\begin{array}{cc}
1\\
0
\end{array}\right)-
\frac{1}{z}\left(\begin{array}{cc}
0\\
iq_{-}
\end{array}\right)
-\sum_{j=1}^{J}\frac{b(z_{j})e^{i\big(z_{j}+\frac{q_{0}^{2}}{z_{j}}\big)x}\cdot N(x,z_{j})}{(z-z_{j})z_{j}a'(z_{j})}\right]\\
&=\rho(z)e^{i\big(z+\frac{q_{0}^{2}}{z}\big)x}\cdot \frac{N(x,z)}{z}
\end{split}
\end{equation}
and
\begin{equation}
\label{E:jump 2'}
\begin{split}
&\left[\frac{\overline{M}(x,z)}{z\overline{a}(z)}-\left(\begin{array}{cc}
0\\
1
\end{array}\right)-
\frac{1}{z}\left(\begin{array}{cc}
-iq_{+}\\
0
\end{array}\right)
-\sum_{j=1}^{\overline{J}}\frac{\overline{M}(x,\overline{z}_{j})}{(z-\overline{z}_{j})\overline{z}_{j}a'(\overline{z}_{j})}\right]\\
&-\left[\frac{N(x,z)}{z}-\left(\begin{array}{cc}
0\\
1
\end{array}\right)-
\frac{1}{z}\left(\begin{array}{cc}
-iq_{+}\\
0
\end{array}\right)
-\sum_{j=1}^{\overline{J}}\frac{\overline{b}(\overline{z}_{j})e^{-i\big(\overline{z}_{j}+\frac{q_{0}^{2}}{\overline{z}_{j}}\big)x}\cdot \overline{N}(x,\overline{z}_{j})}{(z-\overline{z}_{j})\overline{z}_{j}\overline{a}'(\overline{z}_{j})}\right]\\
&=\overline{\rho}(z)e^{-i\big(z+\frac{q_{0}^{2}}{z}\big)x}\cdot \frac{\overline{N}(x,z)}{z}.
\end{split}
\end{equation}
We now introduce the projection operators
\begin{equation}
P_{\pm}(f)(z)=\frac{1}{2\pi i}\int_{\Sigma}\frac{f(\xi)}{\xi-(z\pm i0)}d\xi,
\end{equation}
where $z$ lies in the $\pm$ regions and $\Sigma:=(-\infty,-q_{0})\cup (q_{0}, +\infty)\cup \overrightarrow{(q_{0}, -q_{0})}\cup \{q_{0}e^{i\theta}, \pi\leq \theta\leq 2\pi\}_{clockwise, upper \ circle}\cup \{q_{0}e^{i\theta}, -\pi\leq \theta\leq 0\}_{anticlockwise, lower \ circle}$.

If $f_{\pm}(\xi)$ is analytic in $D^{\pm}$ and $f_{\pm}(\xi)$ is decaying at large $\xi$, then
\begin{equation}
P_{\pm}(f_{\pm})(z)=\pm f_{\pm}(z), \ \ \ P_{\mp}(f_{\pm})(z)=0.
\end{equation}
Applying $P_{-}$ to (\ref{E:jump 1'}) and $P_{+}$ to (\ref{E:jump 2'}), we can obtain
\begin{equation}
\label{E:eigenfunction 1'}
\begin{split}
\overline{N}(x,z)&=\left(\begin{array}{cc}
z\\
iq_{-}
\end{array}\right)
+\sum_{j=1}^{J}\frac{z\cdot b(z_{j})e^{i\big(z_{j}+\frac{q_{0}^{2}}{z_{j}}\big)x}\cdot N(x,z_{j})}{(z-z_{j})z_{j}a'(z_{j})}\\
&+\frac{z}{2\pi i}\int_{\Sigma}\frac{\rho(\xi)}{\xi(\xi-z)}\cdot e^{i\big(\xi+\frac{q_{0}^{2}}{\xi}\big)x}\cdot N(x,\xi)d\xi
\end{split}
\end{equation}
and
\begin{equation}
\label{E:eigenfunction 2'}
\begin{split}
N(x,z)&=\left(\begin{array}{cc}
-iq_{+}\\
z
\end{array}\right)
+\sum_{j=1}^{\overline{J}}\frac{z\cdot\overline{b}(\overline{z}_{j})e^{-i\big(\overline{z}_{j}+\frac{q_{0}^{2}}{\overline{z}_{j}}\big)x}\cdot \overline{N}(x,\overline{z}_{j})}{(z-\overline{z}_{j})\overline{z}_{j}\overline{a}'(\overline{z}_{j})}\\
&-\frac{z}{2\pi i}\int_{\Sigma}\frac{\overline{\rho}(\xi)}{\xi(\xi-z)}\cdot e^{-i\big(\xi+\frac{q_{0}^{2}}{\xi}\big)x}\cdot \overline{N}(x,\xi)d\xi.
\end{split}
\end{equation}
Similarly, we could solve the right scattering problem, which reads
\begin{equation}
\label{E:eigenfunction 3'}
\begin{split}
\overline{M}(x,z)&=\left(\begin{array}{cc}
-iq_{-}\\
z
\end{array}\right)+
\sum_{j=1}^{J}\frac{-z\cdot\overline{b}(z_{j})M(x,z_{j})e^{-i\big(z_{j}+\frac{q_{0}^{2}}{z_{j}}\big)x}}{(z-z_{j})z_{j}a'(z_{j})}\\
&+\frac{z}{2\pi i}\int_{\Sigma}\frac{\rho^{*}(-\xi^{*})}{\xi(\xi-z)}\cdot e^{-i\big(\xi+\frac{q_{0}^{2}}{\xi}\big)x}\cdot M(x,\xi)d\xi
\end{split}
\end{equation}
and
\begin{equation}
\label{E:eigenfunction 4'}
\begin{split}
M(x,z)&=\left(\begin{array}{cc}
z\\
iq_{+}
\end{array}\right)+
\sum_{j=1}^{\overline{J}}\frac{-z\cdot b(\overline{z}_{j})\overline{M}(x,\overline{z}_{j})e^{i\big(\overline{z}_{j}+\frac{q_{0}^{2}}{\overline{z}_{j}}\big)x}}
{(z-\overline{z}_{j})\overline{z}_{j}\overline{a}'(\overline{z}_{j})}\\
&-\frac{z}{2\pi i}\int_{\Sigma}\frac{\overline{\rho}^{*}(-\xi^{*})}{\xi(\xi-z)}\cdot e^{i\big(\xi+\frac{q_{0}^{2}}{\xi}\big)x}\cdot \overline{M}(x,\xi)d\xi.
\end{split}
\end{equation}
\subsection{Recovery of the potentials}
Note that
\begin{equation}
\frac{\overline{N}_{1}(x,z)}{z}\sim \frac{q(x)}{q_{+}}
\end{equation}
as $z\rightarrow 0$, and
\begin{equation}
\frac{\overline{N}_{1}(x,z)}{z}\sim 1+\sum_{j=1}^{J}\frac{b(z_{j})e^{i\big(z_{j}+\frac{q_{0}^{2}}{z_{j}}\big)x}}{-z_{j}^{2}a'(z_{j})}\cdot N_{1}(x,z_{j})+\frac{1}{2\pi i}\int_{\Sigma}\frac{\rho(\xi)}{\xi^{2}}\cdot e^{i\big(\xi+\frac{q_{0}^{2}}{\xi}\big)x}\cdot N_{1}(x,\xi)d\xi
\end{equation}
as $z\rightarrow 0$,
we have
\begin{equation}
\label{asympN1c'}
q(x)=q_{+}\cdot\left[1+\sum_{j=1}^{J}\frac{b(z_{j})e^{i\big(z_{j}+\frac{q_{0}^{2}}{z_{j}}\big)x}}{-z_{j}^{2}a'(z_{j})}\cdot N_{1}(x,z_{j})+\frac{1}{2\pi i}\int_{\Sigma}\frac{\rho(\xi)}{\xi^{2}}\cdot e^{i\big(\xi+\frac{q_{0}^{2}}{\xi}\big)x}\cdot N_{1}(x,\xi)d\xi\right].
\end{equation}
\subsection{Closing the system}
Similarly, we can get $J=\overline{J}$ from $a\left(-\frac{q_{0}^{2}}{z}\right)=-e^{2i\theta_{+}}\overline{a}(z)$.
Combining the above integral equations, we have
\begin{equation}
\label{E:closing system 3}
\begin{split}
&\left(\begin{array}{cc}
N_{1}(x,z)\\
N_{2}(x,z)
\end{array}\right)=\left(\begin{array}{cc}
-iq_{+}\\
z
\end{array}\right)
+\sum_{j=1}^{J}\frac{z\cdot\overline{b}(\overline{z}_{j})e^{-i\big(\overline{z}_{j}+\frac{q_{0}^{2}}{\overline{z}_{j}}\big)x} }{(z-\overline{z}_{j})\overline{z}_{j}\overline{a}'(\overline{z}_{j})}\cdot\\
&\left(\begin{array}{cc}
\overline{z}_{j}+\sum_{l=1}^{J}\frac{\overline{z}_{j}\cdot b(z_{l})e^{i\big(z_{l}+\frac{q_{0}^{2}}{z_{l}}\big)x}}{(\overline{z}_{j}-z_{l})z_{l}a'(z_{l})}\cdot N_{1}(x, z_{l})+\frac{\overline{z}_{j}}{2\pi i}\int_{\Sigma}\frac{\rho(\xi)}{\xi(\xi-\overline{z}_{j})}\cdot e^{i\big(\xi+\frac{q_{0}^{2}}{\xi}\big)x}\cdot N_{1}(x, \xi)d\xi\\
iq_{-}+\sum_{l=1}^{J}\frac{\overline{z}_{j}\cdot b(z_{l})e^{i\big(z_{l}+\frac{q_{0}^{2}}{z_{l}}\big)x}}{(\overline{z}_{j}-z_{l})z_{l}a'(z_{l})}\cdot N_{2}(x, z_{l})+\frac{\overline{z}_{j}}{2\pi i}\int_{\Sigma}\frac{\rho(\xi)}{\xi(\xi-\overline{z}_{j})}\cdot e^{i\big(\xi+\frac{q_{0}^{2}}{\xi}\big)x}\cdot N_{2}(x, \xi)d\xi
\end{array}\right)\\
&-\frac{z}{2\pi i}\int_{\Sigma}\frac{\overline{\rho}(\xi)}{\xi(\xi-z)}\cdot e^{-i\big(\xi+\frac{q_{0}^{2}}{\xi}\big)x}\cdot\\
&\left(\begin{array}{cc}
\xi+\sum_{l=1}^{J}\frac{\xi\cdot b(z_{l})e^{i\big(z_{l}-\frac{q_{0}^{2}}{z_{l}}\big)x}}{(\xi-z_{l})z_{l}a'(z_{l})}\cdot N_{1}(x, z_{l})+\frac{\xi}{2\pi i}\int_{\Sigma}\frac{\rho(\eta)}{\eta(\eta-\xi)}\cdot e^{i\big(\eta+\frac{q_{0}^{2}}{\eta}\big)x}\cdot N_{1}(x, \eta)d\eta\\
iq_{-}+\sum_{l=1}^{J}\frac{\xi\cdot b(z_{l})e^{i\big(z_{l}+\frac{q_{0}^{2}}{z_{l}}\big)x}}{(\xi-z_{l})z_{l}a'(z_{l})}\cdot N_{2}(x, z_{l})+\frac{\xi}{2\pi i}\int_{\Sigma}\frac{\rho(\eta)}{\eta(\eta-\xi)}\cdot e^{i\big(\eta+\frac{q_{0}^{2}}{\eta}\big)x}\cdot N_{2}(x, \eta)d\eta
\end{array}\right)d\xi,
\end{split}
\end{equation}
\begin{equation}
\label{E:closing system 4}
\begin{split}
&\left(\begin{array}{cc}
\overline{M}_{1}(x,z)\\
\overline{M}_{2}(x,z)
\end{array}\right)=\left(\begin{array}{cc}
-iq_{-}\\
z
\end{array}\right)+
\sum_{j=1}^{J}\frac{-z\cdot\overline{b}(z_{j})e^{-i\big(z_{j}+\frac{q_{0}^{2}}{z_{j}}\big)x}}{(z-z_{j})z_{j}a'(z_{j})}\cdot\\
& \left(\begin{array}{cc}
z_{j}+
\sum_{l=1}^{J}\frac{-z_{j}\cdot b(\overline{z}_{l})e^{i\big(\overline{z}_{l}+\frac{q_{0}^{2}}{\overline{z}_{l}}\big)x}}
{(z_{j}-\overline{z}_{l})\overline{z}_{l}\overline{a}'(\overline{z}_{l})}\cdot \overline{M}_{1}(x, \overline{z}_{l})-\frac{z_{j}}{2\pi i}\int_{\Sigma}\frac{\overline{\rho}^{*}(-\xi^{*})}{\xi(\xi-z_{j})}\cdot e^{i\big(\xi+\frac{q_{0}^{2}}{\xi}\big)x}\cdot\overline{M}_{1}(x,\xi)d\xi\\
iq_{+}+
\sum_{l=1}^{J}\frac{-z_{j}\cdot b(\overline{z}_{l})e^{i\big(\overline{z}_{l}+\frac{q_{0}^{2}}{\overline{z}_{l}}\big)x}}
{(z_{j}-\overline{z}_{l})\overline{z}_{l}\overline{a}'(\overline{z}_{l})}\cdot \overline{M}_{2}(x, \overline{z}_{l})-\frac{z_{j}}{2\pi i}\int_{\Sigma}\frac{\overline{\rho}^{*}(-\xi^{*})}{\xi(\xi-z_{j})}\cdot e^{i\big(\xi+\frac{q_{0}^{2}}{\xi}\big)x}\cdot\overline{M}_{2}(x,\xi)d\xi
\end{array}\right)\\
&+\frac{z}{2\pi i}\int_{\Sigma}\frac{\rho^{*}(-\xi^{*})}{\xi(\xi-z)}\cdot e^{-i\big(\xi+\frac{q_{0}^{2}}{\xi}\big)x}\cdot\\
&\left(\begin{array}{cc}
\xi+
\sum_{l=1}^{J}\frac{-\xi\cdot b(\overline{z}_{l})e^{i\big(\overline{z}_{l}+\frac{q_{0}^{2}}{\overline{z}_{l}}\big)x}}
{(\xi-\overline{z}_{l})\overline{z}_{l}\overline{a}'(\overline{z}_{l})}\cdot\overline{M}_{1}(x,\overline{z}_{l})-\frac{\xi}{2\pi i}\int_{\Sigma}\frac{\overline{\rho}^{*}(-\eta^{*})}{\eta(\eta-\xi)}\cdot e^{i\big(\eta+\frac{q_{0}^{2}}{\eta}\big)x}\cdot \overline{M}_{1}(x,\eta)d\eta\\
iq_{+}+
\sum_{l=1}^{J}\frac{-\xi\cdot b(\overline{z}_{l})e^{i\big(\overline{z}_{l}+\frac{q_{0}^{2}}{\overline{z}_{l}}\big)x}}
{(\xi-\overline{z}_{l})\overline{z}_{l}\overline{a}'(\overline{z}_{l})}\cdot\overline{M}_{2}(x,\overline{z}_{l})-\frac{\xi}{2\pi i}\int_{\Sigma}\frac{\overline{\rho}^{*}(-\eta^{*})}{\eta(\eta-\xi)}\cdot e^{i\big(\eta+\frac{q_{0}^{2}}{\eta}\big)x}\cdot \overline{M}_{2}(x,\eta)d\eta
\end{array}\right)d\xi.
\end{split}
\end{equation}
We can reconstruct the potential  from (\ref{asympN1c'}).
\subsection{Trace formula}
In a similar manner to the prior section, we can get the Trace formula as follows.
\begin{equation}
\log a(z)=\log\left(\prod_{j=1}^{J}\frac{z-z_{j}}{z+\frac{q_{0}^{2}}{z_{j}}}\right)+\frac{1}{2\pi i}\int_{\Sigma}\frac{\log (1+b^{2}(\xi))}{\xi-z}d\xi, \ \ \ z\in D^{+}
\end{equation}
and
\begin{equation}
\log \overline{a}(z)=\log\left(\prod_{j=1}^{J} \frac{z+\frac{q_{0}^{2}}{z_{j}}}{z-z_{j}}\right)-\frac{1}{2\pi i}\int_{\Sigma}\frac{\log (1+b^{2}(\xi))}{\xi-z}d\xi, \ \ \ z\in D^{-}.
\end{equation}
\subsection{Discrete scattering data and their symmetries}
In order to find reflectionless potentials and solitons we need to be able to calculate the relevant discrete scattering data.
The coefficients
\[ b(z_j) ~\mbox{and} ~~~\bar{b}(\bar{z}_j) ,~~j=1,2,...J.\]
can be calculated in the same way as in previous section, finding
\begin{equation}
\overline{b}(z_{j})=b(z_{j})=\pm i, \ \ \ b(\overline{z}_{j})=\overline{b}(\overline{z}_{j})=\pm i.
\end{equation}
Since $a(z)\sim -e^{2i\theta_{+}}$ as $z\rightarrow 0$, from the trace formula when $b(\xi)=0$ in $\Sigma$, we have the following constraint for the reflectionless potentials
\begin{equation}
\prod_{j=1}^{J}z_{j}=\pm (-1)^{\frac{J+1}{2}}q_{0}^{J}e^{i\theta_{+}}.
\end{equation}
We claim that $J\geq 2$. Otherwise, if $J=1$, then $z_{1}=\pm q_{0}e^{i\theta_{+}}$. It implies that the eigenvalue $z_{1}$ lies on the circle, which in this case is the continuous spectrum. Such eigenvalues are not proper; they are not considered here.

\subsection{Reflectionless scattering data 2-eigenvalues}
In this subsection, we consider scattering data associated with 2-eigenvalues ; i.e.  $J=2$, with no reflection.
Note that
$|z_{1}|\cdot|z_{2}|=q_{0}^{2}$. Now let $\pi<\theta_{+}<\frac{3\pi}{2}$ and $z_{1}=q_{1}e^{i\theta_{1}}$, where $q_{1}>q_{0}$ and $0<\theta_{1}<\frac{\pi}{2}$, then $z_{2}=\frac{q_{0}^{2}}{q_{1}}e^{i(\theta_{+}-\theta_{1}+\frac{\pi}{2})}$, where $\pi<\arg z_{2}<2\pi$. By the trace formula when $b(\xi)=0$ in $\Sigma$, we have
\begin{equation}
a(z)=\frac{z-q_{1}e^{i\theta_{1}}}{z+\frac{q_{0}^{2}}{q_{1}}e^{-i\theta_{1}}}\cdot \frac{z-\frac{q_{0}^{2}}{q_{1}}e^{i(\theta_{+}-\theta_{1}+\frac{\pi}{2})}}{z+q_{1}e^{-i(\theta_{+}-\theta_{1}+\frac{\pi}{2})}}
\end{equation}
and
\begin{equation}
\overline{a}(z)=\frac{z+\frac{q_{0}^{2}}{q_{1}}e^{-i\theta_{1}}}{z-q_{1}e^{i\theta_{1}}}
\cdot\frac{z+q_{1}e^{-i(\theta_{+}-\theta_{1}+\frac{\pi}{2})}}{z-\frac{q_{0}^{2}}{q_{1}}e^{i(\theta_{+}-\theta_{1}+\frac{\pi}{2})}}.
\end{equation}
Then
\begin{equation}
a'(q_{1}e^{i\theta_{1}})=\frac{q_{0}^{2}e^{2i\theta_{+}}+iq_{1}^{2}e^{(\theta_{+}+2\theta_{1})i}}{q_{1}e^{i\theta_{1}}(1+ie^{i\theta_{+}})
(q_{0}^{2}+q_{1}^{2}e^{2i\theta_{1}})},
\end{equation}
\begin{equation}
a'\left(\frac{q_{0}^{2}}{q_{1}}e^{i(\theta_{+}-\theta_{1}+\frac{\pi}{2})}\right)
=\frac{q_{1}}{q_{0}^{2}}e^{i(\theta_{+}+\theta_{1})}
\cdot\frac{-q_{0}^{2}e^{i\theta_{+}}-iq_{1}^{2}e^{2i\theta_{1}}}{(1+ie^{i\theta_{+}})(-q_{0}^{2}e^{2i\theta_{+}}+q_{1}^{2}e^{2i\theta_{1}})},
\end{equation}
\begin{equation}
\overline{a}'\left(-\frac{q_{0}^{2}}{q_{1}}e^{-i\theta_{1}}\right)
=\frac{-iq_{1}\left(-iq_{0}^{2}e^{i(\theta_{1}+\theta_{+})}+q_{1}^{2}e^{3i\theta_{1}}\right)}{q_{0}^{2}e^{i\theta_{+}}(1+ie^{i\theta_{+}})
(q_{0}^{2}+q_{1}^{2}e^{2i\theta_{1}})},
\end{equation}
\begin{equation}
\overline{a}'\left(-q_{1}e^{-i(\theta_{+}-\theta_{1}+\frac{\pi}{2})}\right)
=\frac{-q_{0}^{2}e^{2i\theta_{+}}-iq_{1}^{2}e^{i(2\theta_{1}+\theta_{+})}}
{q_{1}e^{i\theta_{1}}(1+ie^{i\theta_{+}})(q_{1}^{2}e^{2i\theta_{1}}-q_{0}^{2}e^{2i\theta_{+}})}.
\end{equation}
Moreover, from the symmetry relation
\begin{equation}
b\left(-\frac{q_{0}^{2}}{z}\right)=-\overline{b}(z),
\end{equation}
we obtain
\begin{equation}
\overline{b}\left(-\frac{q_{0}^{2}}{q_{1}}e^{-i\theta_{1}}\right)=-b(q_{1}e^{i\theta_{1}}), \ \ \ \ \
\overline{b}\left(-q_{1}e^{-i(\theta_{+}-\theta_{1}+\frac{\pi}{2})}\right)=-b\left(\frac{q_{0}^{2}}{q_{1}}e^{i(\theta_{+}-\theta_{1}+\frac{\pi}{2})}\right).
\end{equation}
Since $b(q_{1}e^{i\theta_{1}})=\delta_{1}i$, $b\left(\frac{q_{0}^{2}}{q_{1}}e^{i(\theta_{+}-\theta_{1}+\frac{\pi}{2})}\right)=\delta_{2}i$, where $\delta_{1}=\pm 1$ and $\delta_{2}=\pm 1$,
we get
\begin{equation}
\overline{b}\left(-\frac{q_{0}^{2}}{q_{1}}e^{-i\theta_{1}}\right)=-\delta_{1}i, \ \ \ \overline{b}\left(-q_{1}e^{-i(\theta_{+}-\theta_{1}+\frac{\pi}{2})}\right)=-\delta_{2}i.
\end{equation}
In particular, we can choose $z_{1}=iq_{1}$, $z_{2}=-i\frac{q_{0}^{2}}{q_{1}}$ and $\theta_{+}=\frac{\pi}{2}$, then $\overline{z}_{1}=i\frac{q_{0}^{2}}{q_{1}}$, $\overline{z}_{2}=-iq_{1}$ and $z_{1}z_{2}=q_{0}^{2}$. Thus,
\begin{equation}
a(z)=\frac{z-iq_{1}}{z-i\frac{q_{0}^{2}}{q_{1}}}\cdot\frac{z+i\frac{q_{0}^{2}}{q_{1}}}{z+iq_{1}}
\end{equation}
and
\begin{equation}
\overline{a}(z)=\frac{z-i\frac{q_{0}^{2}}{q_{1}}}{z-iq_{1}}\cdot \frac{z+iq_{1}}{z+i\frac{q_{0}^{2}}{q_{1}}}.
\end{equation}
We have
\begin{equation}
a'(iq_{1})=\frac{-i(q_{1}^{2}+q_{0}^{2})}{2q_{1}(q_{1}^{2}-q_{0}^{2})}, \ \ \ a'\left(-i\frac{q_{0}^{2}}{q_{1}}\right)=\frac{iq_{1}(q_{0}^{2}+q_{1}^{2})}{2q_{0}^{2}(q_{0}^{2}-q_{1}^{2})},
\end{equation}
\begin{equation}
\overline{a}'\left(i\frac{q_{0}^{2}}{q_{1}}\right)=-\frac{iq_{1}(q_{0}^{2}+q_{1}^{2})}{2q_{0}^{2}(q_{0}^{2}-q_{1}^{2})}, \ \ \
\overline{a}'(-iq_{1})=\frac{i(q_{1}^{2}+q_{0}^{2})}{2q_{1}(q_{1}^{2}-q_{0}^{2})}.
\end{equation}
Moreover,
\begin{equation}
b(iq_{1})=\delta_{1}i,\ \ \ b\left(-i\frac{q_{0}^{2}}{q_{1}}\right)=\delta_{2}i, \ \ \ \overline{b}\left(i\frac{q_{0}^{2}}{q_{1}}\right)=-\delta_{1}i, \ \ \ \overline{b}(-iq_{1})=-\delta_{2}i.
\end{equation}
\subsection{Time evolution}
In the same manner as in Section 3 we can deduce both $a(t)$ and $\overline{a}(t)$ are time independent and
\begin{equation}
b(z;t)=b(z;0)e^{-i\left(\alpha-\frac{\alpha \lambda}{k}\right)t}=b(z;0)e^{\frac{2\alpha q_{0}^{2}i}{z^{2}-q_{0}^{2}}t},
\end{equation}
and
\begin{equation}
\overline{b}(z;t)=\overline{b}(z;0)e^{i\left(\alpha-\frac{\alpha \lambda}{k}\right)t}=\overline{b}(z;0)e^{-\frac{2\alpha q_{0}^{2}i}{z^{2}-q_{0}^{2}}t}.
\end{equation}

Thus,
\begin{equation}
b(q_{1}e^{i\theta_{1}};t)=\delta_{1}i\cdot e^{\frac{2\alpha q_{0}^{2}i}{q_{1}^{2}e^{2i\theta_{1}}-q_{0}^{2}}t}, \ \ \
b\left(\frac{q_{0}^{2}}{q_{1}}e^{i(\theta_{+}-\theta_{1}+\frac{\pi}{2})};t\right)=\delta_{2}i\cdot e^{-\frac{2\alpha q_{1}^{2}i}{q_{0}^{2}e^{2i(\theta_{+}-\theta_{1})}+q_{1}^{2}}t},
\end{equation}
\begin{equation}
\overline{b}\left(-\frac{q_{0}^{2}}{q_{1}}e^{-i\theta_{1}}; t\right)=-\delta_{1}i\cdot e^{-\frac{2\alpha q_{0}^{2}q_{1}^{2}i}{q_{0}^{4}e^{-2i\theta_{1}}-q_{0}^{2}q_{1}^{2}}t}, \ \ \
\overline{b}\left(-q_{1}e^{-i(\theta_{+}-\theta_{1}+\frac{\pi}{2})}\right)=-\delta_{2}i\cdot e^{\frac{2\alpha q_{0}^{2}i}{q_{1}^{2}e^{2i(\theta_{1}-\theta_{+})}+q_{0}^{2}}t}.
\end{equation}
In particular,
\begin{equation}
b(iq_{1};t)=\delta_{1}i\cdot e^{-\frac{2\alpha q_{0}^{2}i}{q_{0}^{2}+q_{1}^{2}}t}, \ \ \
b\left(-i\frac{q_{0}^{2}}{q_{1}};t\right)=\delta_{2}i\cdot e^{-\frac{2\alpha q_{1}^{2}i}{q_{0}^{2}+q_{1}^{2}}t},
\end{equation}
\begin{equation}
\overline{b}\left(i\frac{q_{0}^{2}}{q_{1}};t\right)=-\delta_{1}i\cdot e^{\frac{2\alpha q_{1}^{2}i}{q_{0}^{2}+q_{1}^{2}}t}, \ \ \
\overline{b}(-iq_{1};t)=-\delta_{2}i\cdot e^{\frac{2\alpha q_{0}^{2}i}{q_{0}^{2}+q_{1}^{2}}t}.
\end{equation}

\subsection{ Pure 2-Soliton Solution}
With $J=2$,  and  $b(\xi,t)=0$ (reflectionless potential), solving the corresponding discrete system (\ref{E:closing system 3}) and from (\ref{asympN1c'}), one
yields nonsingular 2-soliton solutions with $\delta_{1}\delta_{2}=-1$.


(i) When $\delta_{1}=1$ and $\delta_{2}=-1$,
\begin{equation}
\begin{split}
&q(x,t)=\Big(i\cdot e^{i\alpha t}\Big(-4q_{0}^{5}q_{1}\cdot e^{\frac{2q_{0}^{2}x}{q_{1}}+2q_{1}\left(x+\frac{2iq_{1}\alpha t}{q_{0}^{2}+q_{1}^{2}}\right)}-4q_{0}q_{1}^{5}\cdot x^{2q_{1}x+q_{0}^{2}\left(\frac{2x}{q_{1}}+\frac{4i\alpha t}{q_{0}^{2}+q_{1}^{2}}\right)}\\
&-2q_{0}q_{1}(q_{0}^{2}-q_{1}^{2})^{2}\cdot e^{\frac{2q_{0}^{2}x}{q_{1}}+2q_{1}x+2i\alpha t}+q_{0}q_{1}(q_{0}^{2}+q_{1}^{2})^{2}\cdot e^{\frac{4q_{0}^{2}x}{q_{1}}+2i\alpha t}+q_{0}q_{1}(q_{0}^{2}+q_{1}^{2})^{2}\cdot e^{4q_{1}x+2i\alpha t}\\
&+2q_{0}^{2}(q_{0}^{4}-q_{1}^{4})\cdot e^{\frac{q_{0}^{2}x}{q_{1}}+3q_{1}x+\frac{i\alpha\left(q_{0}^{2}+3q_{1}^{2}\right)t}{q_{0}^{2}+q_{1}^{2}}}
+2q_{0}^{2}(q_{0}^{4}-q_{1}^{4})\cdot e^{\frac{3q_{0}^{2}x}{q_{1}}+i\alpha t+q_{1}\left(x+\frac{2iq_{1}\alpha t}{q_{0}^{2}+q_{1}^{2}}\right)}\\
&+2q_{1}^{2}(-q_{0}^{4}+q_{1}^{4})\cdot e^{\frac{q_{0}^{2}x}{q_{1}}+3q_{1}x+\frac{i\alpha\left(q_{1}^{2}+3q_{0}^{2}\right)t}{q_{0}^{2}+q_{1}^{2}}}
+2q_{1}^{2}(-q_{0}^{4}+q_{1}^{4})\cdot e^{\frac{(3q_{0}^{2}+q_{1}^{2})\left(\left(q_{0}^{2}+q_{1}^{2}\right)x+iq_{1}\alpha t\right)}{q_{1}(q_{0}^{2}+q_{1}^{2})}}\Big)\Big)/\\
&\Big(q_{1}\Big(-4q_{0}^{2}q_{1}^{2}\cdot e^{2q_{1}x+q_{0}^{2}\left(\frac{2x}{q_{1}}+\frac{4i\alpha t}{q_{0}^{2}+q_{1}^{2}}\right)}-4q_{0}^{2}q_{1}^{2}\cdot e^{\frac{2q_{0}^{2}x}{q_{1}}+2q_{1}\left(x+\frac{2iq_{1}\alpha t}{q_{0}^{2}+q_{1}^{2}}\right)}+2(q_{0}^{2}-q_{1}^{2})^{2}\cdot e^{\frac{2q_{0}^{2}x}{q_{1}}+2q_{1}x+2i\alpha t}\\
&+(q_{0}^{2}+q_{1}^{2})^{2}\cdot e^{\frac{4q_{0}^{2}x}{q_{1}}+2i\alpha t}+(q_{0}^{2}+q_{1}^{2})^{2}\cdot e^{4q_{1}x+2i\alpha t}\Big)\Big).
\end{split}
\end{equation}
We take $t=0$, if the above solution is singular, then \\
$(q_{0}^{2}+q_{1}^{2})(e^{\frac{2q_{0}^{2}}{q_{1}}x}+e^{2q_{1}x})=4q_{0}q_{1}e^{q_{1}x+\frac{q_{0}^{2}}{q_{1}}x}$, but $(q_{0}^{2}+q_{1}^{2})(e^{\frac{2q_{0}^{2}}{q_{1}}x}+e^{2q_{1}x})>4q_{0}q_{1}e^{q_{1}x+\frac{q_{0}^{2}}{q_{1}}x}$ since $q_{1}>q_{0}$. Hence, $q(x,0)$ is nonsingular. Similarly, we can deduce $q(x,t)$ is nonsingular.

(ii) When $\delta_{1}=-1$ and $\delta_{2}=1$,
\begin{equation}
\begin{split}
&q(x,t)=\Big(i\cdot e^{i\alpha t}\Big(-4q_{0}^{5}q_{1}\cdot e^{\frac{2q_{0}^{2}x}{q_{1}}+2q_{1}\left(x+\frac{2iq_{1}\alpha t}{q_{0}^{2}+q_{1}^{2}}\right)}-4q_{0}q_{1}^{5}\cdot e^{2q_{1}x+q_{0}^{2}\left(\frac{2x}{q_{1}}+\frac{4i\alpha t}{q_{0}^{2}+q_{1}^{2}}\right)}\\
&-2q_{0}q_{1}(q_{0}^{2}-q_{1}^{2})^{2}\cdot e^{\frac{2q_{0}^{2}x}{q_{1}}+2q_{1}x+2i\alpha t}+q_{0}q_{1}(q_{0}^{2}+q_{1}^{2})^{2}\cdot e^{\frac{4q_{0}^{2}x}{q_{1}}+2i\alpha t}+q_{0}q_{1}(q_{0}^{2}+q_{1}^{2})^{2}\cdot e^{4q_{1}x+2i\alpha t}\\
&-2q_{0}^{2}(q_{0}^{4}-q_{1}^{4})\cdot e^{\frac{q_{0}^{2}x}{q_{1}}+3q_{1}x+\frac{i\alpha\left(q_{0}^{2}+3q_{1}^{2}\right)t}{q_{0}^{2}+q_{1}^{2}}}
-2q_{0}^{2}(q_{0}^{4}-q_{1}^{4})\cdot e^{\frac{3q_{0}^{2}x}{q_{1}}+i\alpha t+q_{1}\left(x+\frac{2iq_{1}\alpha t}{q_{0}^{2}+q_{1}^{2}}\right)}\\
&+2q_{1}^{2}(q_{0}^{4}-q_{1}^{4})\cdot e^{\frac{q_{0}^{2}x}{q_{1}}+3q_{1}x+\frac{i\alpha\left(q_{1}^{2}+3q_{0}^{2}\right)t}{q_{0}^{2}+q_{1}^{2}}}
-2q_{1}^{2}(-q_{0}^{4}+q_{1}^{4})\cdot e^{\frac{(3q_{0}^{2}+q_{1}^{2})\left(\left(q_{0}^{2}+q_{1}^{2}\right)x+iq_{1}\alpha t\right)}{q_{1}(q_{0}^{2}+q_{1}^{2})}}\Big)\Big)/\\
&\Big(q_{1}\Big(-4q_{0}^{2}q_{1}^{2}\cdot e^{2q_{1}x+q_{0}^{2}\left(\frac{2x}{q_{1}}+\frac{4i\alpha t}{q_{0}^{2}+q_{1}^{2}}\right)}-4q_{0}^{2}q_{1}^{2}\cdot e^{\frac{2q_{0}^{2}x}{q_{1}}+2q_{1}\left(x+\frac{2iq_{1}\alpha t}{q_{0}^{2}+q_{1}^{2}}\right)}+2(q_{0}^{2}-q_{1}^{2})^{2}\cdot e^{\frac{2q_{0}^{2}x}{q_{1}}+2q_{1}x+2i\alpha t}\\
&+(q_{0}^{2}+q_{1}^{2})^{2}\cdot e^{\frac{4q_{0}^{2}x}{q_{1}}+2i\alpha t}+(q_{0}^{2}+q_{1}^{2})^{2}\cdot e^{4q_{1}x+2i\alpha t}\Big)\Big).
\end{split}
\end{equation}
Similarly, we have the above 2-soliton solution, which is also nonsingular.

\begin{figure}[h]
\begin{tabular}{cc}
\includegraphics[width=0.5\textwidth]{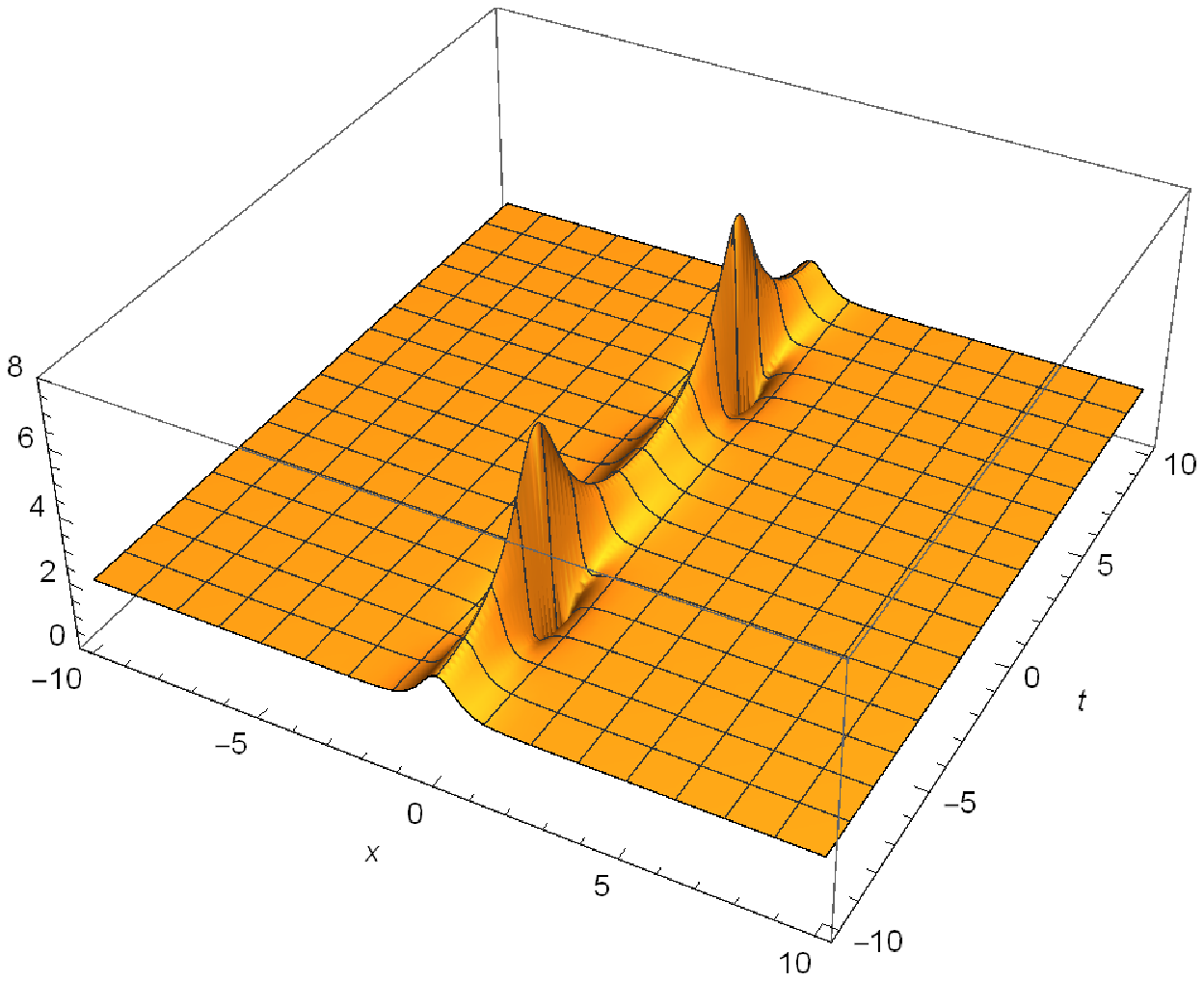}&
\includegraphics[width=0.5\textwidth]{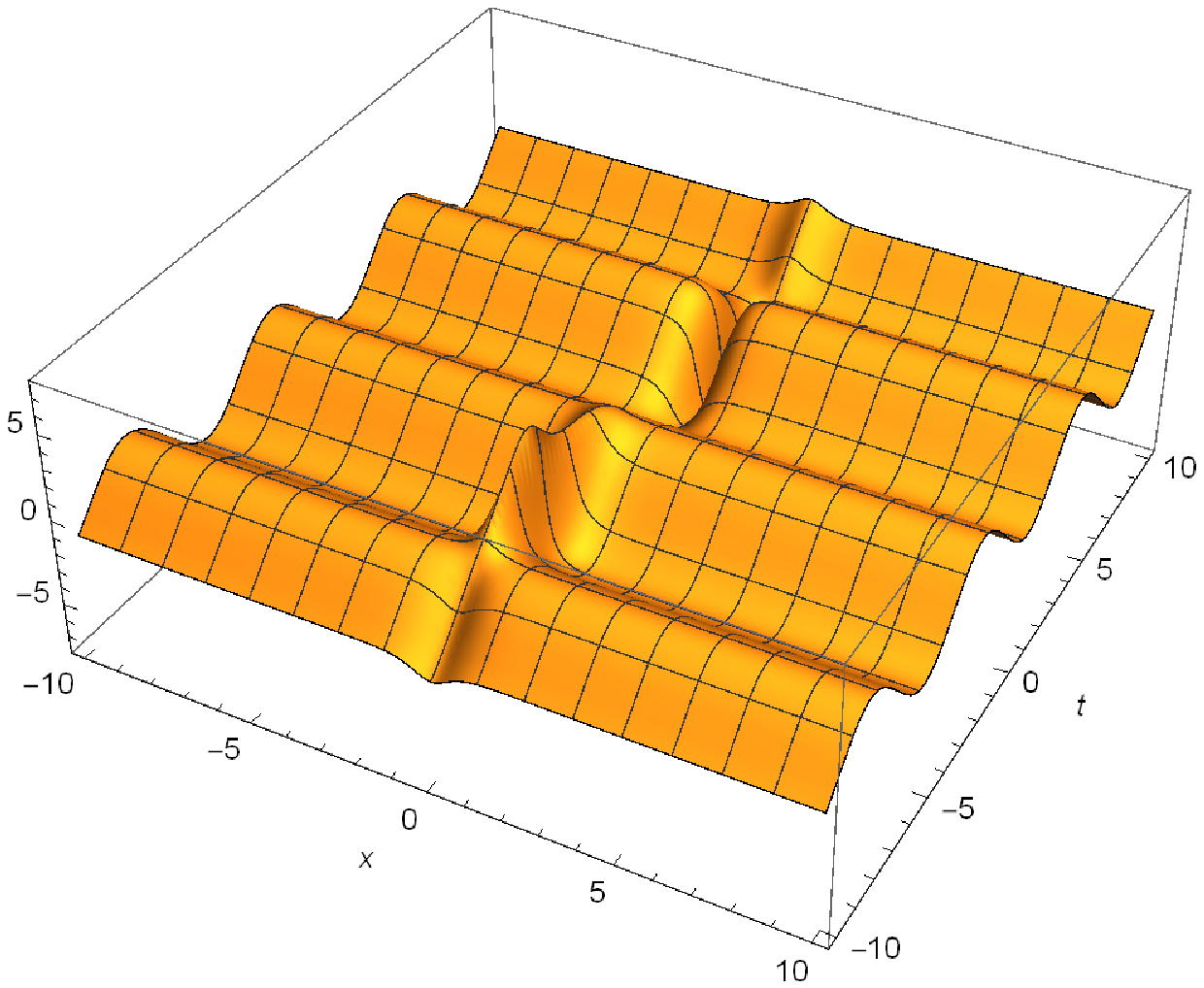}\\
(a) & (b)
\end{tabular}
\caption{(a) The amplitude of $q(x,t)$ with $\delta_{1}=1$, $\delta_{2}=-1$, $q_{1}=4$, $\alpha=1$ and $q_{0}=2$.  (b) The real part of $q(x,t)$ with $\delta_{1}=1$, $\delta_{2}=-1$, $q_{1}=4$, $\alpha=1$ and $q_{0}=2$.}
\label{Sinh 3}
\end{figure}

In Fig. \ref{Sinh 3} below we give a nonsingular 2-soliton solution.

When $\delta_{1}\delta_{2}=1$, we find the 2-soliton solutions are singular along certain space-time  lines (See Section 10).


\section{Nonlocal Sine-Gordon equation}
The nonlocal Sine-Gordon equation
\begin{equation}
q_{xt}(x,t)+2s(x,t)q(x,t)=0,
\end{equation}
where $s_{x}(x,t)=(q(x,t)q(-x,-t))_{t}$, $s(x,t) \rightarrow 0$ as $|x| \rightarrow \infty$
and $s(-x,-t)=s(x,t)$, is associated with the following $2\times2$
compatible systems:
\begin{equation}
\label{E:nonlocal sine scattering}
v_{x}=Xv=
\left(\begin{array}{cc}
-ik& q(x,t)\\
-q(-x,-t)& ik
\end{array}\right)v,
\end{equation}
\begin{equation}
v_{t}=Tv=
\left(\begin{array}{cc}
-\frac{is(x,t)}{2k}& -\frac{iq_{t}(x,t)}{2k}\\
-\frac{iq_{t}(-x,-t)}{2k}& \frac{is(x,t)}{2k}
\end{array}\right)v,
\end{equation}
where $q(x,t)$ is a complex-valued function of the real variables $x$ and $t$ with the corresponding linear operators given by
\begin{equation}
X=
\left(\begin{array}{cc}
-ik& q(x,t)\\
-q(-x,-t)& ik
\end{array}\right),\ \
T=
\left(\begin{array}{cc}
-\frac{is(x,t)}{2k}& -\frac{iq_{t}(x,t)}{2k}\\
-\frac{iq_{t}(-x,-t)}{2k}& \frac{is(x,t)}{2k}
\end{array}\right).
\end{equation}
Alternatively the space part of the compatible system may be written in the form
\begin{equation}
v_{x}=(ikJ+Q)v, \ \ \ x\in \mathbb{R},
\end{equation}
where
\begin{equation}J=
\left(\begin{matrix}
-1& 0\\
0& 1
\end{matrix}\right),\ \ \
Q=
\left(\begin{matrix}
0& q(x,t)\\
-q(-x,-t)& 0
\end{matrix}\right).
\end{equation}
Here, $q(x,t)$ is called the potential and $k$ is a complex spectral parameter.

As $x\rightarrow\pm\infty$, the eigenfunctions of the scattering problem asymptotically satisfies
\begin{equation}
\label{E:sine-Gordon eigenvalue asymptotic}
\left(\begin{array}{cc}
v_{1}\\
v_{2}
\end{array}\right)_{x}
=
\left(\begin{array}{cc}
-ik& q_{0}e^{i(\alpha t+\theta_{\pm})}\\
-q_{0}e^{i(-\alpha t+\theta_{\mp})}& ik
\end{array}\right)
\left(\begin{array}{cc}
v_{1}\\
v_{2}
\end{array}\right),
\end{equation}
i.e.,
\begin{equation}
v_{x}=(ikJ+Q_{\pm}(t))v, \ \ \ Q_{\pm}(t)=\left(\begin{array}{cc}
0& q_{0}e^{i(\alpha t+\theta_{\pm})}\\
-q_{0}e^{i(-\alpha t+\theta_{\mp})}& 0
\end{array}\right).
\end{equation}
\section{Nonlocal Sine-Gordon equation with $\theta_{+}+\theta_{-}=\pi$}
In this section, we consider the nonzero boundary conditions (NZBCs) given above in (\ref{E:NZBC}) and $\theta_{+}+\theta_{-}=\pi$. With this condition, equation (\ref{E:sine-Gordon eigenvalue asymptotic}) conveniently reduces to
\begin{equation}
\frac{\partial^{2}v_{j}}{\partial x^{2}}=-(k^{2}-q_{0}^{2})v_{j}, \ \ j=1,2.
\end{equation}
Each of the two equations has two linearly independent solutions $e^{i\lambda x}$ and $e^{-i\lambda x}$ as $|x|\rightarrow\infty$, where
$\lambda=\sqrt{k^{2}-q_{0}^{2}}$.
This is similar to Case 1; Section 3.
\subsection{Eigenfunctions}
Introducing the eigenfunctions defined by the following boundary
conditions
\begin{equation}
\label{E:asymptotic 1'}
\phi(x,k)\sim w e^{-i\lambda x}, \ \ \ \overline{\phi}(x,k)\sim \overline{w}e^{i\lambda x}
\end{equation}
as $x\rightarrow-\infty$,
\begin{equation}
\label{E:asymptotic 2'}
\psi(x,k)\sim v e^{i\lambda x}, \ \ \ \overline{\psi}(x,k)\sim \overline{v}e^{-i\lambda x}
\end{equation}
as $x\rightarrow +\infty$,and substituting into (\ref{E:eigenvalue asymptotic}), one obtains
\begin{equation}
\label{E:boundary conditions 1'}
w=\left(\begin{array}{cc}
\lambda+k\\
-iq_{+}
\end{array}\right), \ \ \
\overline{w}=\left(\begin{array}{cc}
-iq_{-}\\
\lambda+k
\end{array}\right),
\end{equation}
\begin{equation}
\label{E:boundary conditions 2'}
v=\left(\begin{array}{cc}
-iq_{+}\\
\lambda+k
\end{array}\right), \ \ \
\overline{v}=
\left(\begin{array}{cc}
\lambda+k\\
-iq_{-}
\end{array}\right),
\end{equation}
which satisfy the boundary conditions, which are not unique.

To consider functions with constant boundary conditions, we define the bounded
eigenfunctions as follows:
\begin{equation}
\label{E:definition 1'}
M(x,k)=e^{i\lambda x}\phi(x,k), \ \ \ \overline{M}(x,k)=e^{-i\lambda x}\overline{\phi}(x,k),
\end{equation}
\begin{equation}
\label{E:definition 2'}
N(x,k)=e^{-i\lambda x}\psi(x,k), \ \ \ \overline{N}(x,k)=e^{i\lambda x}\overline{\psi}(x,k).
\end{equation}
The eigenfunctions can be represented by means of the following integral equations
\begin{equation}
M(x,k)=
\left(\begin{array}{cc}
\lambda+k\\
-iq_{+}
\end{array}\right)
+\int_{-\infty}^{+\infty}G_{-}(x-x',k)((Q-Q_{-})M)(x',k)dx',
\end{equation}
\begin{equation}
\overline{M}(x,k)=
\left(\begin{array}{cc}
-iq_{-}\\
\lambda+k
\end{array}\right)
+\int_{-\infty}^{+\infty}\overline{G}_{-}(x-x',k)((Q-Q_{-})M)(x',k)dx',
\end{equation}
\begin{equation}
N(x,k)=
\left(\begin{array}{cc}
-iq_{+}\\
\lambda+k
\end{array}\right)
+\int_{-\infty}^{+\infty}G_{+}(x-x',k)((Q-Q_{+})M)(x',k)dx',
\end{equation}
\begin{equation}
\overline{N}(x,k)=\left(\begin{array}{cc}
\lambda+k\\
-iq_{-}
\end{array}\right)
+\int_{-\infty}^{+\infty}\overline{G}_{+}(x-x',k)((Q-Q_{+})M)(x',k)dx'.
\end{equation}
Using the Fourier transform method, we get
\begin{equation}
G_{-}(x,k)=\frac{\theta(x)}{2\lambda}[(1+e^{2i\lambda x})\lambda I-i(e^{2i\lambda x}-1)(ikJ+Q_{-})],
\end{equation}
\begin{equation}
\overline{G}_{-}(x,k)=\frac{\theta(x)}{2\lambda}[(1+e^{-2i\lambda x})\lambda I+i(e^{-2i\lambda x}-1)(ikJ+Q_{-})],
\end{equation}
\begin{equation}
G_{+}(x,k)=-\frac{\theta(-x)}{2\lambda}[(1+e^{-2i\lambda x})\lambda I+i(e^{-2i\lambda x}-1)(ikJ+Q_{+})],
\end{equation}
\begin{equation}
\overline{G}_{+}(x,k)=-\frac{\theta(-x)}{2\lambda}[(1+e^{2i\lambda x})\lambda I-i(e^{2i\lambda x}-1)(ikJ+Q_{+})],
\end{equation}
where $\theta(x)$ is the Heaviside function, i.e., $\theta(x)=1$ if $x>0$ and $\theta(x)=0$ if $x<0$.

The analyticity of eigenfunctions and definitions of scattering data are the same as Case 1, Section 3.
\subsection{Symmetry reductions}
Taking into account boundary conditions, we can obtain
\begin{equation}
\psi(x,k)=\left(\begin{array}{cc}
0& 1\\
1& 0
\end{array}\right)\phi(-x,k)
\end{equation}
and
\begin{equation}
\overline{\psi}(x,k)=\left(\begin{array}{cc}
0& 1\\
1& 0
\end{array}\right)\overline{\phi}(-x,k).
\end{equation}
Similarly, we can get the symmetry relations of the eigenfunctions, i.e.,
\begin{equation}
N(x,k)=\left(\begin{array}{cc}
0& 1\\
1& 0
\end{array}\right)M(-x,k)
\end{equation}
and
\begin{equation}
\overline{N}(x,k)=\left(\begin{array}{cc}
0& 1\\
1& 0
\end{array}\right)\overline{M}(-x,k).
\end{equation}
Moreover,
\begin{equation}
\overline{b}(k)=-b(k).
\end{equation}
\subsection{Uniformization coordinates}
We introduce a uniformization variable $z$, defined by the conformal mapping:
\begin{equation}
z=z(k)=k+\lambda(k),
\end{equation}
where $\lambda= \sqrt{k^2-q_0^2}$ and the inverse mapping is given by
\begin{equation}
k=k(z)=\frac{1}{2}\left(z+\frac{q_{0}^{2}}{z}\right).
\end{equation}
Then
\begin{equation}
\lambda(z)=\frac{1}{2}\left(z-\frac{q_{0}^{2}}{z}\right).
\end{equation}
\subsection{Symmetries via uniformization coordinates}
From the above eigenfunction symmetries, we have
\begin{equation}
\psi(x,z)=\left(\begin{array}{cc}
0& 1\\
1& 0
\end{array}\right)\phi(-x,z),
\end{equation}
\begin{equation}
\overline{\psi}(x,z)=\left(\begin{array}{cc}
0& 1\\
1& 0
\end{array}\right)\overline{\phi}(-x,z).
\end{equation}
Further, when $z\rightarrow \frac{q_{0}^{2}}{z}$, then $(k, \lambda)\rightarrow (k, -\lambda)$ we have
\begin{equation}
\phi\left(x,\frac{q_{0}^{2}}{z}\right)=\frac{\frac{q_{0}^{2}}{z}}{-iq_{-}}\overline{\phi}(x,z), \ \ \
\psi\left(x, \frac{q_{0}^{2}}{z}\right)=\frac{-iq_{+}}{z}\overline{\psi}(x,z), \ \ \ \Im z<0.
\end{equation}
Similarly, we can get
\begin{equation}
N(x,z)=\left(\begin{array}{cc}
0& 1\\
1& 0
\end{array}\right)M(-x,z),
\end{equation}
\begin{equation}
\overline{N}(x,z)=\left(\begin{array}{cc}
0& 1\\
1& 0
\end{array}\right)\overline{M}(-x,z),
\end{equation}
\begin{equation}
\overline{b}(z)=-b(z),
\end{equation}
\begin{equation}
a\left(\frac{q_{0}^{2}}{z}\right)=-e^{2i\theta_{+}}\overline{a}(z),\ \ \ \Im z<0; \ \ \ b\left(\frac{q_{0}^{2}}{z}\right)=\overline{b}(z).
\end{equation}
\subsection{Asymptotic behavior of eigenfunctions and scattering data}
In order to solve the inverse problem, one has to determine the asymptotic behavior of eigenfunctions and scattering data both as $z\rightarrow\infty$ and as $z\rightarrow 0$. From the integral equations (in terms of Green's functions), we have
\begin{equation}
M(x,z)\sim\left\{\begin{array}{ll}
\left(\begin{array}{cc}
z\\
-iq(-x)
\end{array}\right), \ \ \ z\rightarrow\infty\\
\left(\begin{array}{cc}
z\cdot\frac{q(x)}{q_{-}}\\
-iq_{+}
\end{array}\right), \ \ \ z\rightarrow 0,\\
\end{array}\right.
\end{equation}

\begin{equation}
N(x,z)\sim\left\{\begin{array}{ll}
\left(\begin{array}{cc}
-iq(x)\\
z
\end{array}\right), \ \ \ z\rightarrow\infty\\
\left(\begin{array}{cc}
-iq_{+}\\
z\cdot \frac{q(-x)}{q_{-}}
\end{array}\right), \ \ \ z\rightarrow 0,\\
\end{array}\right.
\end{equation}

\begin{equation}
\overline{M}(x,z)\sim\left\{\begin{array}{ll}
\left(\begin{array}{cc}
-iq(x)\\
z
\end{array}\right), \ \ \ z\rightarrow\infty\\
\left(\begin{array}{cc}
-iq_{-}\\
z\cdot \frac{q(-x)}{q_{+}}
\end{array}\right), \ \ \ z\rightarrow 0,\\
\end{array}\right.
\end{equation}

\begin{equation}
\overline{N}(x,z)\sim\left\{\begin{array}{ll}
\left(\begin{array}{cc}
z\\
-iq(-x)
\end{array}\right), \ \ \ z\rightarrow\infty\\
\left(\begin{array}{cc}
z\cdot\frac{q(x)}{q_{+}}\\
-iq_{-}
\end{array}\right), \ \ \ z\rightarrow 0,\\
\end{array}\right.
\end{equation}

\begin{equation}
a(z)=
\left\{\begin{array}{ll}
1,\ \ \ z\rightarrow\infty,\\
-e^{2i\theta_{+}}, \ \ \ z\rightarrow 0,\\
\end{array}\right.
\end{equation}

\begin{equation}
\overline{a}(z)=
\left\{\begin{array}{ll}
1,\ \ \ z\rightarrow\infty,\\
-e^{-2i\theta_{+}}, \ \ \ z\rightarrow 0,\\
\end{array}\right.
\end{equation}
\begin{equation}
\lim_{z\rightarrow\infty}zb(z)=0, \ \ \ \lim_{z\rightarrow0}\frac{b(z)}{z^{2}}=0.
\end{equation}
\subsection{Left and right scattering problems.}
Using similar methods in Case 1, we find
\begin{equation}
\begin{split}
\overline{N}(x,z)&=\left(\begin{array}{cc}
z\\
-iq_{-}
\end{array}\right)
+\sum_{j=1}^{J}\frac{z\cdot b(z_{j})e^{i\big(z_{j}-\frac{q_{0}^{2}}{z_{j}}\big)x}\cdot N(x,z_{j})}{(z-z_{j})z_{j}a'(z_{j})}\\
&+\frac{z}{2\pi i}\int_{-\infty}^{+\infty}\frac{\rho(\xi)}{\xi(\xi-z)}\cdot e^{i\big(\xi-\frac{q_{0}^{2}}{\xi}\big)x}\cdot N(x,\xi)d\xi,
\end{split}
\end{equation}
\begin{equation}
\begin{split}
N(x,z)&=\left(\begin{array}{cc}
-iq_{+}\\
z
\end{array}\right)
+\sum_{j=1}^{\overline{J}}\frac{z\cdot\overline{b}(\overline{z}_{j})e^{-i\big(\overline{z}_{j}-\frac{q_{0}^{2}}{\overline{z}_{j}}\big)x}\cdot \overline{N}(x,\overline{z}_{j})}{(z-\overline{z}_{j})\overline{z}_{j}\overline{a}'(\overline{z}_{j})}\\
&-\frac{z}{2\pi i}\int_{-\infty}^{+\infty}\frac{\overline{\rho}(\xi)}{\xi(\xi-z)}\cdot e^{-i\big(\xi-\frac{q_{0}^{2}}{\xi}\big)x}\cdot \overline{N}(x,\xi)d\xi,
\end{split}
\end{equation}
\begin{equation}
\begin{split}
\overline{M}(x,z)&=\left(\begin{array}{cc}
-iq_{-}\\
z
\end{array}\right)+
\sum_{j=1}^{J}\frac{-z\cdot\overline{b}(z_{j})M(x,z_{j})e^{-i\big(z_{j}-\frac{q_{0}^{2}}{z_{j}}\big)x}}{(z-z_{j})z_{j}a'(z_{j})}\\
&+\frac{z}{2\pi i}\int_{-\infty}^{+\infty}\frac{\rho^{*}(-\xi)}{\xi(\xi-z)}\cdot e^{-i\big(\xi-\frac{q_{0}^{2}}{\xi}\big)x}\cdot M(x,\xi)d\xi
\end{split}
\end{equation}
and
\begin{equation}
\begin{split}
M(x,z)&=\left(\begin{array}{cc}
z\\
-iq_{+}
\end{array}\right)+
\sum_{j=1}^{\overline{J}}\frac{-z\cdot b(\overline{z}_{j})\overline{M}(x,\overline{z}_{j})e^{i\big(\overline{z}_{j}-\frac{q_{0}^{2}}{\overline{z}_{j}}\big)x}}
{(z-\overline{z}_{j})\overline{z}_{j}\overline{a}'(\overline{z}_{j})}\\
&-\frac{z}{2\pi i}\int_{-\infty}^{+\infty}\frac{\overline{\rho}^{*}(-\xi)}{\xi(\xi-z)}\cdot e^{i\big(\xi-\frac{q_{0}^{2}}{\xi}\big)x}\cdot \overline{M}(x,\xi)d\xi.
\end{split}
\end{equation}
\subsection{Recovery of the potentials}
In order to reconstruct the potential, we use asymptotics. Note that
\begin{equation}
\frac{\overline{N}_{1}(x,z)}{z}\sim \frac{q(x)}{q_{+}}
\end{equation}
as $z\rightarrow 0$, and
\begin{equation}
\frac{\overline{N}_{1}(x,z)}{z}\sim 1+\sum_{j=1}^{J}\frac{b(z_{j})e^{i\big(z_{j}-\frac{q_{0}^{2}}{z_{j}}\big)x}}{-z_{j}^{2}a'(z_{j})}\cdot N_{1}(x,z_{j})+\frac{1}{2\pi i}\int_{-\infty}^{+\infty}\frac{\rho(\xi)}{\xi^{2}}\cdot e^{i\big(\xi-\frac{q_{0}^{2}}{\xi}\big)x}\cdot N_{1}(x,\xi)d\xi
\end{equation}
as $z\rightarrow 0$, we have
\begin{equation}
\label{asympN1c''}
q(x)=q_{+}\cdot\left[1+\sum_{j=1}^{J}\frac{b(z_{j})e^{i\big(z_{j}-\frac{q_{0}^{2}}{z_{j}}\big)x}}{-z_{j}^{2}a'(z_{j})}\cdot N_{1}(x,z_{j})+\frac{1}{2\pi i}\int_{-\infty}^{+\infty}\frac{\rho(\xi)}{\xi^{2}}\cdot e^{i\big(\xi-\frac{q_{0}^{2}}{\xi}\big)x}\cdot N_{1}(x,\xi)d\xi\right].
\end{equation}
\subsection{Closing the system}
We can find $J=\overline{J}$ from $a\left(\frac{q_{0}^{2}}{z}\right)=-e^{2i\theta_{+}}\overline{a}(z)$.
Combining the above integral equations, we have
\begin{equation}
\label{E:closing system 5}
\begin{split}
&\left(\begin{array}{cc}
N_{1}(x,z)\\
N_{2}(x,z)
\end{array}\right)=\left(\begin{array}{cc}
-iq_{+}\\
z
\end{array}\right)
+\sum_{j=1}^{J}\frac{z\cdot\overline{b}(\overline{z}_{j})e^{-i\big(\overline{z}_{j}-\frac{q_{0}^{2}}{\overline{z}_{j}}\big)x} }{(z-\overline{z}_{j})\overline{z}_{j}\overline{a}'(\overline{z}_{j})}\cdot\\
&\left(\begin{array}{cc}
\overline{z}_{j}+\sum_{l=1}^{J}\frac{\overline{z}_{j}\cdot b(z_{l})e^{i\big(z_{l}-\frac{q_{0}^{2}}{z_{l}}\big)x}}{(\overline{z}_{j}-z_{l})z_{l}a'(z_{l})}\cdot N_{1}(x, z_{l})+\frac{\overline{z}_{j}}{2\pi i}\int_{-\infty}^{+\infty}\frac{\rho(\xi)}{\xi(\xi-\overline{z}_{j})}\cdot e^{i\big(\xi-\frac{q_{0}^{2}}{\xi}\big)x}\cdot N_{1}(x, \xi)d\xi\\
-iq_{-}+\sum_{l=1}^{J}\frac{\overline{z}_{j}\cdot b(z_{l})e^{i\big(z_{l}-\frac{q_{0}^{2}}{z_{l}}\big)x}}{(\overline{z}_{j}-z_{l})z_{l}a'(z_{l})}\cdot N_{2}(x, z_{l})+\frac{\overline{z}_{j}}{2\pi i}\int_{-\infty}^{+\infty}\frac{\rho(\xi)}{\xi(\xi-\overline{z}_{j})}\cdot e^{i\big(\xi-\frac{q_{0}^{2}}{\xi}\big)x}\cdot N_{2}(x, \xi)d\xi
\end{array}\right)\\
&-\frac{z}{2\pi i}\int_{-\infty}^{+\infty}\frac{\overline{\rho}(\xi)}{\xi(\xi-z)}\cdot e^{-i\big(\xi-\frac{q_{0}^{2}}{\xi}\big)x}\cdot\\
&\left(\begin{array}{cc}
\xi+\sum_{l=1}^{J}\frac{\xi\cdot b(z_{l})e^{i\big(z_{l}-\frac{q_{0}^{2}}{z_{l}}\big)x}}{(\xi-z_{l})z_{l}a'(z_{l})}\cdot N_{1}(x, z_{l})+\frac{\xi}{2\pi i}\int_{-\infty}^{+\infty}\frac{\rho(\eta)}{\eta(\eta-\xi)}\cdot e^{i\big(\eta-\frac{q_{0}^{2}}{\eta}\big)x}\cdot N_{1}(x, \eta)d\eta\\
-iq_{-}+\sum_{l=1}^{J}\frac{\xi\cdot b(z_{l})e^{i\big(z_{l}-\frac{q_{0}^{2}}{z_{l}}\big)x}}{(\xi-z_{l})z_{l}a'(z_{l})}\cdot N_{2}(x, z_{l})+\frac{\xi}{2\pi i}\int_{-\infty}^{+\infty}\frac{\rho(\eta)}{\eta(\eta-\xi)}\cdot e^{i\big(\eta-\frac{q_{0}^{2}}{\eta}\big)x}\cdot N_{2}(x, \eta)d\eta
\end{array}\right)d\xi,
\end{split}
\end{equation}
\begin{equation}
\begin{split}
&\left(\begin{array}{cc}
\overline{M}_{1}(x,z)\\
\overline{M}_{2}(x,z)
\end{array}\right)=\left(\begin{array}{cc}
-iq_{-}\\
z
\end{array}\right)+
\sum_{j=1}^{J}\frac{-z\cdot\overline{b}(z_{j})e^{-i\big(z_{j}-\frac{q_{0}^{2}}{z_{j}}\big)x}}{(z-z_{j})z_{j}a'(z_{j})}\cdot\\
& \left(\begin{array}{cc}
z_{j}+
\sum_{l=1}^{J}\frac{-z_{j}\cdot b(\overline{z}_{l})e^{i\big(\overline{z}_{l}-\frac{q_{0}^{2}}{\overline{z}_{l}}\big)x}}
{(z_{j}-\overline{z}_{l})\overline{z}_{l}\overline{a}'(\overline{z}_{l})}\cdot \overline{M}_{1}(x, \overline{z}_{l})-\frac{z_{j}}{2\pi i}\int_{-\infty}^{+\infty}\frac{\overline{\rho}^{*}(-\xi)}{\xi(\xi-z_{j})}\cdot e^{i\big(\xi-\frac{q_{0}^{2}}{\xi}\big)x}\cdot\overline{M}_{1}(x,\xi)d\xi\\
-iq_{+}+
\sum_{l=1}^{J}\frac{-z_{j}\cdot b(\overline{z}_{l})e^{i\big(\overline{z}_{l}-\frac{q_{0}^{2}}{\overline{z}_{l}}\big)x}}
{(z_{j}-\overline{z}_{l})\overline{z}_{l}\overline{a}'(\overline{z}_{l})}\cdot \overline{M}_{2}(x, \overline{z}_{l})-\frac{z_{j}}{2\pi i}\int_{-\infty}^{+\infty}\frac{\overline{\rho}^{*}(-\xi)}{\xi(\xi-z_{j})}\cdot e^{i\big(\xi-\frac{q_{0}^{2}}{\xi}\big)x}\cdot\overline{M}_{2}(x,\xi)d\xi
\end{array}\right)\\
&+\frac{z}{2\pi i}\int_{-\infty}^{+\infty}\frac{\rho^{*}(-\xi)}{\xi(\xi-z)}\cdot e^{-i\big(\xi-\frac{q_{0}^{2}}{\xi}\big)x}\cdot\\
&\left(\begin{array}{cc}
\xi+
\sum_{l=1}^{J}\frac{-\xi\cdot b(\overline{z}_{l})e^{i\big(\overline{z}_{l}-\frac{q_{0}^{2}}{\overline{z}_{l}}\big)x}}
{(\xi-\overline{z}_{l})\overline{z}_{l}\overline{a}'(\overline{z}_{l})}\cdot\overline{M}_{1}(x,\overline{z}_{l})-\frac{\xi}{2\pi i}\int_{-\infty}^{+\infty}\frac{\overline{\rho}^{*}(-\eta)}{\eta(\eta-\xi)}\cdot e^{i\big(\eta-\frac{q_{0}^{2}}{\eta}\big)x}\cdot \overline{M}_{1}(x,\eta)d\eta\\
-iq_{+}+
\sum_{l=1}^{J}\frac{-\xi\cdot b(\overline{z}_{l})e^{i\big(\overline{z}_{l}-\frac{q_{0}^{2}}{\overline{z}_{l}}\big)x}}
{(\xi-\overline{z}_{l})\overline{z}_{l}\overline{a}'(\overline{z}_{l})}\cdot\overline{M}_{2}(x,\overline{z}_{l})-\frac{\xi}{2\pi i}\int_{-\infty}^{+\infty}\frac{\overline{\rho}^{*}(-\eta)}{\eta(\eta-\xi)}\cdot e^{i\big(\eta-\frac{q_{0}^{2}}{\eta}\big)x}\cdot \overline{M}_{2}(x,\eta)d\eta
\end{array}\right)d\xi.
\end{split}
\end{equation}
We can reconstruct the potential from (\ref{asympN1c''}).
\subsection{Trace formula}
Using $\overline{b}(z)=-b(z)$ and following the analysis in Case 1, we obtain
\begin{equation}
\log a(z)=\log\left(\prod_{j=1}^{J}\frac{z-z_{j}}{z-\frac{q_{0}^{2}}{z_{j}}}\right)+\frac{1}{2\pi i}\int_{-\infty}^{+\infty}\frac{\log (1-b^{2}(\xi))}{\xi-z}d\xi, \ \ \ \Im z>0,
\end{equation}

\begin{equation}
\log \overline{a}(z)=\log\left(\prod_{j=1}^{J} \frac{z-\frac{q_{0}^{2}}{z_{j}}}{z-z_{j}}\right)-\frac{1}{2\pi i}\int_{-\infty}^{+\infty}\frac{\log (1-b^{2}(\xi))}{\xi-z}d\xi, \ \ \ \Im z<0.
\end{equation}
Since $a(z)\sim -e^{2i\theta_{+}}$ as $z\rightarrow 0$ from the trace formula when $b(\xi)=0$ in the real axis, we have the following constraint for the reflectionless potentials
\begin{equation}
\prod_{j=1}^{J}z_{j}=i q_{0}^{J}e^{i\theta_{+}}
\end{equation}
or
\begin{equation}
\prod_{j=1}^{J}z_{j}=-i q_{0}^{J}e^{i\theta_{+}}.
\end{equation}
\subsection{Discrete scattering data and their symmetries}

In order to find reflectionless potentials/solitons, we calculate the relevant scattering data:

\[ b(z_j) ~\mbox{and} ~~~\bar{b}(\bar{z}_j) ,~~j=1,2,...J.\]
and

\[a'(z_j), ~~\bar{a}'(z_j) \]
We calculate the former first, the latter functions can be calculated via the trace
formulae.

Since
\begin{equation}
N_{1}(x,z)=M_{2}(-x,z), \ \ \ N_{2}(x,z)=M_{1}(-x,z),
\end{equation}
\begin{equation}
M_{1}(x,z_{j})=b(z_{j})e^{i\big(z_{j}-\frac{q_{0}^{2}}{z_{j}}\big)x}\cdot N_{1}(x,z_{j})
\end{equation}
and
\begin{equation}
M_{2}(x,z_{j})=b(z_{j})e^{i\big(z_{j}-\frac{q_{0}^{2}}{z_{j}}\big)x}\cdot N_{2}(x,z_{j}),
\end{equation}
we have
\begin{equation}
\label{E:N1'}
N_{1}(x,z_{j})=b(z_{j})\cdot e^{-i\left(z_{j}-\frac{q_{0}^{2}}{z_{j}}\right)x}\cdot N_{2}(-x,z_{j}),
\end{equation}
\begin{equation}
\label{E:N2'}
N_{2}(x,z_{j})=b(z_{j})\cdot e^{-\left(z_{j}-\frac{q_{0}^{2}}{z_{j}}\right)x}\cdot N_{1}(-x,z_{j}).
\end{equation}
By rewriting (\ref{E:N2'}), we obtain
\begin{equation}
N_{2}(-x,z_{j})=b(z_{j})\cdot e^{\left(z_{j}-\frac{q_{0}^{2}}{z_{j}}\right)x}\cdot N_{1}(x,z_{j}).
\end{equation}
Combining (\ref{E:N1'}), we can deduce the following symmetry condition on the discrete data $b(z_j)$
\begin{equation}
\label{E:b'}
b^{2}(z_{j})=1.
\end{equation}

Similar analysis shows that $\bar{b}(\bar{z}_j) $ satisfies an analogous equation

\[ \bar{b}^{2}(\bar{z}_j)  =1,\]
i.e.,
\begin{equation}
b(z_{j})=\pm 1, \ \ \ \overline{b}(\overline{z}_{j})=\pm 1.
\end{equation}
By the symmetry relation
\begin{equation}
\overline{b}(z)=-b(z),
\end{equation}
we have
\begin{equation}
\overline{b}(z_{j})=-b(z_{j}), \ \ \ b(\overline{z}_{j})=-\overline{b}(\overline{z}_{j}).
\end{equation}
When $J=1$
we may assume that $-\frac{\pi}{2}<\theta_{+}<\frac{\pi}{2}$, then $z_{1}=q_{0}e^{i(\theta_{+}+\frac{\pi}{2})}$. By the trace formula when $b(\xi)=0$ in the real axis, we can get
\begin{equation}
a'(z_{1})=a'(q_{0}e^{i(\theta_{+}+\frac{\pi}{2})})=\frac{1}{q_{0}(e^{i(\theta_{+}+\frac{\pi}{2})}-e^{-i(\theta_{+}+\frac{\pi}{2})})},
\end{equation}
\begin{equation}
\overline{a}'(\overline{z}_{1})=a'(q_{0}e^{-i(\theta_{+}+\frac{\pi}{2})})=\frac{1}{q_{0}(e^{-i(\theta_{+}+\frac{\pi}{2})}-e^{i(\theta_{+}+\frac{\pi}{2})})}.
\end{equation}
Moreover, from the symmetry relation
\begin{equation}
b\left(\frac{q_{0}^{2}}{z}\right)=\overline{b}(z),
\end{equation}
we have
\begin{equation}
\overline{b}(q_{0}e^{-i(\theta_{+}+\frac{\pi}{2})})=b(q_{0}e^{i(\theta_{+}+\frac{\pi}{2})}).
\end{equation}
For convenience, we assume $b(q_{0}e^{i(\theta_{+}+\frac{\pi}{2})})=\delta$, then $\overline{b}(q_{0}e^{-i(\theta_{+}+\frac{\pi}{2})})=\delta$,
where $\delta=\pm 1$
\subsection{Time evolution}
From the time evolution of the eigenfunctions, we can deduce that $a(t)$ and $\overline{a}(t)$ are time independent and
\begin{equation}
b(z;t)=b(z;0)e^{-i\left(\alpha-\frac{\alpha \lambda}{k}\right)t}=b(z;0)e^{-\frac{2\alpha q_{0}^{2}i}{z^{2}+q_{0}^{2}}t},
\end{equation}
and
\begin{equation}
\overline{b}(z;t)=\overline{b}(z;0)e^{i\left(\alpha-\frac{\alpha \lambda}{k}\right)t}=\overline{b}(z;0)e^{\frac{2\alpha q_{0}^{2}i}{z^{2}+q_{0}^{2}}t}.
\end{equation}
Then we can write
\begin{equation}
b(q_{0}e^{i(\theta_{+}+\frac{\pi}{2})};t)=\delta e^{\frac{2\alpha i t}{e^{2i\theta_{+}}-1}}, \ \ \
\overline{b}(q_{0}e^{-i(\theta_{+}+\frac{\pi}{2})};t)=\delta e^{-\frac{2\alpha i t}{e^{-2i\theta_{+}}-1}}.
\end{equation}

\subsection{Pure 1-Soliton Solution}
With $J=1$ and $b(\xi,t)=0$, we have from (\ref{asympN1c''}) and (\ref{E:closing system 5})
\begin{equation}
q(x,t)=q_{0}e^{i(\alpha t+\theta_{+})}\left[1-\frac{2i\sin(\theta_{+}+\frac{\pi}{2})\cdot b(q_{0}e^{i(\theta_{+}+\frac{\pi}{2})})e^{-2q_{0}x\sin(\theta_{+}+\frac{\pi}{2})}}{q_{0}e^{2i(\theta_{+}+\frac{\pi}{2})}}\cdot N_{1}(x, q_{0}e^{i(\theta_{+}+\frac{\pi}{2})}, t)\right],
\end{equation}
where
\begin{equation}
N_{1}(x, q_{0}e^{i(\theta_{+}+\frac{\pi}{2})}, t)=\frac{-iq_{0}e^{i(\alpha t+\theta_{+})}-\overline{b}(q_{0}e^{-i(\theta_{+}+\frac{\pi}{2})})q_{0}e^{i(\theta_{+}+\frac{\pi}{2})}e^{-2q_{0}x\sin(\theta_{+}+\frac{\pi}{2})}}
{1-b(q_{0}e^{i(\theta_{+}+\frac{\pi}{2})})\overline{b}(q_{0}e^{-i(\theta_{+}+\frac{\pi}{2})})e^{-4q_{0}x\sin(\theta_{+}+\frac{\pi}{2})}}.
\end{equation}
%
Note that $\delta$ can be chosen either $1$ or $-1$; we find there is a nonsingular "dark" 1-soliton solution with $\delta=-1$ (see the following), and there is a "bright" 1-soliton solution which is singular along a space-time line with $\delta=1$ (see the Section 10).

\begin{figure}[h]
\begin{tabular}{cc}
\includegraphics[width=0.5\textwidth]{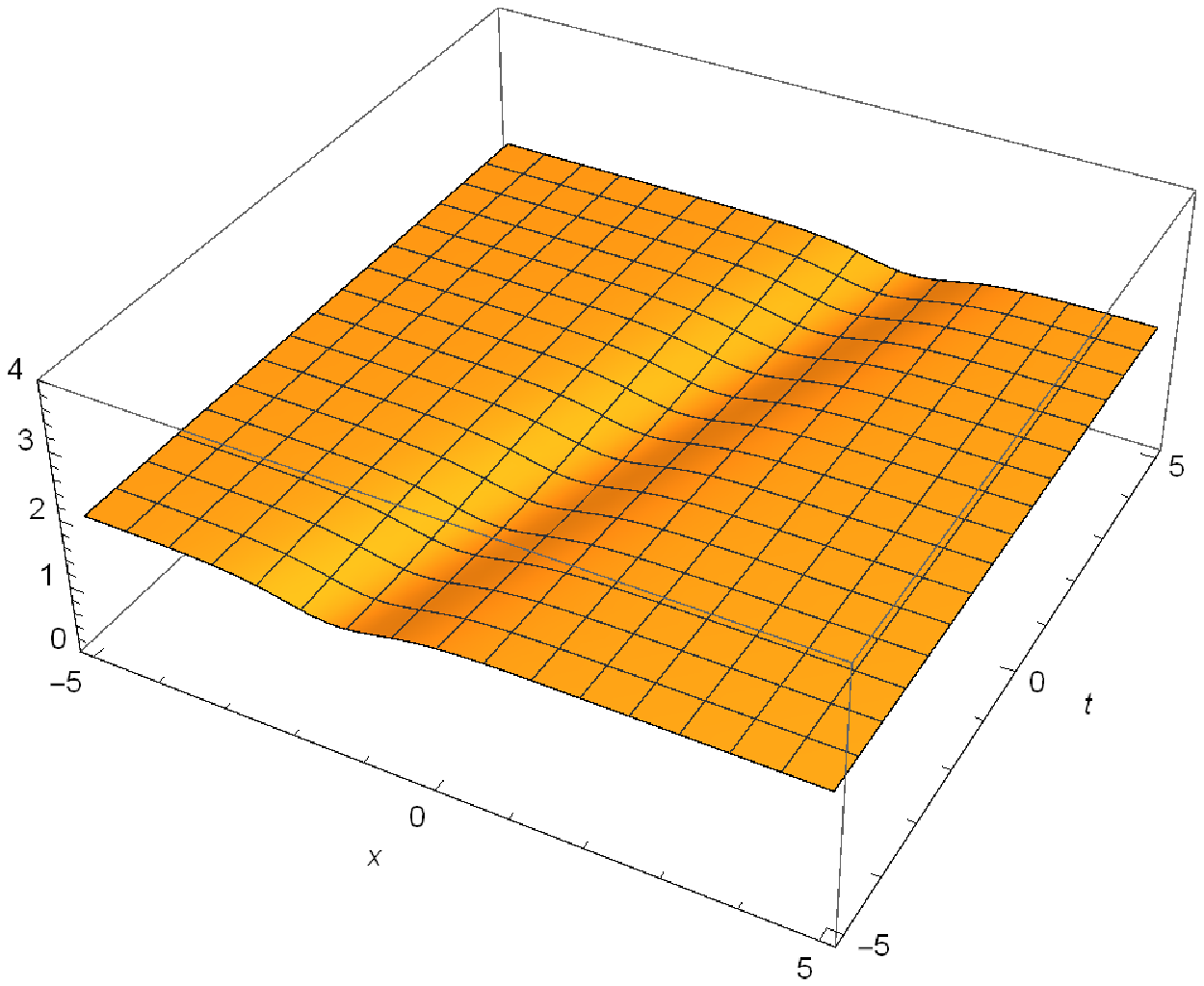}&
\includegraphics[width=0.5\textwidth]{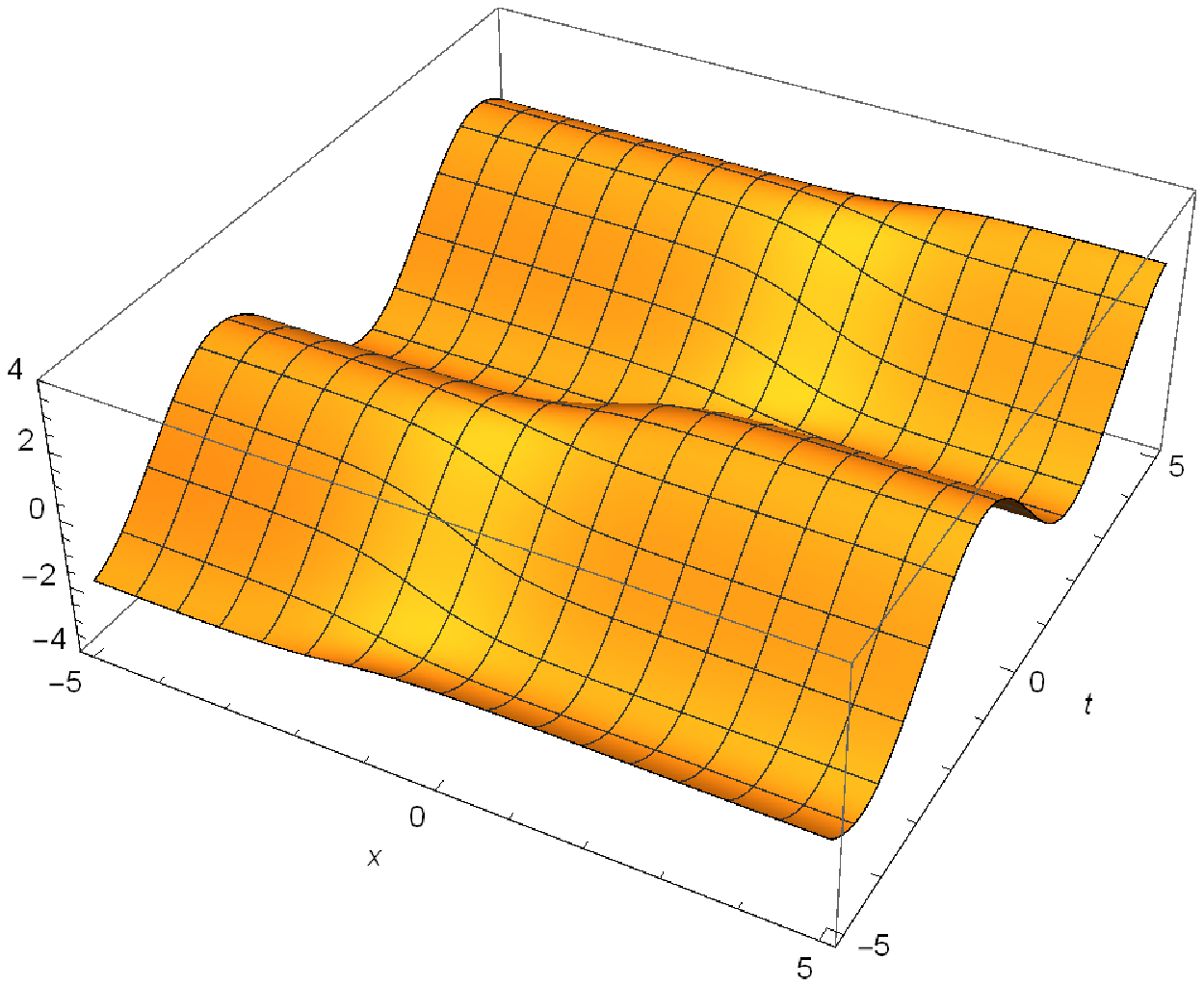}\\
(a) & (b)
\end{tabular}
\caption{(a) The amplitude of $q(x,t)$ for $\delta=-1$ with $\theta_{+}=\frac{\pi}{3}$, $\alpha=1$ and $q_{0}=2$.  (b) The real part of $q(x,t)$ for $\delta=-1$ with $\theta_{+}=\frac{\pi}{3}$, $\alpha=1$ and $q_{0}=2$.}
\label{Sine 2}
\end{figure}

When $\delta=-1$, we obtain
\begin{equation}
q(x,t)=q_{0}e^{i\alpha t}\left[i\sin\theta_{+}+\cos\theta_{+}\tanh\left(q_{0}x\cos\theta_{+}-\frac{\alpha t}{2}\cot\theta_{+}\right)\right]
\end{equation}
and
\begin{equation}
s(x,t)=-\frac{1}{2}q_{0}\alpha\cos\theta_{+}\cot\theta_{+} \text{sech}^{2}\left[q_{0}x\cos\theta_{+}-\frac{\alpha t}{2}\cot\theta_{+}\right].
\end{equation}
We see that $s(x,t) \rightarrow 0$ as $|x| \rightarrow \infty$.
This solution is not singular because $e^{q_{0}x\cos\theta_{+}-\frac{\alpha t}{2}\cot\theta_{+}}>0$.
In Fig. \ref{Sine 2} below we give a typical `dark' 1-soliton solution.

\section{Nonlocal Sine-Gordon equation with $\theta_{+}+\theta_{-}=0$}
In this section, we consider the nonzero boundary conditions (NZBCs) given above in (\ref{E:NZBC}) and $\theta_{+}+\theta_{-}=0$. With this condition, equation (\ref{E:sine-Gordon eigenvalue asymptotic}) conveniently reduces to
\begin{equation}
\frac{\partial^{2}v_{j}}{\partial x^{2}}=-(k^{2}+q_{0}^{2})v_{j}, \ \ j=1,2.
\end{equation}
Each of the two equations has two linearly independent solutions $e^{i\lambda x}$ and $e^{-i\lambda x}$ as $|x|\rightarrow\infty$, where
$\lambda=\sqrt{k^{2}+q_{0}^{2}}$. This is similar to Case 2 discussed in Section 4.
\subsection{Eigenfunctions}
Introducing the eigenfunctions defined by the following boundary
conditions
\begin{equation}
\phi(x,k)\sim w e^{-i\lambda x}, \ \ \ \overline{\phi}(x,k)\sim \overline{w}e^{i\lambda x}
\end{equation}
as $x\rightarrow-\infty$,
\begin{equation}
\psi(x,k)\sim v e^{i\lambda x}, \ \ \ \overline{\psi}(x,k)\sim \overline{v}e^{-i\lambda x}
\end{equation}
as $x\rightarrow +\infty$, and substituting into (\ref{E:eigenvalue asymptotic}), one obtains
\begin{equation}
w=\left(\begin{array}{cc}
\lambda+k\\
-iq_{+}
\end{array}\right), \ \ \
\overline{w}=\left(\begin{array}{cc}
-iq_{-}\\
\lambda+k
\end{array}\right),
\end{equation}
\begin{equation}
v=\left(\begin{array}{cc}
-iq_{+}\\
\lambda+k
\end{array}\right), \ \ \
\overline{v}=
\left(\begin{array}{cc}
\lambda+k\\
-iq_{-}
\end{array}\right),
\end{equation}
which satisfy the boundary conditions, which are not unique.

Further, in order to consider functions with constant boundary conditions, we
define the bounded eigenfunctions as follows:
\begin{equation}
M(x,k)=e^{i\lambda x}\phi(x,k), \ \ \ \overline{M}(x,k)=e^{-i\lambda x}\overline{\phi}(x,k),
\end{equation}
\begin{equation}
N(x,k)=e^{-i\lambda x}\psi(x,k), \ \ \ \overline{N}(x,k)=e^{i\lambda x}\overline{\psi}(x,k).
\end{equation}
The eigenfunctions can be represented by means of the following integral equations
\begin{equation}
M(x,k)=
\left(\begin{array}{cc}
\lambda+k\\
-iq_{+}
\end{array}\right)
+\int_{-\infty}^{+\infty}G_{-}(x-x',k)((Q-Q_{-})M)(x',k)dx',
\end{equation}
\begin{equation}
\overline{M}(x,k)=
\left(\begin{array}{cc}
-iq_{-}\\
\lambda+k
\end{array}\right)
+\int_{-\infty}^{+\infty}\overline{G}_{-}(x-x',k)((Q-Q_{-})M)(x',k)dx',
\end{equation}
\begin{equation}
N(x,k)=
\left(\begin{array}{cc}
-iq_{+}\\
\lambda+k
\end{array}\right)
+\int_{-\infty}^{+\infty}G_{+}(x-x',k)((Q-Q_{+})M)(x',k)dx',
\end{equation}
\begin{equation}
\overline{N}(x,k)=\left(\begin{array}{cc}
\lambda+k\\
-iq_{-}
\end{array}\right)
+\int_{-\infty}^{+\infty}\overline{G}_{+}(x-x',k)((Q-Q_{+})M)(x',k)dx'.
\end{equation}
Using the Fourier transform method, we get
\begin{equation}
G_{-}(x,k)=\frac{\theta(x)}{2\lambda}[(1+e^{2i\lambda x})\lambda I-i(e^{2i\lambda x}-1)(ikJ+Q_{-})],
\end{equation}
\begin{equation}
\overline{G}_{-}(x,k)=\frac{\theta(x)}{2\lambda}[(1+e^{-2i\lambda x})\lambda I+i(e^{-2i\lambda x}-1)(ikJ+Q_{-})],
\end{equation}
\begin{equation}
G_{+}(x,k)=-\frac{\theta(-x)}{2\lambda}[(1+e^{-2i\lambda x})\lambda I+i(e^{-2i\lambda x}-1)(ikJ+Q_{+})],
\end{equation}
\begin{equation}
\overline{G}_{+}(x,k)=-\frac{\theta(-x)}{2\lambda}[(1+e^{2i\lambda x})\lambda I-i(e^{2i\lambda x}-1)(ikJ+Q_{+})],
\end{equation}
where $\theta(x)$ is the Heaviside function, i.e., $\theta(x)=1$ if $x>0$ and $\theta(x)=0$ if $x<0$.

The analyticity of eigenfunctions and definitions of scattering data are the same as Case 2, Section 4.
\subsection{Symmetry reductions}
Taking into account boundary conditions, we can obtain
\begin{equation}
\psi(x,k)=\left(\begin{array}{cc}
0& 1\\
1& 0
\end{array}\right)\phi(-x,k)
\end{equation}
and
\begin{equation}
\overline{\psi}(x,k)=\left(\begin{array}{cc}
0& 1\\
1& 0
\end{array}\right)\overline{\phi}(-x,k).
\end{equation}
Similarly, we can get the symmetry relations of the eigenfunctions, i.e.,
\begin{equation}
N(x,k)=\left(\begin{array}{cc}
0& 1\\
1& 0
\end{array}\right)M(-x,k)
\end{equation}
and
\begin{equation}
\overline{N}(x,k)=\left(\begin{array}{cc}
0& 1\\
1& 0
\end{array}\right)\overline{M}(-x,k).
\end{equation}
Moreover,
\begin{equation}
\overline{b}(k)=-b(k).
\end{equation}
\subsection{Uniformization coordinates}
Prior to discussing the properties of scattering data and solving the inverse problem, we introduce a uniformization variable $z$, defined by the conformal mapping:
\begin{equation}
z=z(k)=k+\lambda(k),
\end{equation}
where $ \lambda= \sqrt{k^2+q_0^2}$ and the inverse mapping is given by
\begin{equation}
k=k(z)=\frac{1}{2}\left(z-\frac{q_{0}^{2}}{z}\right).
\end{equation}
Then
\begin{equation}
\lambda(z)=\frac{1}{2}\left(z+\frac{q_{0}^{2}}{z}\right).
\end{equation}
\subsection{Symmetries via uniformization coordinates}
From the eigenfunction symmetries, one has
\begin{equation}
\psi(x,z)=\left(\begin{array}{cc}
0& 1\\
1& 0
\end{array}\right)\phi(-x,z),
\end{equation}
\begin{equation}
\overline{\psi}(x,z)=\left(\begin{array}{cc}
0& 1\\
1& 0
\end{array}\right)\overline{\phi}(-x,z).
\end{equation}
Furthermore, when $z\rightarrow -\frac{q_{0}^{2}}{z}$, then $(k, \lambda)\rightarrow (k, -\lambda)$. Hence,
\begin{equation}
\phi\left(x,-\frac{q_{0}^{2}}{z}\right)=\frac{\frac{q_{0}^{2}}{z}}{iq_{-}}\overline{\phi}(x,z), \ \ \ \psi\left(x,-\frac{q_{0}^{2}}{z}\right)=\frac{-iq_{+}}{z}\overline{\psi}(x,z), \ \ \ z\in D^{-}.
\end{equation}
Similarly, we can get
\begin{equation}
N(x,z)=\left(\begin{array}{cc}
0& 1\\
1& 0
\end{array}\right)M(-x,z),
\end{equation}
\begin{equation}
\overline{N}(x,z)=\left(\begin{array}{cc}
0& 1\\
1& 0
\end{array}\right)\overline{M}(-x,z),
\end{equation}
\begin{equation}
\overline{b}(z)=-b(z),
\end{equation}
\begin{equation}
a\left(-\frac{q_{0}^{2}}{z}\right)=e^{2i\theta_{+}}\overline{a}(z),\ \ \ z\in D^{-}; \ \ \ b\left(-\frac{q_{0}^{2}}{z}\right)=\overline{b}(z).
\end{equation}
\subsection{Asymptotic behavior of eigenfunctions and scattering data}
In order to solve the inverse problem, one has to determine the asymptotic behavior of eigenfunctions and scattering data both as $z\rightarrow\infty$ and as $z\rightarrow 0$. From the integral equations (in terms of Green's functions), we have
\begin{equation}
M(x,z)\sim\left\{\begin{array}{ll}
\left(\begin{array}{cc}
z\\
-iq(-x)
\end{array}\right), \ \ \ z\rightarrow\infty\\
\left(\begin{array}{cc}
z\cdot\frac{q(x)}{q_{-}}\\
-iq_{+}
\end{array}\right), \ \ \ z\rightarrow 0,\\
\end{array}\right.
\end{equation}

\begin{equation}
N(x,z)\sim\left\{\begin{array}{ll}
\left(\begin{array}{cc}
-iq(x)\\
z
\end{array}\right), \ \ \ z\rightarrow\infty\\
\left(\begin{array}{cc}
-iq_{+}\\
z\cdot \frac{q(-x)}{q_{-}}
\end{array}\right), \ \ \ z\rightarrow 0,\\
\end{array}\right.
\end{equation}

\begin{equation}
\overline{M}(x,z)\sim\left\{\begin{array}{ll}
\left(\begin{array}{cc}
-iq(x)\\
z
\end{array}\right), \ \ \ z\rightarrow\infty\\
\left(\begin{array}{cc}
-iq_{-}\\
z\cdot \frac{q(-x)}{q_{+}}
\end{array}\right), \ \ \ z\rightarrow 0,\\
\end{array}\right.
\end{equation}

\begin{equation}
\overline{N}(x,z)\sim\left\{\begin{array}{ll}
\left(\begin{array}{cc}
z\\
-iq(-x)
\end{array}\right), \ \ \ z\rightarrow\infty\\
\left(\begin{array}{cc}
z\cdot\frac{q(x)}{q_{+}}\\
-iq_{-}
\end{array}\right), \ \ \ z\rightarrow 0,\\
\end{array}\right.
\end{equation}

\begin{equation}
a(z)=
\left\{\begin{array}{ll}
1,\ \ \ z\rightarrow\infty,\\
e^{2i\theta_{+}}, \ \ \ z\rightarrow 0,\\
\end{array}\right.
\end{equation}

\begin{equation}
\overline{a}(z)=
\left\{\begin{array}{ll}
1,\ \ \ z\rightarrow\infty,\\
e^{-2i\theta_{+}}, \ \ \ z\rightarrow 0,\\
\end{array}\right.
\end{equation}
\begin{equation}
\lim_{z\rightarrow\infty}zb(z)=0, \ \ \ \lim_{z\rightarrow0}\frac{b(z)}{z^{2}}=0.
\end{equation}
\subsection{Left and right scattering problems.}
Using similar methods in Case 2, we find
\begin{equation}
\begin{split}
\overline{N}(x,z)&=\left(\begin{array}{cc}
z\\
-iq_{-}
\end{array}\right)
+\sum_{j=1}^{J}\frac{z\cdot b(z_{j})e^{i\big(z_{j}+\frac{q_{0}^{2}}{z_{j}}\big)x}\cdot N(x,z_{j})}{(z-z_{j})z_{j}a'(z_{j})}\\
&+\frac{z}{2\pi i}\int_{\Sigma}\frac{\rho(\xi)}{\xi(\xi-z)}\cdot e^{i\big(\xi+\frac{q_{0}^{2}}{\xi}\big)x}\cdot N(x,\xi)d\xi,
\end{split}
\end{equation}
\begin{equation}
\begin{split}
N(x,z)&=\left(\begin{array}{cc}
-iq_{+}\\
z
\end{array}\right)
+\sum_{j=1}^{\overline{J}}\frac{z\cdot\overline{b}(\overline{z}_{j})e^{-i\big(\overline{z}_{j}+\frac{q_{0}^{2}}{\overline{z}_{j}}\big)x}\cdot \overline{N}(x,\overline{z}_{j})}{(z-\overline{z}_{j})\overline{z}_{j}\overline{a}'(\overline{z}_{j})}\\
&-\frac{z}{2\pi i}\int_{\Sigma}\frac{\overline{\rho}(\xi)}{\xi(\xi-z)}\cdot e^{-i\big(\xi+\frac{q_{0}^{2}}{\xi}\big)x}\cdot \overline{N}(x,\xi)d\xi,
\end{split}
\end{equation}
\begin{equation}
\begin{split}
\overline{M}(x,z)&=\left(\begin{array}{cc}
-iq_{-}\\
z
\end{array}\right)+
\sum_{j=1}^{J}\frac{-z\cdot\overline{b}(z_{j})M(x,z_{j})e^{-i\big(z_{j}+\frac{q_{0}^{2}}{z_{j}}\big)x}}{(z-z_{j})z_{j}a'(z_{j})}\\
&+\frac{z}{2\pi i}\int_{\Sigma}\frac{\rho^{*}(-\xi^{*})}{\xi(\xi-z)}\cdot e^{-i\big(\xi+\frac{q_{0}^{2}}{\xi}\big)x}\cdot M(x,\xi)d\xi
\end{split}
\end{equation}
and
\begin{equation}
\begin{split}
M(x,z)&=\left(\begin{array}{cc}
z\\
-iq_{+}
\end{array}\right)+
\sum_{j=1}^{\overline{J}}\frac{-z\cdot b(\overline{z}_{j})\overline{M}(x,\overline{z}_{j})e^{i\big(\overline{z}_{j}+\frac{q_{0}^{2}}{\overline{z}_{j}}\big)x}}
{(z-\overline{z}_{j})\overline{z}_{j}\overline{a}'(\overline{z}_{j})}\\
&-\frac{z}{2\pi i}\int_{\Sigma}\frac{\overline{\rho}^{*}(-\xi^{*})}{\xi(\xi-z)}\cdot e^{i\big(\xi+\frac{q_{0}^{2}}{\xi}\big)x}\cdot \overline{M}(x,\xi)d\xi.
\end{split}
\end{equation}
\subsection{Recovery of the potentials}
Note that
\begin{equation}
\frac{\overline{N}_{1}(x,z)}{z}\sim \frac{q(x)}{q_{+}}
\end{equation}
as $z\rightarrow 0$, and
\begin{equation}
\frac{\overline{N}_{1}(x,z)}{z}\sim 1+\sum_{j=1}^{J}\frac{b(z_{j})e^{i\big(z_{j}+\frac{q_{0}^{2}}{z_{j}}\big)x}}{-z_{j}^{2}a'(z_{j})}\cdot N_{1}(x,z_{j})+\frac{1}{2\pi i}\int_{\Sigma}\frac{\rho(\xi)}{\xi^{2}}\cdot e^{i\big(\xi+\frac{q_{0}^{2}}{\xi}\big)x}\cdot N_{1}(x,\xi)d\xi
\end{equation}
as $z\rightarrow 0$,
we have
\begin{equation}
\label{asympN1c'''}
q(x)=q_{+}\cdot\left[1+\sum_{j=1}^{J}\frac{b(z_{j})e^{i\big(z_{j}+\frac{q_{0}^{2}}{z_{j}}\big)x}}{-z_{j}^{2}a'(z_{j})}\cdot N_{1}(x,z_{j})+\frac{1}{2\pi i}\int_{\Sigma}\frac{\rho(\xi)}{\xi^{2}}\cdot e^{i\big(\xi+\frac{q_{0}^{2}}{\xi}\big)x}\cdot N_{1}(x,\xi)d\xi\right].
\end{equation}
\subsection{Closing the system}
We can find $J=\overline{J}$ from $a\left(\frac{q_{0}^{2}}{z}\right)=e^{2i\theta_{+}}\overline{a}(z)$.
Combining the above integral equations, we have
\begin{equation}
\label{E:closing system 7}
\begin{split}
&\left(\begin{array}{cc}
N_{1}(x,z)\\
N_{2}(x,z)
\end{array}\right)=\left(\begin{array}{cc}
-iq_{+}\\
z
\end{array}\right)
+\sum_{j=1}^{J}\frac{z\cdot\overline{b}(\overline{z}_{j})e^{-i\big(\overline{z}_{j}+\frac{q_{0}^{2}}{\overline{z}_{j}}\big)x} }{(z-\overline{z}_{j})\overline{z}_{j}\overline{a}'(\overline{z}_{j})}\cdot\\
&\left(\begin{array}{cc}
\overline{z}_{j}+\sum_{l=1}^{J}\frac{\overline{z}_{j}\cdot b(z_{l})e^{i\big(z_{l}+\frac{q_{0}^{2}}{z_{l}}\big)x}}{(\overline{z}_{j}-z_{l})z_{l}a'(z_{l})}\cdot N_{1}(x, z_{l})+\frac{\overline{z}_{j}}{2\pi i}\int_{\Sigma}\frac{\rho(\xi)}{\xi(\xi-\overline{z}_{j})}\cdot e^{i\big(\xi+\frac{q_{0}^{2}}{\xi}\big)x}\cdot N_{1}(x, \xi)d\xi\\
-iq_{-}+\sum_{l=1}^{J}\frac{\overline{z}_{j}\cdot b(z_{l})e^{i\big(z_{l}+\frac{q_{0}^{2}}{z_{l}}\big)x}}{(\overline{z}_{j}-z_{l})z_{l}a'(z_{l})}\cdot N_{2}(x, z_{l})+\frac{\overline{z}_{j}}{2\pi i}\int_{\Sigma}\frac{\rho(\xi)}{\xi(\xi-\overline{z}_{j})}\cdot e^{i\big(\xi+\frac{q_{0}^{2}}{\xi}\big)x}\cdot N_{2}(x, \xi)d\xi
\end{array}\right)\\
&-\frac{z}{2\pi i}\int_{\Sigma}\frac{\overline{\rho}(\xi)}{\xi(\xi-z)}\cdot e^{-i\big(\xi+\frac{q_{0}^{2}}{\xi}\big)x}\cdot\\
&\left(\begin{array}{cc}
\xi+\sum_{l=1}^{J}\frac{\xi\cdot b(z_{l})e^{i\big(z_{l}-\frac{q_{0}^{2}}{z_{l}}\big)x}}{(\xi-z_{l})z_{l}a'(z_{l})}\cdot N_{1}(x, z_{l})+\frac{\xi}{2\pi i}\int_{\Sigma}\frac{\rho(\eta)}{\eta(\eta-\xi)}\cdot e^{i\big(\eta+\frac{q_{0}^{2}}{\eta}\big)x}\cdot N_{1}(x, \eta)d\eta\\
-iq_{-}+\sum_{l=1}^{J}\frac{\xi\cdot b(z_{l})e^{i\big(z_{l}+\frac{q_{0}^{2}}{z_{l}}\big)x}}{(\xi-z_{l})z_{l}a'(z_{l})}\cdot N_{2}(x, z_{l})+\frac{\xi}{2\pi i}\int_{\Sigma}\frac{\rho(\eta)}{\eta(\eta-\xi)}\cdot e^{i\big(\eta+\frac{q_{0}^{2}}{\eta}\big)x}\cdot N_{2}(x, \eta)d\eta
\end{array}\right)d\xi,
\end{split}
\end{equation}
\begin{equation}
\begin{split}
&\left(\begin{array}{cc}
\overline{M}_{1}(x,z)\\
\overline{M}_{2}(x,z)
\end{array}\right)=\left(\begin{array}{cc}
-iq_{-}\\
z
\end{array}\right)+
\sum_{j=1}^{J}\frac{-z\cdot\overline{b}(z_{j})e^{-i\big(z_{j}+\frac{q_{0}^{2}}{z_{j}}\big)x}}{(z-z_{j})z_{j}a'(z_{j})}\cdot\\
& \left(\begin{array}{cc}
z_{j}+
\sum_{l=1}^{J}\frac{-z_{j}\cdot b(\overline{z}_{l})e^{i\big(\overline{z}_{l}+\frac{q_{0}^{2}}{\overline{z}_{l}}\big)x}}
{(z_{j}-\overline{z}_{l})\overline{z}_{l}\overline{a}'(\overline{z}_{l})}\cdot \overline{M}_{1}(x, \overline{z}_{l})-\frac{z_{j}}{2\pi i}\int_{\Sigma}\frac{\overline{\rho}^{*}(-\xi^{*})}{\xi(\xi-z_{j})}\cdot e^{i\big(\xi+\frac{q_{0}^{2}}{\xi}\big)x}\cdot\overline{M}_{1}(x,\xi)d\xi\\
-iq_{+}+
\sum_{l=1}^{J}\frac{-z_{j}\cdot b(\overline{z}_{l})e^{i\big(\overline{z}_{l}+\frac{q_{0}^{2}}{\overline{z}_{l}}\big)x}}
{(z_{j}-\overline{z}_{l})\overline{z}_{l}\overline{a}'(\overline{z}_{l})}\cdot \overline{M}_{2}(x, \overline{z}_{l})-\frac{z_{j}}{2\pi i}\int_{\Sigma}\frac{\overline{\rho}^{*}(-\xi^{*})}{\xi(\xi-z_{j})}\cdot e^{i\big(\xi+\frac{q_{0}^{2}}{\xi}\big)x}\cdot\overline{M}_{2}(x,\xi)d\xi
\end{array}\right)\\
&+\frac{z}{2\pi i}\int_{\Sigma}\frac{\rho^{*}(-\xi^{*})}{\xi(\xi-z)}\cdot e^{-i\big(\xi+\frac{q_{0}^{2}}{\xi}\big)x}\cdot\\
&\left(\begin{array}{cc}
\xi+
\sum_{l=1}^{J}\frac{-\xi\cdot b(\overline{z}_{l})e^{i\big(\overline{z}_{l}+\frac{q_{0}^{2}}{\overline{z}_{l}}\big)x}}
{(\xi-\overline{z}_{l})\overline{z}_{l}\overline{a}'(\overline{z}_{l})}\cdot\overline{M}_{1}(x,\overline{z}_{l})-\frac{\xi}{2\pi i}\int_{\Sigma}\frac{\overline{\rho}^{*}(-\eta^{*})}{\eta(\eta-\xi)}\cdot e^{i\big(\eta+\frac{q_{0}^{2}}{\eta}\big)x}\cdot \overline{M}_{1}(x,\eta)d\eta\\
-iq_{+}+
\sum_{l=1}^{J}\frac{-\xi\cdot b(\overline{z}_{l})e^{i\big(\overline{z}_{l}+\frac{q_{0}^{2}}{\overline{z}_{l}}\big)x}}
{(\xi-\overline{z}_{l})\overline{z}_{l}\overline{a}'(\overline{z}_{l})}\cdot\overline{M}_{2}(x,\overline{z}_{l})-\frac{\xi}{2\pi i}\int_{\Sigma}\frac{\overline{\rho}^{*}(-\eta^{*})}{\eta(\eta-\xi)}\cdot e^{i\big(\eta+\frac{q_{0}^{2}}{\eta}\big)x}\cdot \overline{M}_{2}(x,\eta)d\eta
\end{array}\right)d\xi.
\end{split}
\end{equation}
We can reconstruct the potential  from (\ref{asympN1c'''}).
\subsection{Trace formula}
Using $\overline{b}(z)=-b(z)$ and following the analysis in Case 2, Section 4, we obtain
\begin{equation}
\log a(z)=\log\left(\prod_{j=1}^{J}\frac{z-z_{j}}{z+\frac{q_{0}^{2}}{z_{j}}}\right)+\frac{1}{2\pi i}\int_{\Sigma}\frac{\log (1-b^{2}(\xi))}{\xi-z}d\xi, \ \ \ z\in D^{+}
\end{equation}
and
\begin{equation}
\log \overline{a}(z)=\log\left(\prod_{j=1}^{J} \frac{z+\frac{q_{0}^{2}}{z_{j}}}{z-z_{j}}\right)-\frac{1}{2\pi i}\int_{\Sigma}\frac{\log (1-b^{2}(\xi))}{\xi-z}d\xi, \ \ \ z\in D^{-}.
\end{equation}
Since $a(z)\sim e^{2i\theta_{+}}$ as $z\rightarrow 0$, from the trace formula when $b(\xi)=0$ in $\Sigma$, we have the following constraint for the reflectionless potentials
\begin{equation}
\prod_{j=1}^{J}z_{j}=\pm (-1)^{\frac{J}{2}}q_{0}^{J}e^{i\theta_{+}}.
\end{equation}
We claim that $J\geq 2$. Otherwise, if $J=1$, then $z_{1}=\pm q_{0}e^{i(\theta_{+}+\frac{\pi}{2})}$. It implies that the eigenvalue $z_{1}$ lies on the circle, which in this case is the continuous spectrum. Such eigenvalues are not proper, which are not considered here.
\subsection{Discrete scattering data and their symmetries}
Using similar methods in Case 3, we have
\begin{equation}
b(z_{j})=\pm 1, \ \ \ \overline{b}(\overline{z}_{j})=\pm 1,
\end{equation}
\begin{equation}
\overline{b}(z_{j})=-b(z_{j}), \ \ \ b(\overline{z}_{j})=-\overline{b}(\overline{z}_{j}).
\end{equation}


Next we give the scattering data for a pure 2-eigenvalue/ solution. 
Note that $|z_{1}|\cdot|z_{2}|=q_{0}^{2}$. In particular, we choose $z_{1}=iq_{1}$, $z_{2}=-i\frac{q_{0}^{2}}{q_{1}}$ and $\theta_{+}=0$, then $\overline{z}_{1}=i\frac{q_{0}^{2}}{q_{1}}$, $\overline{z}_{2}=-iq_{1}$ and $z_{1}z_{2}=q_{0}^{2}$. Thus,
\begin{equation}
a(z)=\frac{z-iq_{1}}{z-i\frac{q_{0}^{2}}{q_{1}}}\cdot\frac{z+i\frac{q_{0}^{2}}{q_{1}}}{z+iq_{1}}
\end{equation}
and
\begin{equation}
\overline{a}(z)=\frac{z-i\frac{q_{0}^{2}}{q_{1}}}{z-iq_{1}}\cdot \frac{z+iq_{1}}{z+i\frac{q_{0}^{2}}{q_{1}}}.
\end{equation}
We have
\begin{equation}
a'(iq_{1})=\frac{-i(q_{1}^{2}+q_{0}^{2})}{2q_{1}(q_{1}^{2}-q_{0}^{2})}, \ \ \ a'\left(-i\frac{q_{0}^{2}}{q_{1}}\right)=\frac{iq_{1}(q_{0}^{2}+q_{1}^{2})}{2q_{0}^{2}(q_{0}^{2}-q_{1}^{2})},
\end{equation}
\begin{equation}
\overline{a}'\left(i\frac{q_{0}^{2}}{q_{1}}\right)=-\frac{iq_{1}(q_{0}^{2}+q_{1}^{2})}{2q_{0}^{2}(q_{0}^{2}-q_{1}^{2})}, \ \ \
\overline{a}'(-iq_{1})=\frac{i(q_{1}^{2}+q_{0}^{2})}{2q_{1}(q_{1}^{2}-q_{0}^{2})}.
\end{equation}
Since
\begin{equation}
b(iq_{1})=\delta_{1}, \ \ \ b\left(-i\frac{q_{0}^{2}}{q_{1}}\right)=\delta_{2},
\end{equation}
where $\delta_{1}=\pm 1$ and $\delta_{2}=\pm 1$, then
\begin{equation}
\overline{b}\left(i\frac{q_{0}^{2}}{q_{1}}\right)=\delta_{1}, \ \ \ \overline{b}(-iq_{1})=\delta_{2}.
\end{equation}
\subsection{Time evolution}
Similarly, we can deduce both $a(t)$ and $\overline{a}(t)$ are time independent,
\begin{equation}
b(z;t)=b(z;0)e^{-i\left(\alpha-\frac{\alpha \lambda}{k}\right)t}=b(z;0)e^{\frac{2\alpha q_{0}^{2}i}{z^{2}-q_{0}^{2}}t},
\end{equation}
and
\begin{equation}
\overline{b}(z;t)=\overline{b}(z;0)e^{i\left(\alpha-\frac{\alpha \lambda}{k}\right)t}=\overline{b}(z;0)e^{-\frac{2\alpha q_{0}^{2}i}{z^{2}-q_{0}^{2}}t}.
\end{equation}
Thus,
\begin{equation}
b(iq_{1};t)=\delta_{1} e^{-\frac{2\alpha q_{0}^{2}i}{q_{0}^{2}+q_{1}^{2}}t}, \ \ \
b\left(-i\frac{q_{0}^{2}}{q_{1}};t\right)=\delta_{2} e^{-\frac{2\alpha q_{1}^{2}i}{q_{0}^{2}+q_{1}^{2}}t},
\end{equation}
\begin{equation}
\overline{b}\left(i\frac{q_{0}^{2}}{q_{1}};t\right)=\delta_{1} e^{\frac{2\alpha q_{1}^{2}i}{q_{0}^{2}+q_{1}^{2}}t}, \ \ \
\overline{b}(-iq_{1};t)=\delta_{2} e^{\frac{2\alpha q_{0}^{2}i}{q_{0}^{2}+q_{1}^{2}}t}.
\end{equation}

\subsection{Pure 2-Soltion Solution}

There are different situations to investigate. There are nonsingular 2-soliton solutions with $\delta_{1}\delta_{2}=-1$ (see the following), and there are 2-soliton solutions with $\delta_{1}\delta_{2}=1$, which are singular along some  lines (see Section 10).

(i): When $\delta_{1}=1$ and $\delta_{2}=-1$,
\begin{equation}
\begin{split}
&q(x,t)=\Big(e^{i\alpha t}\Big(-4q_{0}^{5}q_{1}\cdot e^{\frac{2q_{0}^{2}x}{q_{1}}+2q_{1}\left(x+\frac{2iq_{1}\alpha t}{q_{0}^{2}+q_{1}^{2}}\right)}-4q_{0}q_{1}^{5}\cdot e^{2q_{1}x+q_{0}^{2}\left(\frac{2x}{q_{1}}+\frac{4i\alpha t}{q_{0}^{2}+q_{1}^{2}}\right)}\\
&-2q_{0}q_{1}(q_{0}^{2}-q_{1}^{2})^{2}\cdot e^{\frac{2q_{0}^{2}x}{q_{1}}+2q_{1}x+2i\alpha t}+q_{0}q_{1}(q_{0}^{2}+q_{1}^{2})^{2}\cdot e^{\frac{4q_{0}^{2}x}{q_{1}}+2i\alpha t}+q_{0}q_{1}(q_{0}^{2}+q_{1}^{2})^{2}\cdot e^{4q_{1}x+2i\alpha t}\\
&-2q_{0}^{2}(q_{0}^{4}-q_{1}^{4})\cdot e^{\frac{q_{0}^{2}x}{q_{1}}+3q_{1}x+\frac{i\alpha\left(q_{0}^{2}+3q_{1}^{2}\right)t}{q_{0}^{2}+q_{1}^{2}}}
-2q_{0}^{2}(q_{0}^{4}-q_{1}^{4})\cdot e^{\frac{3q_{0}^{2}x}{q_{1}}+i\alpha t+q_{1}\left(x+\frac{2iq_{1}\alpha t}{q_{0}^{2}+q_{1}^{2}}\right)}\\
&+2q_{1}^{2}(q_{0}^{4}-q_{1}^{4})\cdot e^{\frac{q_{0}^{2}x}{q_{1}}+3q_{1}x+\frac{i\alpha\left(q_{1}^{2}+3q_{0}^{2}\right)t}{q_{0}^{2}+q_{1}^{2}}}
-2q_{1}^{2}(-q_{0}^{4}+q_{1}^{4})\cdot e^{\frac{(3q_{0}^{2}+q_{1}^{2})\left(\left(q_{0}^{2}+q_{1}^{2}\right)x+iq_{1}\alpha t\right)}{q_{1}(q_{0}^{2}+q_{1}^{2})}}\Big)\Big)/\\
&\Big(q_{1}\Big(-4q_{0}^{2}q_{1}^{2}\cdot e^{2q_{1}x+q_{0}^{2}\left(\frac{2x}{q_{1}}+\frac{4i\alpha t}{q_{0}^{2}+q_{1}^{2}}\right)}-4q_{0}^{2}q_{1}^{2}\cdot e^{\frac{2q_{0}^{2}x}{q_{1}}+2q_{1}\left(x+\frac{2iq_{1}\alpha t}{q_{0}^{2}+q_{1}^{2}}\right)}+2(q_{0}^{2}-q_{1}^{2})^{2}\cdot e^{\frac{2q_{0}^{2}x}{q_{1}}+2q_{1}x+2i\alpha t}\\
&+(q_{0}^{2}+q_{1}^{2})^{2}\cdot e^{\frac{4q_{0}^{2}x}{q_{1}}+2i\alpha t}+(q_{0}^{2}+q_{1}^{2})^{2}\cdot e^{4q_{1}x+2i\alpha t}\Big)\Big).
\end{split}
\end{equation}
(ii):  When $\delta_{1}=-1$ and $\delta_{2}=1$,
\begin{equation}
\begin{split}
&q(x,t)=\Big(e^{i\alpha t}\Big(-4q_{0}^{5}q_{1}\cdot e^{\frac{2q_{0}^{2}x}{q_{1}}+2q_{1}\left(x+\frac{2iq_{1}\alpha t}{q_{0}^{2}+q_{1}^{2}}\right)}-4q_{0}q_{1}^{5}\cdot e^{2q_{1}x+q_{0}^{2}\left(\frac{2x}{q_{1}}+\frac{4i\alpha t}{q_{0}^{2}+q_{1}^{2}}\right)}\\
&-2q_{0}q_{1}(q_{0}^{2}-q_{1}^{2})^{2}\cdot e^{\frac{2q_{0}^{2}x}{q_{1}}+2q_{1}x+2i\alpha t}+q_{0}q_{1}(q_{0}^{2}+q_{1}^{2})^{2}\cdot e^{\frac{4q_{0}^{2}x}{q_{1}}+2i\alpha t}+q_{0}q_{1}(q_{0}^{2}+q_{1}^{2})^{2}\cdot e^{4q_{1}x+2i\alpha t}\\
&+2q_{0}^{2}(q_{0}^{4}-q_{1}^{4})\cdot e^{\frac{q_{0}^{2}x}{q_{1}}+3q_{1}x+\frac{i\alpha\left(q_{0}^{2}+3q_{1}^{2}\right)t}{q_{0}^{2}+q_{1}^{2}}}
+2q_{0}^{2}(q_{0}^{4}-q_{1}^{4})\cdot e^{\frac{3q_{0}^{2}x}{q_{1}}+i\alpha t+q_{1}\left(x+\frac{2iq_{1}\alpha t}{q_{0}^{2}+q_{1}^{2}}\right)}\\
&+2q_{1}^{2}(-q_{0}^{4}+q_{1}^{4})\cdot e^{\frac{q_{0}^{2}x}{q_{1}}+3q_{1}x+\frac{i\alpha\left(q_{1}^{2}+3q_{0}^{2}\right)t}{q_{0}^{2}+q_{1}^{2}}}
+2q_{1}^{2}(-q_{0}^{4}+q_{1}^{4})\cdot e^{\frac{(3q_{0}^{2}+q_{1}^{2})\left(\left(q_{0}^{2}+q_{1}^{2}\right)x+iq_{1}\alpha t\right)}{q_{1}(q_{0}^{2}+q_{1}^{2})}}\Big)\Big)/\\
&\Big(q_{1}\Big(-4q_{0}^{2}q_{1}^{2}\cdot e^{2q_{1}x+q_{0}^{2}\left(\frac{2x}{q_{1}}+\frac{4i\alpha t}{q_{0}^{2}+q_{1}^{2}}\right)}-4q_{0}^{2}q_{1}^{2}\cdot e^{\frac{2q_{0}^{2}x}{q_{1}}+2q_{1}\left(x+\frac{2iq_{1}\alpha t}{q_{0}^{2}+q_{1}^{2}}\right)}+2(q_{0}^{2}-q_{1}^{2})^{2}\cdot e^{\frac{2q_{0}^{2}x}{q_{1}}+2q_{1}x+2i\alpha t}\\
&+(q_{0}^{2}+q_{1}^{2})^{2}\cdot e^{\frac{4q_{0}^{2}x}{q_{1}}+2i\alpha t}+(q_{0}^{2}+q_{1}^{2})^{2}\cdot e^{4q_{1}x+2i\alpha t}\Big)\Big).
\end{split}
\end{equation}
We can use the similar method in Section 4 to verify the above two solutions are nonsingular.
For all these solutions  we require that $s(x,t)=-\int_{x}^{\infty}(q(x',t)q(-x',-t))_{t}dx' \rightarrow 0 ~\text{as}~ |x| \rightarrow \infty$.



\begin{figure}[h]
\begin{tabular}{cc}
\includegraphics[width=0.5\textwidth]{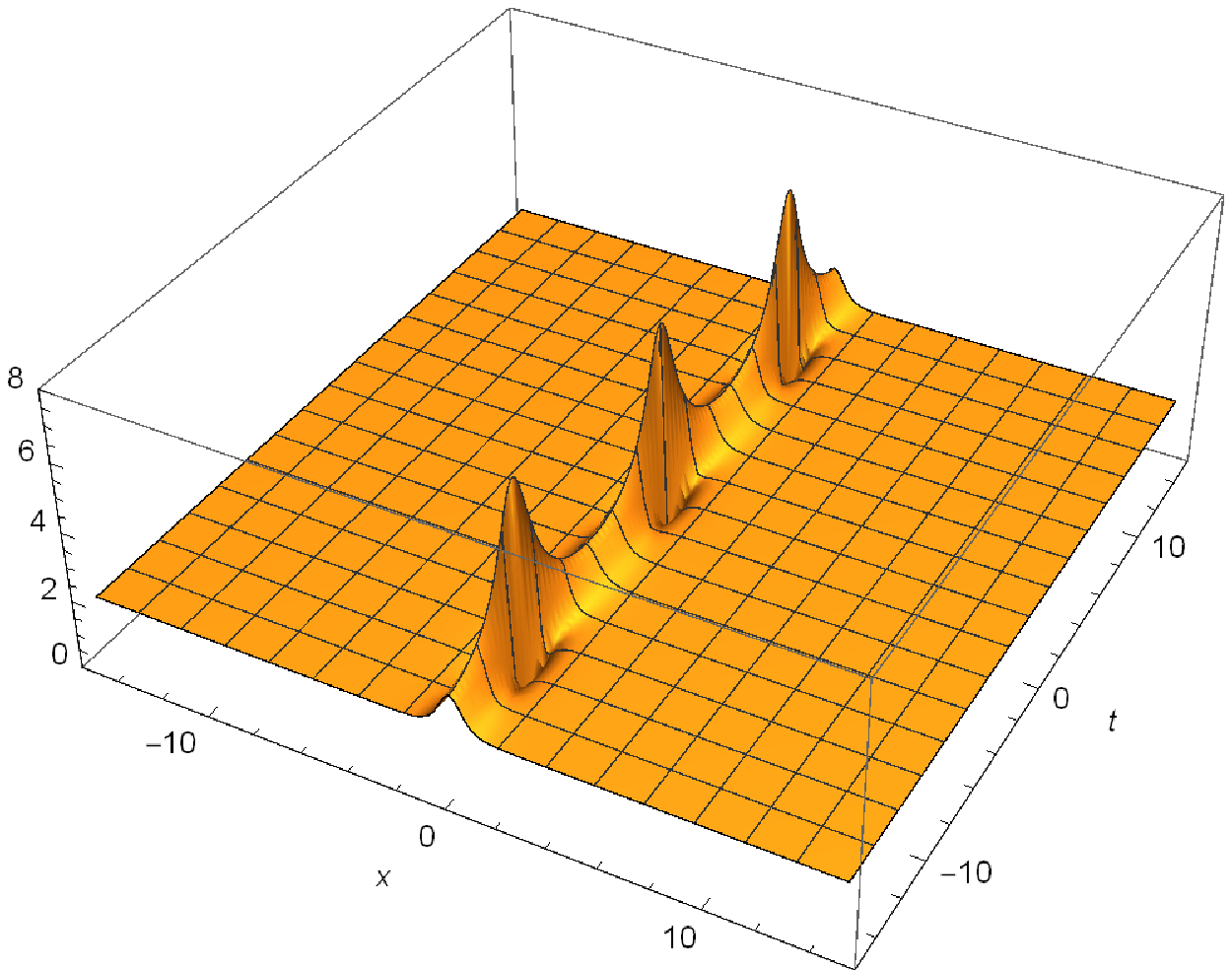}&
\includegraphics[width=0.5\textwidth]{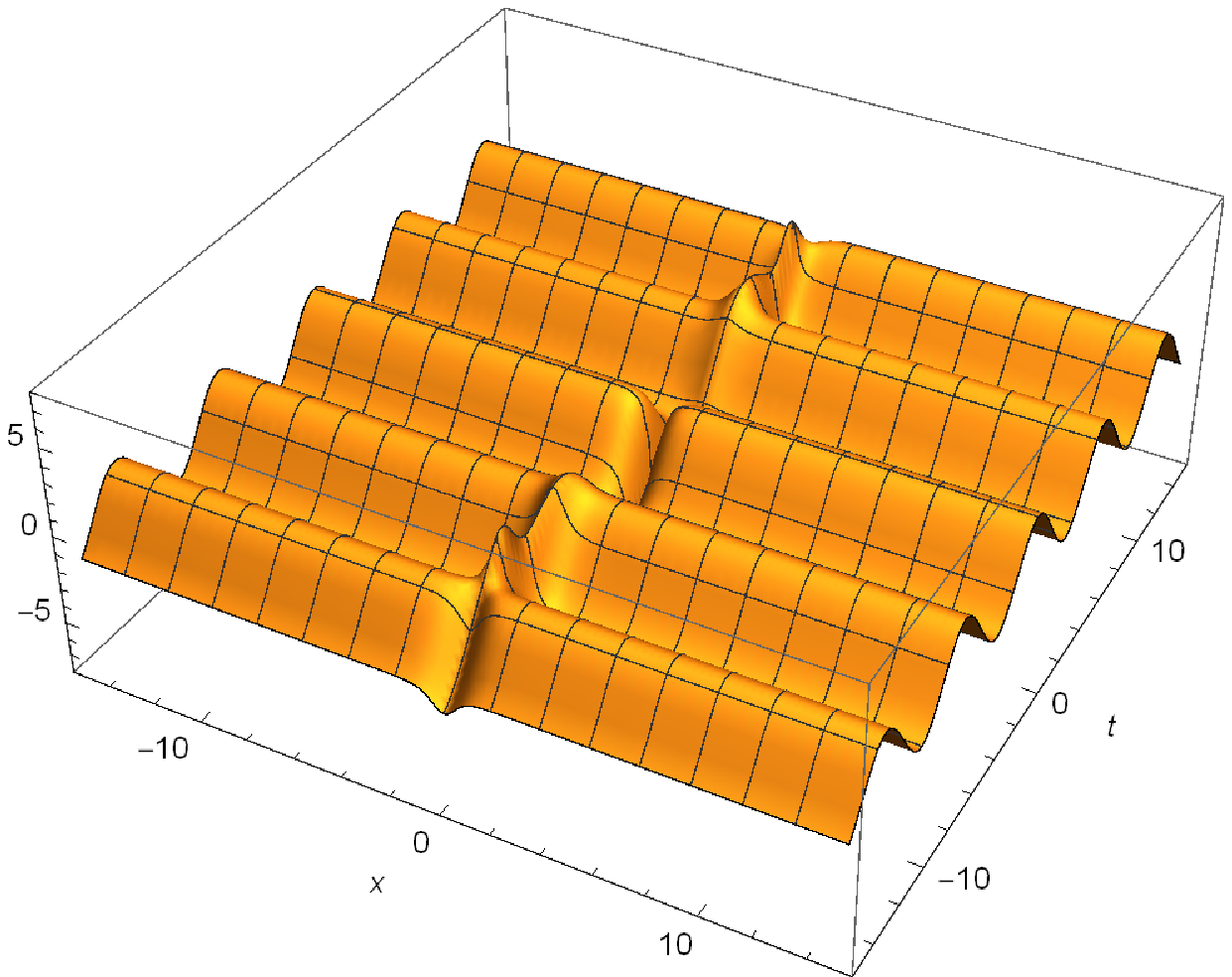}\\
(a) & (b)
\end{tabular}
\caption{(a) The amplitude of $q(x,t)$ with $\delta_{1}=1$, $\delta_{2}=-1$, $q_{1}=4$, $\alpha=1$ and $q_{0}=2$.  (b) The real part of $q(x,t)$ with $\delta_{1}=1$, $\delta_{2}=-1$, $q_{1}=4$, $\alpha=1$ and $q_{0}=2$.}
\label{Sine 3}
\end{figure}

In Fig. \ref{Sine 3} below we give a typical 2-soliton solution.

\section{The nonlocal reverse space-time NLS equation}
The scattering problem of the nonlocal reverse space-time NLS equation is the same as nonlocal Sine/Sinh-Gordon equations, the only difference between them is the time evolution equations governing the eigenfunctions, which lead to different time evolutions for scattering data ($a(t), \overline{a}(t), b(t), \overline{b}(t)$).

The nonlocal reverse space-time NLS equation is given by
\begin{equation}
iq_{t}(x,t)=q_{xx}(x,t)-2\sigma q^{2}(x,t)q(-x,-t),
\end{equation}
whose time evolution equation is
\begin{equation}
v_{t}=
\left(\begin{array}{cc}
2ik^{2}+ i\sigma q(x,t)q(-x,-t)& -2kq(x,t)-iq_{x}(x,t)\\
-2\sigma kq(-x,-t)-\sigma iq_{x}(-x,-t)& -2ik^{2}-i\sigma q(x,t)q(-x,-t)
\end{array}\right)v.
\end{equation}
The nonzero boundary conditions
\begin{equation}
q(x,t)\rightarrow q_{\pm}(t)=q_{0}e^{i(\alpha t+\theta_{\pm})}, \ \ as\  x\rightarrow\pm\infty,
\end{equation}
yields $\alpha=2\sigma q_{0}^{2}$ when $\theta_{+}+\theta_{-}=0$ and $\alpha=-2\sigma q_{0}^{2}$ when $\theta_{+}+\theta_{-}=\pi$.
By such nonzero boundary conditions
we can deduce the following.

If $\theta_{+}+\theta_{-}=0$, then both $a(t)$ and $\overline{a}(t)$ are time independent, and
\begin{equation}
b(t)=b(0)e^{-2i(\sigma q_{0}^{2}+2\lambda k)t}, \ \ \ \overline{b}(t)=\overline{b}(0)e^{2i(\sigma q_{0}^{2}+2\lambda k)t},
\end{equation}
where $\lambda=\sqrt{k^{2}-\sigma q_{0}^{2}}$.

If $\theta_{+}+\theta_{-}=\pi$, then both $a(t)$ and $\overline{a}(t)$ are time independent, and
\begin{equation}
b(t)=b(0)e^{-2i(-\sigma q_{0}^{2}+2\lambda k)t}, \ \ \ \overline{b}(t)=\overline{b}(0)e^{2i(-\sigma q_{0}^{2}+2\lambda k)t},
\end{equation}
where $\lambda=\sqrt{k^{2}+\sigma q_{0}^{2}}$. Then we make use of the scattering problems of nonlocal Sine/Sinh-Gordon equations to obtain the pure soliton solutions of the nonlocal reverse space-time NLS equation as follows.

\begin{figure}[h]
\begin{tabular}{cc}
\includegraphics[width=0.5\textwidth]{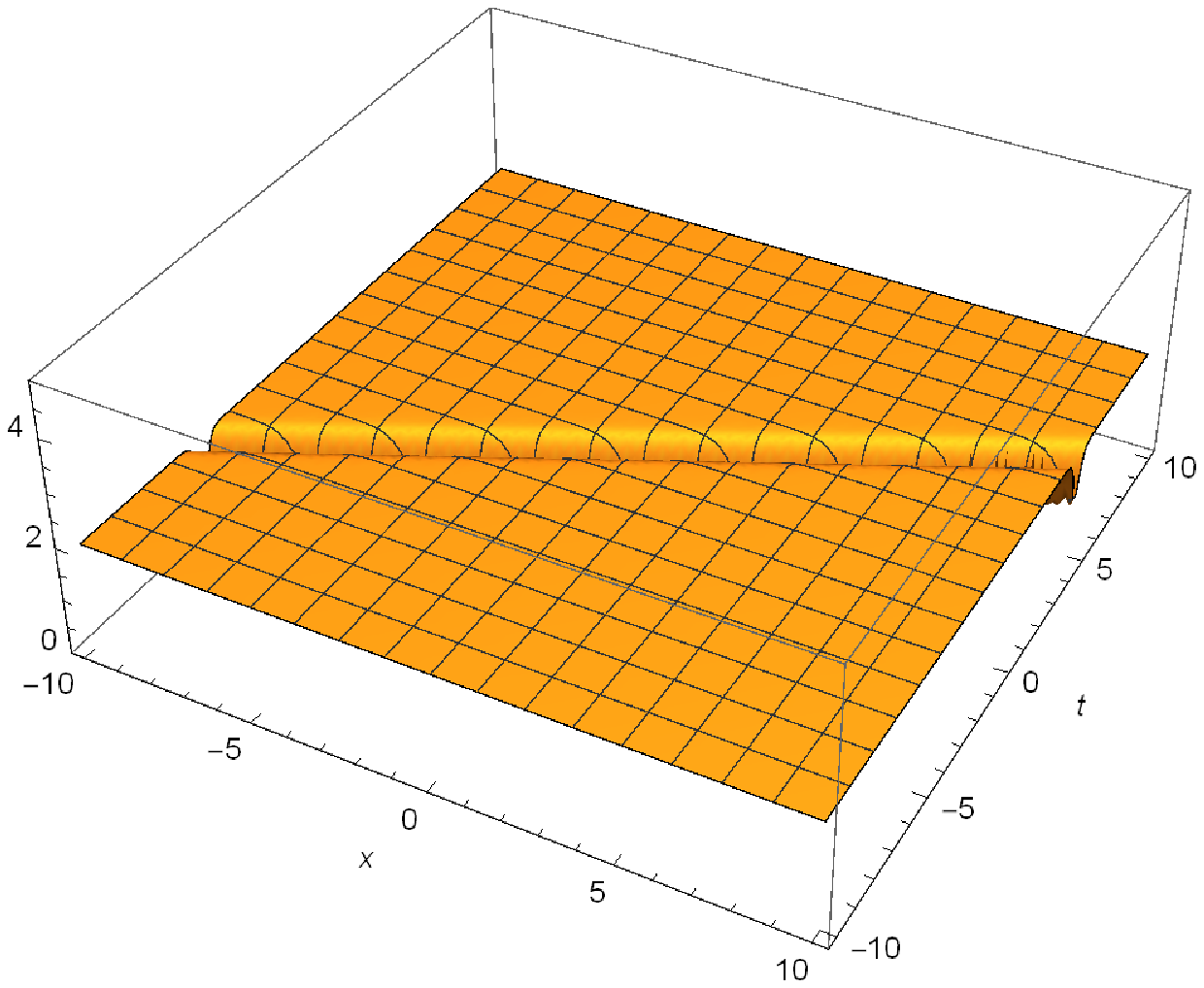}&
\includegraphics[width=0.5\textwidth]{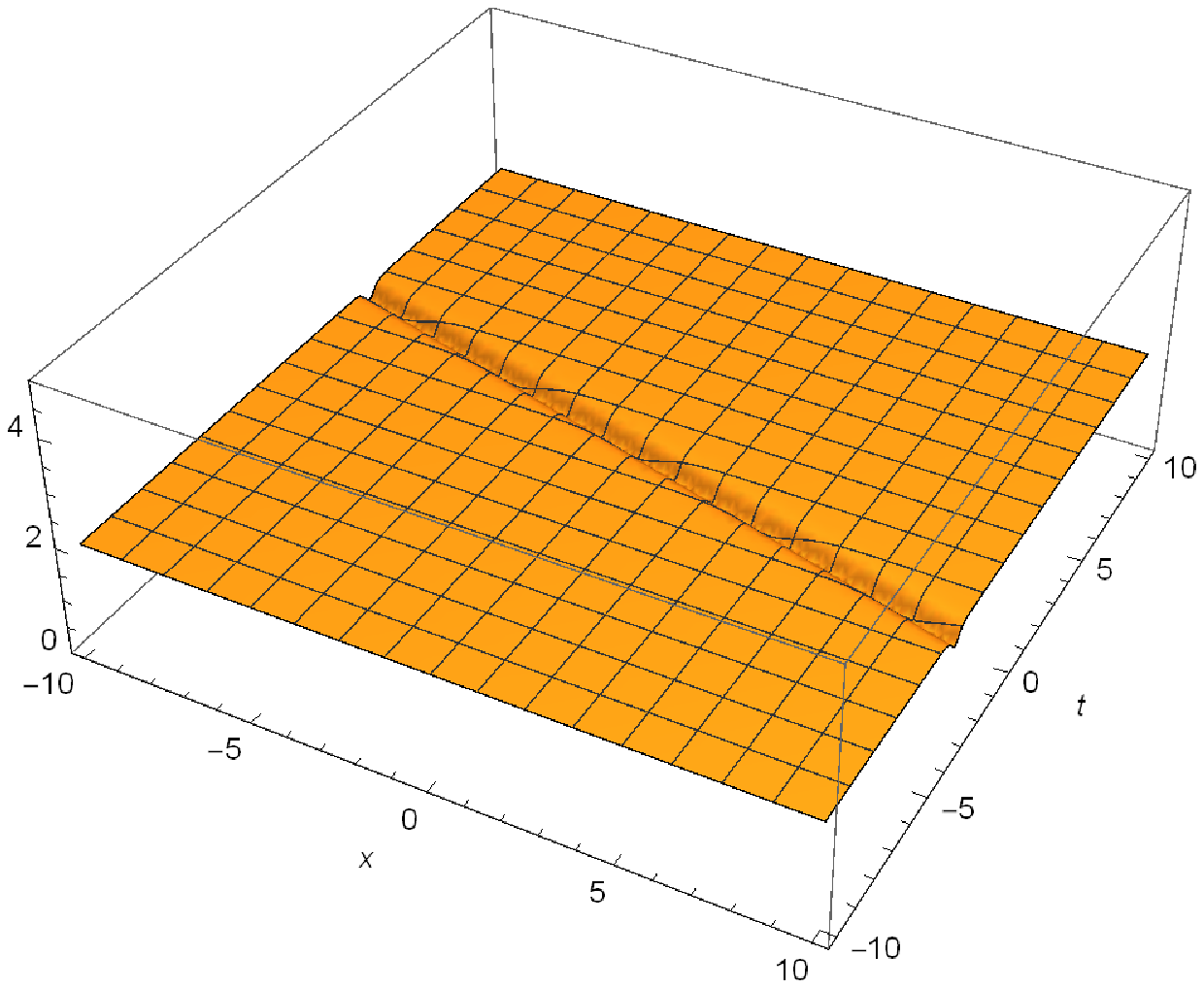}\\
(a) & (b)
\end{tabular}
\caption{1-soliton solutions. (a) Case 1: The amplitude of $q(x,t)$ with $\sigma=1$, $\delta=1$, $\theta_{+}=\frac{\pi}{3}$ and $q_{0}=2$.  (b) Case 3: The amplitude of $q(x,t)$ with $\sigma=-1$, $\delta=-1$, $\theta_{+}=\frac{\pi}{3}$ and $q_{0}=2$.}
\label{RST 1}
\end{figure}

{\bf Case 1.} When $\sigma=1$ and $\theta_{+}+\theta_{-}=0$, we have
\begin{equation}
b(z;t)=b(z;0)e^{-2i\left[q_{0}^{2}+\frac{1}{2}\left(z^{2}-\frac{q_{0}^{4}}{z^{2}}\right)\right]t}, \ \ \
\overline{b}(z;t)=\overline{b}(z;0)e^{2i\left[q_{0}^{2}+\frac{1}{2}\left(z^{2}-\frac{q_{0}^{4}}{z^{2}}\right)\right]t},
\end{equation}
which can be written
\begin{equation}
b(q_{0}e^{i\theta_{+}};t)=\delta i e^{-2iq_{0}^{2}\left[1+i\sin(2\theta_{+})\right]t}, \ \ \
\overline{b}(q_{0}e^{-i\theta_{+}};t)=-\delta i e^{2iq_{0}^{2}\left[1-i\sin(2\theta_{+})\right]t}.
\end{equation}
We find there is a nonsingular pure 1-soliton solution only with $\delta=1$, which is given by
\begin{equation}
\begin{split}
q(x,t)=\frac{q_{0}e^{2iq_{0}^{2} t}\left[e^{i\theta_{+}}\cdot e^{2q_{0}x\sin\theta_{+}}+e^{-i\theta_{+}}\cdot e^{2q_{0}^{2}t\sin(2\theta_{+})}\right]}{e^{2q_{0}x\sin\theta_{+}}+e^{2q_{0}^{2}t\sin(2\theta_{+})}}.
\end{split}
\end{equation}
A typical 1-soliton solution is shown in Fig. \ref{RST 1}.

When $\delta=-1$, there is a 1-soliton solution, which is singular along a space-time line (See Section 10).


%
%
%
%
%

{\bf Case 2.} When $\sigma=1$ and $\theta_{+}+\theta_{-}=\pi$, we have
\begin{equation}
b(z;t)=b(z;0)e^{-2i\left[-q_{0}^{2}+\frac{1}{2}\left(z^{2}-\frac{q_{0}^{4}}{z^{2}}\right)\right]t}, \ \ \
\overline{b}(z;t)=\overline{b}(z;0)e^{2i\left[-q_{0}^{2}+\frac{1}{2}\left(z^{2}-\frac{q_{0}^{4}}{z^{2}}\right)\right]t}.
\end{equation}
Thus,
\begin{equation}
b(iq_{1}; t)=\delta_{1}i\cdot e^{-2i\left[-q_{0}^{2}+\frac{1}{2}\left(-q_{1}^{2}+\frac{q_{0}^{4}}{q_{1}^{2}}\right)\right]t}, \ \ \
b\left(-i\frac{q_{0}^{2}}{q_{1}}; t\right)=\delta_{2}i\cdot e^{-2i\left[-q_{0}^{2}+\frac{1}{2}\left(q_{1}^{2}-\frac{q_{0}^{4}}{q_{1}^{2}}\right)\right]t},
\end{equation}
\begin{equation}
\overline{b}\left(i\frac{q_{0}^{2}}{q_{1}}; t\right)=-\delta_{1}i\cdot e^{2i\left[-q_{0}^{2}+\frac{1}{2}\left(q_{1}^{2}-\frac{q_{0}^{4}}{q_{1}^{2}}\right)\right]t}, \ \ \
\overline{b}\left(-iq_{1}; t\right)=-\delta_{2}i\cdot e^{2i\left[-q_{0}^{2}+\frac{1}{2}\left(-q_{1}^{2}+\frac{q_{0}^{4}}{q_{1}^{2}}\right)\right]t}.
\end{equation}
It is found that there are nonsingular 2-soliton solutions with $\delta_{1}\delta_{2}=-1$ (see the following), and there are 2-soliton solutions with $\delta_{1}\delta_{2}=1$, which are singular along some  lines (see Section 10).

(i) When $\delta_{1}=1$ and $\delta_{2}=-1$, one yields

\begin{equation}
\begin{split}
q(x,t)=\frac{e^{-2iq_{0}^{2}t}\left[i(q_{0}^{4}+q_{1}^{4})\cos\left(\frac{q_{0}^{4}-q_{1}^{4}}{q_{1}^{2}}t\right)
+iq_{0}q_{1}(q_{0}^{2}+q_{1}^{2})\cosh\left(\frac{q_{0}^{2}-q_{1}^{2}}{q_{1}}x\right)+(q_{0}^{4}-q_{1}^{4})\sin\left(\frac{q_{0}^{4}-q_{1}^{4}}{q_{1}^{2}}t\right)\right]}{q_{1}\left[2q_{0}q_{1}\cos\left(\frac{(q_{0}^{4}-q_{1}^{4})t}{q_{1}^{2}}\right)+(q_{0}^{2}+q_{1}^{2})\cosh\left(\frac{(q_{0}^{2}-q_{1}^{2})x}{q_{1}}\right)\right]}.
\end{split}
\end{equation}

(ii) When $\delta_{1}=-1$ and $\delta_{2}=1$, one obtains
\begin{equation}
\begin{split}
q(x,t)=\frac{e^{-2iq_{0}^{2}t}\left[-i(q_{0}^{4}+q_{1}^{4})\cos\left(\frac{q_{0}^{4}-q_{1}^{4}}{q_{1}^{2}}t\right)
+iq_{0}q_{1}(q_{0}^{2}+q_{1}^{2})\cosh\left(\frac{q_{0}^{2}-q_{1}^{2}}{q_{1}}x\right)+(-q_{0}^{4}+q_{1}^{4})\sin\left(\frac{q_{0}^{4}-q_{1}^{4}}{q_{1}^{2}}t\right)\right]}{q_{1}\left[-2q_{0}q_{1}\cos\left(\frac{(q_{0}^{4}-q_{1}^{4})t}{q_{1}^{2}}\right)+(q_{0}^{2}+q_{1}^{2})\cosh\left(\frac{(q_{0}^{2}-q_{1}^{2})x}{q_{1}}\right)\right]}.
\end{split}
\end{equation}
To show the above two solutions are nonsingular, we only need to prove \\
$(q_{0}^{2}+q_{1}^{2})\cosh\left(\frac{(q_{0}^{2}-q_{1}^{2})x}{q_{1}}\right)-2q_{0}q_{1}\neq 0$ for any $x$. Indeed, if $(q_{0}^{2}+q_{1}^{2})\cosh\left(\frac{(q_{0}^{2}-q_{1}^{2})x}{q_{1}}\right)-2q_{0}q_{1}=0$ for some $x$, then
$(q_{0}^{2}+q_{1}^{2}) y^{2}-4q_{0}q_{1}y+(q_{0}^{2}+q_{1}^{2})=0$ has a least one solution for $y>0$, where $y:=e^{\frac{(q_{0}^{2}-q_{1}^{2})x}{2}}$. But it is easy to see that this quadratic does not have real solution, which is a contradiction.

\begin{figure}[h]
\begin{tabular}{cc}
\includegraphics[width=0.5\textwidth]{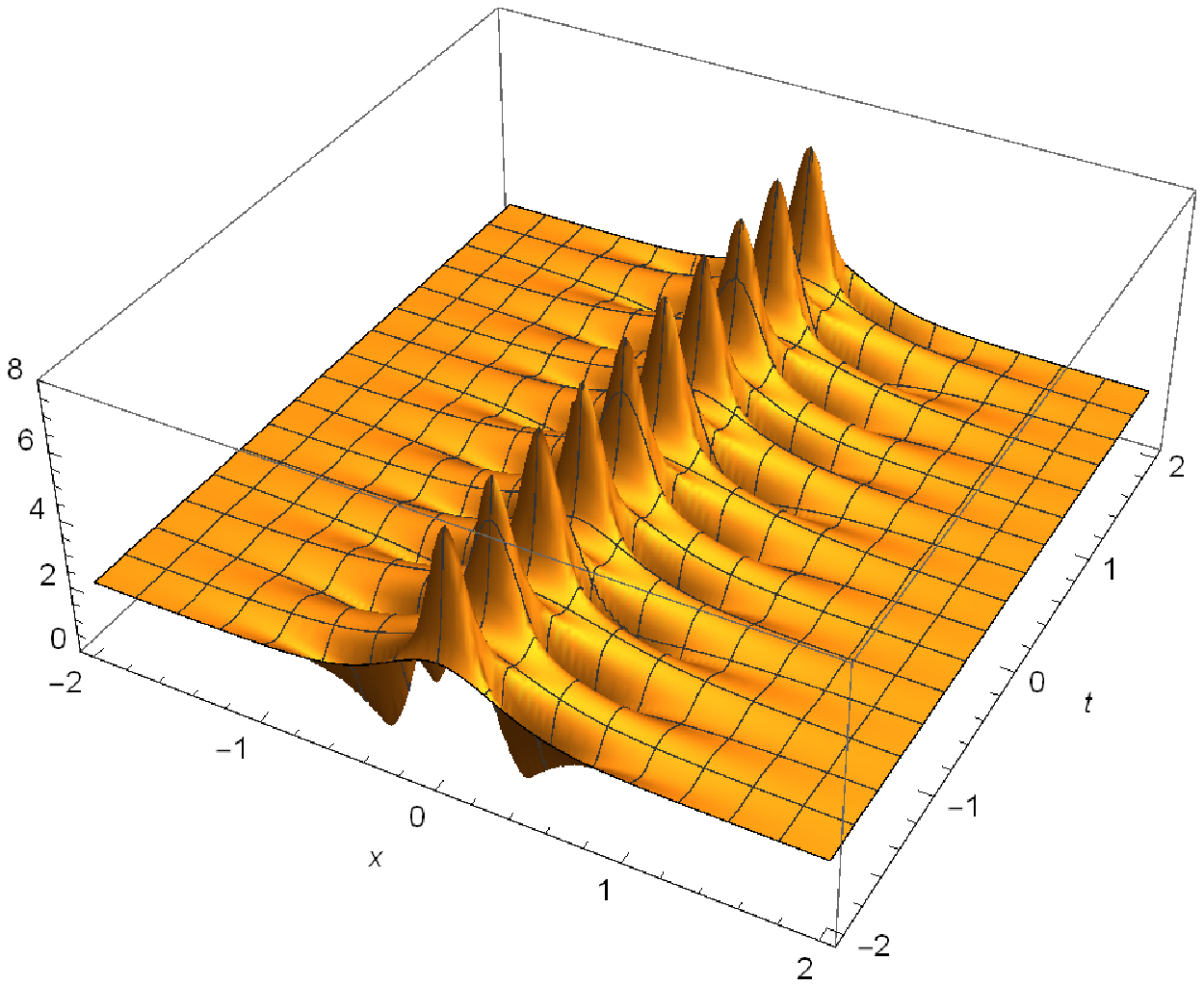}&
\includegraphics[width=0.5\textwidth]{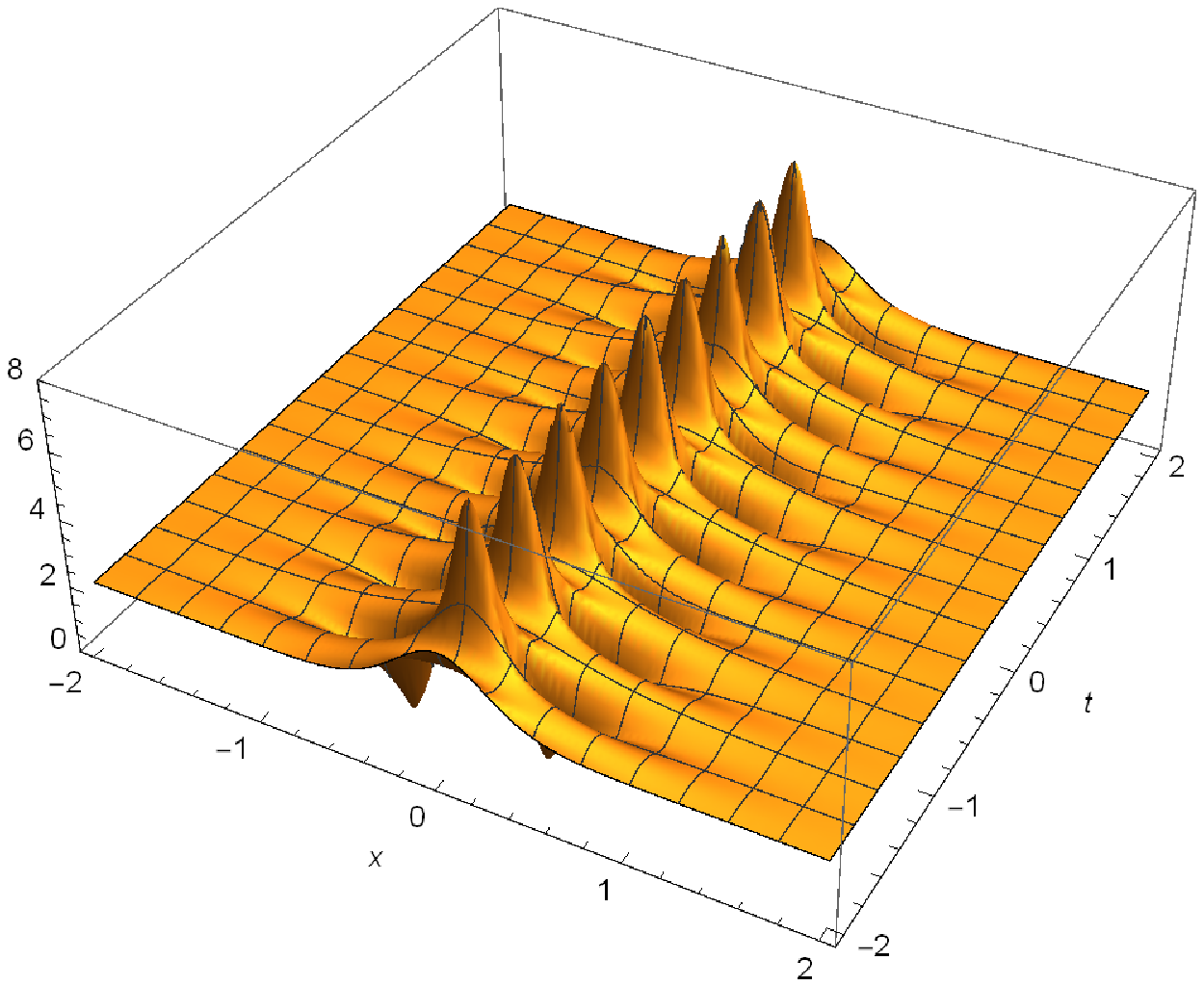}\\
(a) & (b)
\end{tabular}
\caption{(a) The amplitude of $q(x,t)$ for Case 2 with $\delta_{1}=1$, $\delta_{2}=-1$, $q_{1}=4$ and $q_{0}=2$.  (b) The amplitude of $q(x,t)$ for Case 4 with $\delta_{1}=1$, $\delta_{2}=-1$, $q_{1}=4$ and $q_{0}=2$.}
\label{RST 3}
\end{figure}

In Fig. \ref{RST 3} below we give typical 2-soliton solutions, which are of breather type.


{\bf Case 3.} When $\sigma=-1$ and $\theta_{+}+\theta_{-}=\pi$, we have
\begin{equation}
b(z;t)=b(z;0)e^{-2i\left[q_{0}^{2}+\frac{1}{2}\left(z^{2}-\frac{q_{0}^{4}}{z^{2}}\right)\right]t}, \ \ \
\overline{b}(z;t)=\overline{b}(z;0)e^{2i\left[q_{0}^{2}+\frac{1}{2}\left(z^{2}-\frac{q_{0}^{4}}{z^{2}}\right)\right]t}.
\end{equation}
Thus,
\begin{equation}
b\left(q_{0}e^{i(\theta_{+}+\frac{\pi}{2})};t\right)=\delta  e^{-2iq_{0}^{2}\left[1-i\sin(2\theta_{+})\right]t}, \ \ \
\overline{b}(q_{0}e^{-i(\theta_{+}+\frac{\pi}{2})};t)=\delta e^{2iq_{0}^{2}\left[1+i\sin(2\theta_{+})\right]t}.
\end{equation}
Note that $\delta$ can be chosen either $1$ or $-1$, we find there is a nonsingular "dark" 1-soliton solution with $\delta=-1$ (see the following), and there is a "bright" 1-soliton solution which is singular along a space-time line with $\delta=1$ (see Section 10).

When $\delta=-1$,
\begin{equation}
\begin{split}
q(x,t)=q_{0}e^{2iq_{0}^{2}t}[i\sin\theta_{+}+\cos\theta_{+}\tanh(q_{0}\cos\theta_{+}\cdot(x+2q_{0}t\sin\theta_{+}))],
\end{split}
\end{equation}
which is nonsingular because $e^{q_{0}\cos\theta_{+}\cdot(x+2q_{0}t\sin\theta_{+})}>0$.

In Fig. \ref{RST 1}, a typical nonsingular 1-soliton solution is displayed.

{\bf Case 4.} When $\sigma=-1$ and $\theta_{+}+\theta_{-}=0$, we have
\begin{equation}
b(z;t)=b(z;0)e^{-2i\left[-q_{0}^{2}+\frac{1}{2}\left(z^{2}-\frac{q_{0}^{4}}{z^{2}}\right)\right]t}, \ \ \
\overline{b}(z;t)=\overline{b}(z;0)e^{2i\left[-q_{0}^{2}+\frac{1}{2}\left(z^{2}-\frac{q_{0}^{4}}{z^{2}}\right)\right]t}.
\end{equation}
Thus,
\begin{equation}
b(iq_{1}; t)=\delta_{1}\cdot e^{-2i\left[-q_{0}^{2}+\frac{1}{2}\left(-q_{1}^{2}+\frac{q_{0}^{4}}{q_{1}^{2}}\right)\right]t}, \ \ \
b\left(-i\frac{q_{0}^{2}}{q_{1}}; t\right)=\delta_{2}\cdot e^{-2i\left[-q_{0}^{2}+\frac{1}{2}\left(q_{1}^{2}-\frac{q_{0}^{4}}{q_{1}^{2}}\right)\right]t},
\end{equation}
\begin{equation}
\overline{b}\left(i\frac{q_{0}^{2}}{q_{1}}; t\right)=\delta_{1}\cdot e^{2i\left[-q_{0}^{2}+\frac{1}{2}\left(q_{1}^{2}-\frac{q_{0}^{4}}{q_{1}^{2}}\right)\right]t}, \ \ \
\overline{b}\left(-iq_{1}; t\right)=\delta_{2}\cdot e^{2i\left[-q_{0}^{2}+\frac{1}{2}\left(-q_{1}^{2}+\frac{q_{0}^{4}}{q_{1}^{2}}\right)\right]t}.
\end{equation}
As shown previously, there are nonsingular 2-soliton solutions with $\delta_{1}\delta_{2}=-1$ (see the following), and there are 2-soliton solutions with $\delta_{1}\delta_{2}=1$, which are singular along certain space-time  lines (see Section 10).

(i) When $\delta_{1}=1$ and $\delta_{2}=-1$, it yields
\begin{equation}
\begin{split}
&q(x,t)=iq_{0}e^{-2i q_{0}^{2} t}\Big(1-\Big((q_{0}^{2}-q_{1}^{2})
\Big(q_{0}^{2}e^{\frac{i(-q_{0}^{4}+q_{1}^{4})t}{q_{1}^{2}}}-q_{1}^{2}e^{\frac{i(q_{0}^{4}-q_{1}^{4})t}{q_{1}^{2}}}\Big)\Big(2q_{0}q_{1}\cdot\\
&e^{-\frac{it}{2}\Big(\frac{2q_{0}^{4}}{q_{1}^{2}}-2q_{1}^{2}\Big)}+2q_{0}q_{1}
e^{-\frac{it}{2}\cdot\left(2\left(-\frac{q_{0}^{4}}{q_{1}^{2}}+q_{1}^{2}\right)\right)}
+(q_{0}^{2}+q_{1}^{2})e^{-q_{1}x+\frac{q_{0}^{2}(iq_{1}t+x)}{q_{1}}-iq_{0}^{2}t}\\
&+(q_{0}^{2}+q_{1}^{2})e^{q_{1}x+q_{0}^{2}\left(it-\frac{x}{q_{1}}\right)-iq_{0}^{2}t}\Big)\Big)/\Big(q_{0}q_{1}\Big((q_{0}^{2}-q_{1}^{2})^{2}-4q_{0}^{2}q_{1}^{2}\cos\Big(\frac{2(q_{0}^{4}-q_{1}^{4})t}{q_{1}^{2}}\Big)\\
&+(q_{0}^{2}+q_{1}^{2})^{2}\cosh\Big(\frac{2(q_{0}^{2}-q_{1}^{2})x}{q_{1}}\Big)\Big)\Big)\Big)\\
&=\frac{e^{-2iq_{0}^{2}t}\left[-q_{0}^{4}e^{i\frac{q_{1}^{4}-q_{0}^{4}}{q_{1}^{2}}t}-q_{1}^{4}e^{i\frac{q_{0}^{4}-q_{1}^{4}}{q_{1}^{2}}t}
+q_{0}q_{1}(q_{0}^{2}+q_{1}^{2})\cosh\left(\frac{q_{0}^{2}-q_{1}^{2}}{q_{1}}x\right)\right]}{q_{1}\left[-2q_{0}q_{1}\cos\left(\frac{(q_{0}^{4}-q_{1}^{4})t}{q_{1}^{2}}\right)+(q_{0}^{2}+q_{1}^{2})\cosh\left(\frac{(q_{0}^{2}-q_{1}^{2})x}{q_{1}}\right)\right]}.
\end{split}
\end{equation}
(ii) When $\delta_{1}=-1$ and $\delta_{2}=1$, it yields
\begin{equation}
\begin{split}
q(x,t)=\frac{e^{-2iq_{0}^{2}t}\left[q_{0}^{4}e^{i\frac{q_{1}^{4}-q_{0}^{4}}{q_{1}^{2}}t}+q_{1}^{4}e^{i\frac{q_{0}^{4}-q_{1}^{4}}{q_{1}^{2}}t}
+q_{0}q_{1}(q_{0}^{2}+q_{1}^{2})\cosh\left(\frac{q_{0}^{2}-q_{1}^{2}}{q_{1}}x\right)\right]}{q_{1}\left[2q_{0}q_{1}\cos\left(\frac{(q_{0}^{4}-q_{1}^{4})t}{q_{1}^{2}}\right)+(q_{0}^{2}+q_{1}^{2})\cosh\left(\frac{(q_{0}^{2}-q_{1}^{2})x}{q_{1}}\right)\right]}.
\end{split}
\end{equation}

The above solutions have the same characteristics as Case 2. Fig. \ref{RST 3} shows a typical 2-soliton solution of breather type.


\section{Spatial Dependent  Boundary Conditions}

\subsection{Nonlocal Sine-Gordon and Sinh-Gordon equations}


The nonlocal Sine-Gordon and Sinh-Gordon equations are
\begin{equation}
q_{xt}(x,t)+2s(x,t)q(x,t)=0,
\end{equation}
where $s(-x,-t)=s(x,t)$, $s_{x}(x,t)=-\sigma(q(x,t)q(-x,-t))_{t}$ and $\sigma=\mp 1$. In particular, when $\sigma=-1$, (\ref{E:nonlocal Sine/Sinh}) is the nonlocal Sine-Gordon equation; when $\sigma=1$, (\ref{E:nonlocal Sine/Sinh}) is the nonlocal Sinh-Gordon equation.
We will define $s(x,t)$ as


\begin{equation}
s(x,t)= \sigma \int_x^{\infty} \partial_t (q(x',t)q(-x',-t) )dx' + \frac{\alpha \beta}{2}
\end{equation}
and require $s(x,t) \rightarrow \frac{\alpha \beta}{2} ~\text{as} ~ |x| \rightarrow \infty$.
In this section, we consider the boundary condition
\begin{equation}
q(x,t)\rightarrow q_{0}e^{i(\alpha t+\beta x+\theta_{\pm})},\ \ \ s(x,t)\rightarrow \frac{\alpha \beta}{2}
\end{equation}
as $x\rightarrow \pm\infty$, where both $\alpha$ and $\beta$ are real. Setting $q(x,t)=\widetilde{q}(x,t)e^{i\beta x}$, we have
\begin{equation}
\widetilde{q}_{xt}(x,t)+2s(x,t)\widetilde{q}(x,t)+i\beta\widetilde{q}_{t}(x,t)=0
\end{equation}
which is associated with the following $2\times2$
compatible systems:


\begin{equation}
v_{x}=Xv=
\left(\begin{array}{cc}
-ik& \widetilde{q}(x,t)\\
\sigma\widetilde{q}(-x,-t)& ik
\end{array}\right)v,
\end{equation}

\begin{equation}
v_{t}=Tv=
\left(\begin{array}{cc}
\frac{s(x,t)}{2i(k-\frac{\beta}{2})}& \frac{\widetilde{q}_{t}(x,t)}{2i(k-\frac{\beta}{2})}\\
-\frac{\sigma \widetilde{q}_{t}(-x,-t)}{2i(k-\frac{\beta}{2})}& -\frac{s(x,t)}{2i(k-\frac{\beta}{2})}
\end{array}\right)v.
\end{equation}
Then both $a(t)$ and $\overline{a}(t)$ are time independent, and
\begin{equation}
b(z;t)=b(z;0)e^{-\frac{i\alpha(2k-2\lambda-\beta)}{2\left(k-\frac{\beta}{2}\right)}t}
\end{equation}
and
\begin{equation}
\overline{b}(z;t)=\overline{b}(z;0)e^{\frac{i\alpha(2k-2\lambda-\beta)}{2\left(k-\frac{\beta}{2}\right)}t}.
\end{equation}
In particular, when $\sigma=1$, $\theta_{+}+\theta_{-}=0$ and $\delta=1$, we have
\begin{equation}
b(q_{0}e^{i\theta_{+}};t)=i e^{-i\alpha\left(\frac{2q_{0}e^{-i\theta_{+}}-\beta}{2q_{0}\cos\theta_{+}-\beta}\right)t},
\end{equation}
\begin{equation}
\overline{b}(q_{0}e^{-i\theta_{+}};t)=-i e^{i\alpha\left(\frac{2q_{0}e^{i\theta_{+}}-\beta}{2q_{0}\cos\theta_{+}-\beta}\right)t}.
\end{equation}
Substituting the above into (\ref{E:soliton 1}), then the special pure 1-soliton solution is given by


\begin{equation}
q(x,t)=q_{0}e^{i(\alpha t+\beta x)}\left[\cos\theta_{+}+i\sin\theta_{+}\cdot\tanh\left(q_{0}\sin\theta_{+}\left(x-\frac{\alpha t}{\beta-2q_{0}\cos\theta_{+}}\right)\right)\right].
\end{equation}


When $\sigma=-1$, $\theta_{+}+\theta_{-}=\pi$ and $\delta=-1$, we obtain
\begin{equation}
b(q_{0}e^{i(\theta_{+}+\frac{\pi}{2})};t)=-e^{-i\alpha\left(\frac{2q_{0}e^{-i(\theta_{+}+\frac{\pi}{2})}-\beta}{2q_{0}\cos(\theta_{+}+\frac{\pi}{2})-\beta}\right)t},
\end{equation}
\begin{equation}
\overline{b}(q_{0}e^{-i(\theta_{+}+\frac{\pi}{2})};t)=-e^{i\alpha\left(\frac{2q_{0}e^{i(\theta_{+}+\frac{\pi}{2})}-\beta}{2q_{0}\cos(\theta_{+}+\frac{\pi}{2})-\beta}\right)t}.
\end{equation}
Then
\begin{equation}
\begin{split}
&q(x,t)=q_{0}e^{i(\alpha t+\beta x)}\Big[e^{i(\alpha t+\theta_{+})}\left(-e^{\frac{2q_{0}\alpha t\cos\theta_{+}}{\beta+2q_{0}\sin\theta_{+}}}+e^{q_{0}\cos\theta_{+}\left(4x-\frac{2\alpha t}{\beta+2q_{0}\sin\theta_{+}}\right)}\right)\\
&-2\cos\theta_{+}\left(e^{i\alpha t+2q_{0}x\cos\theta_{+}}-e^{\frac{\alpha t(2e^{i\theta_{+}}+i\beta)}{\beta+2q_{0}\sin\theta_{+}}}\right)\Big]/\left[-e^{\frac{\alpha t(2q_{0}e^{i\theta_{+}}+i\beta)}{\beta+2q_{0}\sin\theta_{+}}}+e^{i\alpha t+q_{0}\cos\theta_{+}\left(4x-\frac{2\alpha t}{\beta+2q_{0}\sin\theta_{+}}\right)}\right].
\end{split}
\end{equation}
We can see that $x=\frac{\alpha t}{\beta+2q_{0}\sin\theta_{+}}$ is a removable singularity, so $q(x,t)$ is nonsingular.


\subsection{Complex standard Sine-Gordon and Sinh-Gordon equations}


The complex standard Sine-Gordon and Sinh-Gordon equations are
\begin{equation}
\label{E:complex Sine/Sinh}
q_{xt}(x,t)+2s(x,t)q(x,t)=0,
\end{equation}
where $s(x,t)$ is real, $s_{x}(x,t)=-\sigma(q(x,t)q^{*}(x,t))_{t}$ and $\sigma=\mp 1$. In particular, when $\sigma=-1$, (\ref{E:complex Sine/Sinh}) is the complex Sine-Gordon equation; when $\sigma=1$, (\ref{E:complex Sine/Sinh}) is the complex Sinh-Gordon equation.

We consider the boundary condition
\begin{equation}
q(x,t)\rightarrow q_{0}e^{i(\alpha t+\beta x+\theta_{\pm})},
\end{equation}
as $x\rightarrow \pm\infty$, where both $\alpha$ and $\beta$ are real. Set $q(x,t)=\widetilde{q}(x,t)e^{i\beta x}$, we have
\begin{equation}
\widetilde{q}_{xt}(x,t)+2s(x,t)\widetilde{q}(x,t)+i\beta\widetilde{q}_{t}(x,t)=0
\end{equation}
which is associated with the following $2\times2$
compatible systems:
\begin{equation}
v_{x}=Xv=
\left(\begin{array}{cc}
-ik& \widetilde{q}(x,t)\\
\sigma\widetilde{q}^{*}(x,t)& ik
\end{array}\right)v,
\end{equation}
\begin{equation}
v_{t}=Tv=
\left(\begin{array}{cc}
\frac{s(x,t)}{2i(k-\frac{\beta}{2})}& \frac{\widetilde{q}_{t}(x,t)}{2i(k-\frac{\beta}{2})}\\
-\frac{\sigma \widetilde{q}_{t}^{*}(x,t)}{2i(k-\frac{\beta}{2})}& -\frac{s(x,t)}{2i(k-\frac{\beta}{2})}
\end{array}\right)v.
\end{equation}


Then both $a(t)$ and $\overline{a}(t)$ are time independent,
\begin{equation}
b(z;t)=b(z;0)e^{-\frac{i\alpha(2k-2\lambda-\beta)}{2\left(k-\frac{\beta}{2}\right)}t}
\end{equation}
and
\begin{equation}
\overline{b}(z;t)=\overline{b}(z;0)e^{\frac{i\alpha(2k-2\lambda-\beta)}{2\left(k-\frac{\beta}{2}\right)}t},
\end{equation}
where $\lambda=\sqrt{k^{2}-\sigma q_{0}^{2}}$.

Further analysis follows in the same way as in the prior section.

\subsection{The nonlinear Schr\"{o}dinger equation}


The nonlinear Schr\"{o}dinger equation
\begin{equation}
iq_{t}(x,t)=q_{xx}(x,t)-2\sigma|q(x,t)|^{2}q(x,t),
\end{equation}
where $q^*$ denotes complex conjugate of $q$ and $\sigma=\mp 1$. We consider the boundary condition
\begin{equation}
q(x,t)\rightarrow q_{0}e^{i(\alpha t+\beta x+\theta_{\pm})},
\end{equation}
as $x\rightarrow\pm\infty$ ,where $q_{0}>0$, $0\leq \theta_{\pm}<2\pi$, both $\alpha$ and $\beta$ are real. It is easy to see that $\alpha=\beta^{2}+2\sigma q_{0}^{2}$, then the boundary condition becomes
\begin{equation}
q(x,t)\rightarrow q_{0}e^{i[(\beta^{2}+2\sigma q_{0}^{2}) t+\beta x+\theta_{\pm}]},
\end{equation}
as $x\rightarrow\pm\infty$, where $\beta$ is real.
By setting $q(x,t)=\widetilde{q}(x,t)e^{i\beta x}$, we have
\begin{equation}
\label{E:modified NLS}
i\widetilde{q}_{t}=\widetilde{q}_{xx}+2i\beta\widetilde{q}_{x}-(\beta^{2}+2\sigma|\widetilde{q}|^{2})\widetilde{q},
\end{equation}
which is associated with the following $2\times2$
compatible systems:
\begin{equation}
\label{E:morified NLS scattering}
v_{x}=Xv=
\left(\begin{array}{cc}
-ik& \widetilde{q}(x,t)\\
\sigma\widetilde{q}^{*}(x,t)& ik
\end{array}\right)v,
\end{equation}

\begin{equation}
v_{t}=Tv=
\left(\begin{array}{cc}
2ik^{2}-2\beta ik+i\sigma|\widetilde{q}|^{2}+\frac{\beta^{2}}{2}i& -2\widetilde{q}k+2\beta\widetilde{q}-i\widetilde{q}_{x}\\
-2\sigma\widetilde{q}^{*}k+i\sigma\widetilde{q}^{*}_{x}+2\beta\sigma\widetilde{q}^{*}& -2ik^{2}+2\beta ik-i\sigma|\widetilde{q}|^{2}-\frac{\beta^{2}}{2}i
\end{array}\right)v.
\end{equation}
See \cite{Ablowitz1}.


Then we can deduce both $a(t)$ and $\overline{a}(t)$ are time independent,
\begin{equation}
\begin{split}
b(t)=b(0)e^{-2i[\sigma q_{0}^{2}+\frac{\beta^{2}}{2}-2\lambda(\beta-k)]t}, \ \ \overline{b}(t)=\overline{b}(0)e^{2i[\sigma q_{0}^{2}+\frac{\beta^{2}}{2}-2\lambda(\beta-k)]t}.
\end{split}
\end{equation}
The scattering problem (\ref{E:morified NLS scattering}) is the same as the NLS equation with a nonzero boundary condition


\begin{equation}
\tilde{q}(x,t)\rightarrow q_{0}e^{i(2\sigma q_{0}^{2}t+\theta_{\pm})}
\end{equation}
as $x\rightarrow\pm\infty$, the only difference between the NLS equation and (\ref{E:modified NLS}) is the time evolution. Based on the scattering problem and 1-soliton solution of defocusing NLS (\cite{Demontis}) equation, we deduce that when $\sigma=1$,
\begin{equation}
q(x,t)=\tilde{q}e^{i \beta x}= e^{i\beta x}\left[q_{0}e^{i[(\beta^{2}+2q_{0}^{2})t+\theta_{+}]}+\frac{iC_{1}^{*}(0)\alpha_{1}^{*}
e^{2i(q_{0}^{2}+\frac{\beta^{2}}{2})t-4v_{1}(\beta-k_{1})t-2v_{1}x}}{1+\frac{q_{0}|C_{1}(0)|}{2v_{1}}e^{-2v_{1}x-4v_{1}(\beta-k_{1})t}}\right],
\end{equation}
where $\alpha_{1}=k_{1}+iv_{1}$, $v_{1}=\sqrt{q_{0}^{2}-k_{1}^{2}}$, $-q_{0}<k<q_{0}$, $e^{2v_{1}x_{0}}=\frac{q_{0}|C_{1}(0)|}{2v_{1}}$ and
$C_{1}^{*}(0)= -|C_{1}(0)|e^{i\theta_{+}}$.


We can rewrite it in the form
\begin{equation}
q(x,t)= q_{0}e^{2iq_{0}^{2}t} e^{i(\beta x+\beta^{2}t)} \left[ e^{i\theta_{+}}+\frac{iC_{1}^{*}(0)\alpha_{1}^{*}
e^{-2v_1(x+2\beta t -2 k_1t)}}{1+\frac{q_{0}|C_{1}(0)|}{2v_{1}}e^{-2v_{1}(x+2\beta t -2 k_1t)}}
\right]
\end{equation}

A property of the NLS equation is its Galilean invariance, i.e., if $q_{1}(x,t)$ solves NLS equation and satisfies boundary condition
\begin{equation}
q_{1}(x,t)\rightarrow q_{0}e^{i(2\sigma q_{0}^{2}t+\theta_{\pm})}
\end{equation}
as $x\rightarrow\pm\infty$, then
\begin{equation}
q_{2}(x,t):=q_{1}(x+2\beta t, t) e^{i(\beta x+\beta^{2}t)}
\end{equation}
also solves NLS equation and satisfies boundary condition
\begin{equation}
q_{2}(x,t)\rightarrow q_{0}e^{i[(\beta^{2}+2\sigma q_{0}^{2})t+\beta x+\theta_{\pm}]}
\end{equation}
as $x\rightarrow\pm\infty$.
When $\sigma=1$, we have the 1-soliton solution

\begin{equation}
q_{1}(x,t)= q_{0}e^{2iq_{0}^{2}t}  \left[ e^{i\theta_{+}}+\frac{iC_{1}^{*}(0)\alpha_{1}^{*}
e^{-2v_1(x -2 k_1t)}}{1+\frac{q_{0}|C_{1}(0)|}{2v_{1}}e^{-2v_{1}(x -2 k_1t)}}
\right].
\end{equation}


We have that $q(x,t)=q_{1}(x+2\beta t, t) e^{i(\beta x+\beta^{2}t)}$, which means that the result of IST agrees with the one based on Galilean invariance of NLS equation.

If we ignore the nonlinear term, then the following linear PDE
\begin{equation}
\label{E:linear NLS}
iq_{t}(x,t)-q_{xx}(x,t)=0
\end{equation}
also satisfies the Galilean invariance. Indeed, by Fourier transform,
\begin{equation}
u_{1}(x,t)=\frac{1}{2\pi}\int_{-\infty}^{\infty}b^{(1)}(\xi)e^{i(\xi x+\xi^{2}t)}d\xi
\end{equation}
is a solution of (\ref{E:linear NLS}). Set
\begin{equation}
u_{2}(x,t)=u_{1}(x+2\beta t,t)e^{i(\beta x+\beta^{2}t)},
\end{equation}
thus,
\begin{equation}
u_{2}(x,t)=\frac{1}{2\pi}\int_{-\infty}^{\infty}b^{(1)}(\xi)e^{i(\xi+\beta)x} e^{i(\xi+\beta)^{2}t}d\xi.
\end{equation}
Redefining variables ($\xi'= \xi+\beta$) shows that $u_{2}(x,t)$ is also a solution of (\ref{E:linear NLS}), which implies that the Galilean invariance holds.

\section{Novel Class of Singular Solutions}
In the following, we present solutions of the Sinh/Sine-Gordon equations and the nonlocal reverse space-time NLS equation which are singular along space-time lines.
\subsection{Singular soliton solutions of the nonlocal Sinh-Gordon equation with $\theta_{+}+\theta_{-}=0$}
\begin{figure}[h]
\begin{tabular}{cc}
\includegraphics[width=0.5\textwidth]{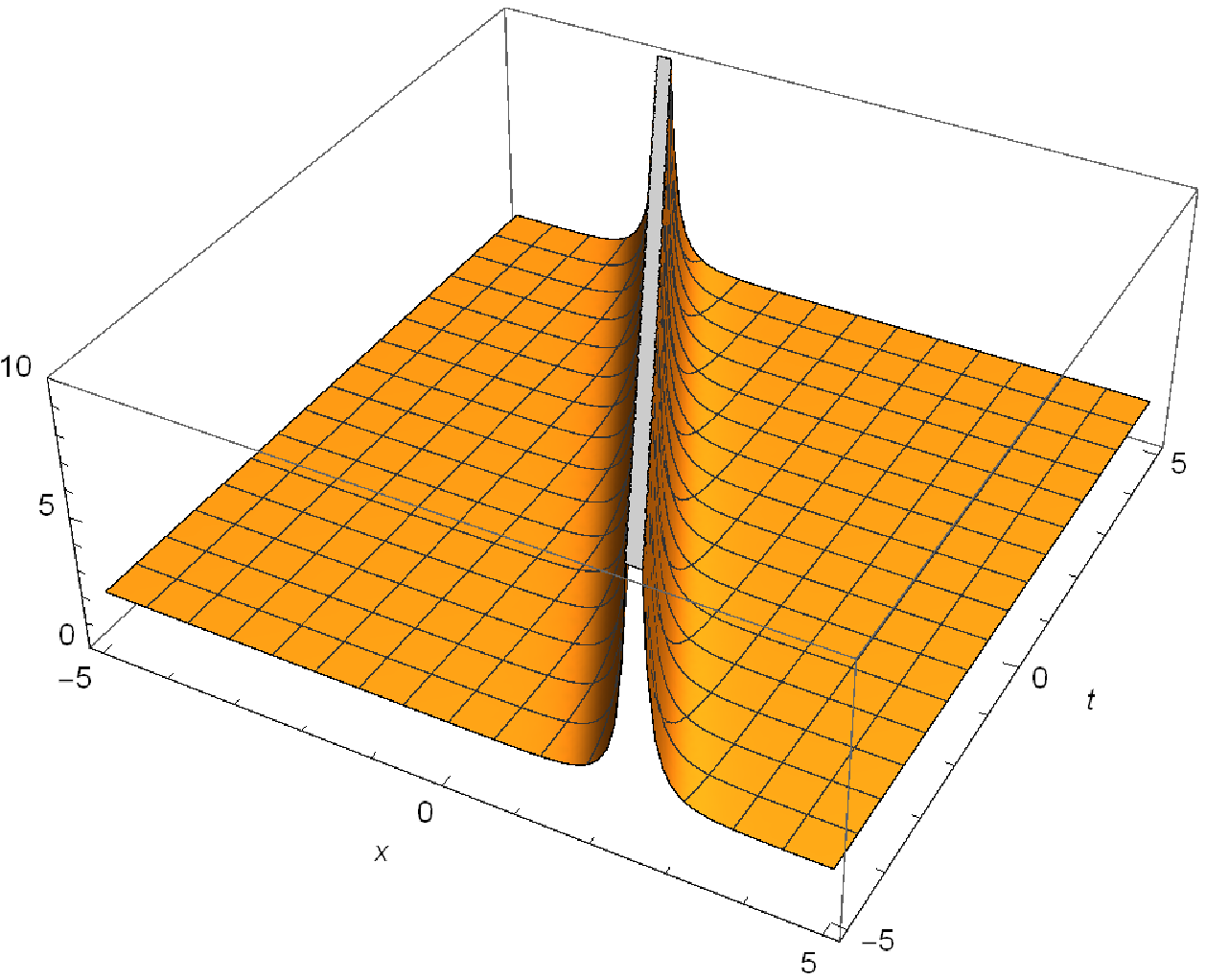}&
\includegraphics[width=0.5\textwidth]{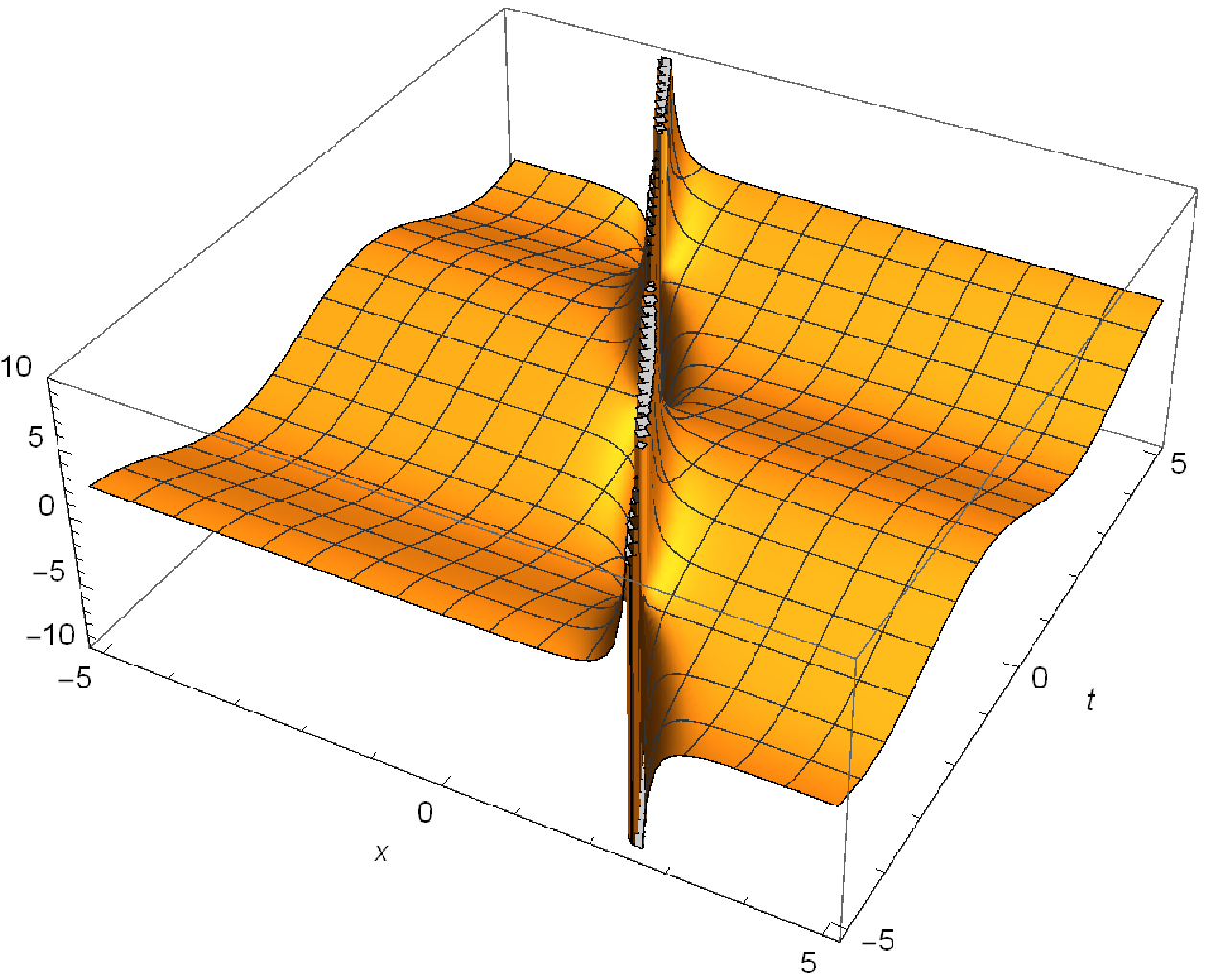}\\
(a) & (b)
\end{tabular}
\caption{(a) The amplitude of $q(x,t)$ with $\delta=-1$, $\theta_{+}=\frac{\pi}{3}$, $\alpha=1$ and $q_{0}=2$.  (b) The real part of $q(x,t)$ with $\delta=-1$, $\theta_{+}=\frac{\pi}{3}$, $\alpha=1$ and $q_{0}=2$.}
\label{Sinh 2}
\end{figure}

When $\delta=-1$, we obtain the following singular "bright" 1-soliton solution
\begin{equation}
q(x,t)=q_{0}e^{i\alpha t}\left[\cos\theta_{+}+i\coth\left(q_{0}x\sin\theta_{+}+\frac{1}{2}\alpha t\tan\theta_{+}\right)\sin\theta_{+}\right],
\end{equation}
and
\begin{equation}
s(x,t)=-\frac{1}{2}q_{0}\alpha\coth^{2}\left(q_{0}x\sin\theta_{+}+\frac{1}{2}\alpha t\tan\theta_{+}\right)\sin\theta_{+}\tan\theta_{+}
+\frac{1}{2}q_{0}\alpha\sin\theta_{+}\tan\theta_{+}.
\end{equation}
We see that $s \rightarrow 0$ as $|x| \rightarrow \infty$. This solution is singular along  the  line: $q_{0}x\sin\theta_{+}+\frac{1}{2}\alpha t\tan\theta_{+}=0.$ In Fig. \ref{Sinh 2} below we give a typical bright 1-soliton solution.

\subsection{Singular soliton solutions of the nonlocal Sinh-Gordon equation with $\theta_{+}+\theta_{-}=\pi$}
We find the following singular 2-soliton solutions with $\delta_{1}\delta_{2}=1$. (i) When $\delta_{1}=\delta_{2}=1$,
\begin{equation}
\begin{split}
&q(x,t)=\Big(i\cdot e^{i\alpha t}\Big(-4q_{0}^{5}q_{1}\cdot e^{\frac{2q_{0}^{2}x}{q_{1}}+2q_{1}\left(x+\frac{2iq_{1}\alpha t}{q_{0}^{2}+q_{1}^{2}}\right)}-4q_{0}q_{1}^{5}\cdot e^{2q_{1}x+q_{0}^{2}\left(\frac{2x}{q_{1}}+\frac{4i\alpha t}{q_{0}^{2}+q_{1}^{2}}\right)}\\
&+2q_{0}q_{1}(q_{0}^{2}-q_{1}^{2})^{2}\cdot e^{\frac{2q_{0}^{2}x}{q_{1}}+2q_{1}x+2i\alpha t}+q_{0}q_{1}(q_{0}^{2}+q_{1}^{2})^{2}\cdot e^{\frac{4q_{0}^{2}x}{q_{1}}+2i\alpha t}+q_{0}q_{1}(q_{0}^{2}+q_{1}^{2})^{2}\cdot e^{4q_{1}x+2i\alpha t}\\
&+2q_{0}^{2}(q_{0}^{4}-q_{1}^{4})\cdot e^{\frac{q_{0}^{2}x}{q_{1}}+3q_{1}x+\frac{i\alpha\left(q_{0}^{2}+3q_{1}^{2}\right)t}{q_{0}^{2}+q_{1}^{2}}}
-2q_{0}^{2}(q_{0}^{4}-q_{1}^{4})\cdot e^{\frac{3q_{0}^{2}x}{q_{1}}+i\alpha t+q_{1}\left(x+\frac{2iq_{1}\alpha t}{q_{0}^{2}+q_{1}^{2}}\right)}\\
&+2q_{1}^{2}(q_{0}^{4}-q_{1}^{4})\cdot e^{\frac{q_{0}^{2}x}{q_{1}}+3q_{1}x+\frac{i\alpha\left(q_{1}^{2}+3q_{0}^{2}\right)t}{q_{0}^{2}+q_{1}^{2}}}
+2q_{1}^{2}(-q_{0}^{4}+q_{1}^{4})\cdot e^{\frac{(3q_{0}^{2}+q_{1}^{2})\left(\left(q_{0}^{2}+q_{1}^{2}\right)x+iq_{1}\alpha t\right)}{q_{1}(q_{0}^{2}+q_{1}^{2})}}\Big)\Big)/\\
&\Big(q_{1}\Big(-4q_{0}^{2}q_{1}^{2}\cdot e^{2q_{1}x+q_{0}^{2}\left(\frac{2x}{q_{1}}+\frac{4i\alpha t}{q_{0}^{2}+q_{1}^{2}}\right)}-4q_{0}^{2}q_{1}^{2}\cdot e^{\frac{2q_{0}^{2}x}{q_{1}}+2q_{1}\left(x+\frac{2iq_{1}\alpha t}{q_{0}^{2}+q_{1}^{2}}\right)}-2(q_{0}^{2}-q_{1}^{2})^{2}\cdot e^{\frac{2q_{0}^{2}x}{q_{1}}+2q_{1}x+2i\alpha t}\\
&+(q_{0}^{2}+q_{1}^{2})^{2}\cdot e^{\frac{4q_{0}^{2}x}{q_{1}}+2i\alpha t}+(q_{0}^{2}+q_{1}^{2})^{2}\cdot e^{4q_{1}x+2i\alpha t}\Big)\Big),
\end{split}
\end{equation}
and $s(x,t)=\int_{x}^{\infty}(q(x',t)q(-x',-t))_{t}dx'$  $\rightarrow 0 ~\text{as}~ |x| \rightarrow \infty.$

(ii)When $\delta_{1}=\delta_{2}=-1$, \begin{equation}
\begin{split}
&q(x,t)=\Big(i\cdot e^{i\alpha t}\Big(-4q_{0}^{5}q_{1}\cdot e^{\frac{2q_{0}^{2}x}{q_{1}}+2q_{1}\left(x+\frac{2iq_{1}\alpha t}{q_{0}^{2}+q_{1}^{2}}\right)}-4q_{0}q_{1}^{5}\cdot e^{2q_{1}x+q_{0}^{2}\left(\frac{2x}{q_{1}}+\frac{4i\alpha t}{q_{0}^{2}+q_{1}^{2}}\right)}\\
&+2q_{0}q_{1}(q_{0}^{2}-q_{1}^{2})^{2}\cdot e^{\frac{2q_{0}^{2}x}{q_{1}}+2q_{1}x+2i\alpha t}+q_{0}q_{1}(q_{0}^{2}+q_{1}^{2})^{2}\cdot e^{\frac{4q_{0}^{2}x}{q_{1}}+2i\alpha t}+q_{0}q_{1}(q_{0}^{2}+q_{1}^{2})^{2}\cdot e^{4q_{1}x+2i\alpha t}\\
&-2q_{0}^{2}(q_{0}^{4}-q_{1}^{4})\cdot e^{\frac{q_{0}^{2}x}{q_{1}}+3q_{1}x+\frac{i\alpha\left(q_{0}^{2}+3q_{1}^{2}\right)t}{q_{0}^{2}+q_{1}^{2}}}
+2q_{0}^{2}(q_{0}^{4}-q_{1}^{4})\cdot e^{\frac{3q_{0}^{2}x}{q_{1}}+i\alpha t+q_{1}\left(x+\frac{2iq_{1}\alpha t}{q_{0}^{2}+q_{1}^{2}}\right)}\\
&+2q_{1}^{2}(-q_{0}^{4}+q_{1}^{4})\cdot e^{\frac{q_{0}^{2}x}{q_{1}}+3q_{1}x+\frac{i\alpha\left(q_{1}^{2}+3q_{0}^{2}\right)t}{q_{0}^{2}+q_{1}^{2}}}
+2q_{1}^{2}(-q_{0}^{4}+q_{1}^{4})\cdot e^{\frac{(3q_{0}^{2}+q_{1}^{2})\left(\left(q_{0}^{2}+q_{1}^{2}\right)x+iq_{1}\alpha t\right)}{q_{1}(q_{0}^{2}+q_{1}^{2})}}\Big)\Big)/\\
&\Big(q_{1}\Big(-4q_{0}^{2}q_{1}^{2}\cdot e^{2q_{1}x+q_{0}^{2}\left(\frac{2x}{q_{1}}+\frac{4i\alpha t}{q_{0}^{2}+q_{1}^{2}}\right)}-4q_{0}^{2}q_{1}^{2}\cdot e^{\frac{2q_{0}^{2}x}{q_{1}}+2q_{1}\left(x+\frac{2iq_{1}\alpha t}{q_{0}^{2}+q_{1}^{2}}\right)}-2(q_{0}^{2}-q_{1}^{2})^{2}\cdot e^{\frac{2q_{0}^{2}x}{q_{1}}+2q_{1}x+2i\alpha t}\\
&+(q_{0}^{2}+q_{1}^{2})^{2}\cdot e^{\frac{4q_{0}^{2}x}{q_{1}}+2i\alpha t}+(q_{0}^{2}+q_{1}^{2})^{2}\cdot e^{4q_{1}x+2i\alpha t}\Big)\Big),
\end{split}
\end{equation}
and $s(x,t)=\int_{x}^{\infty}(q(x',t)q(-x',-t))_{t}dx'$  $\rightarrow 0 ~\text{as}~ |x| \rightarrow \infty.$
\subsection{Singular soliton solutions of the nonlocal Sine-Gordon equation with $\theta_{+}+\theta_{-}=\pi$}
\begin{figure}[h]
\begin{tabular}{cc}
\includegraphics[width=0.5\textwidth]{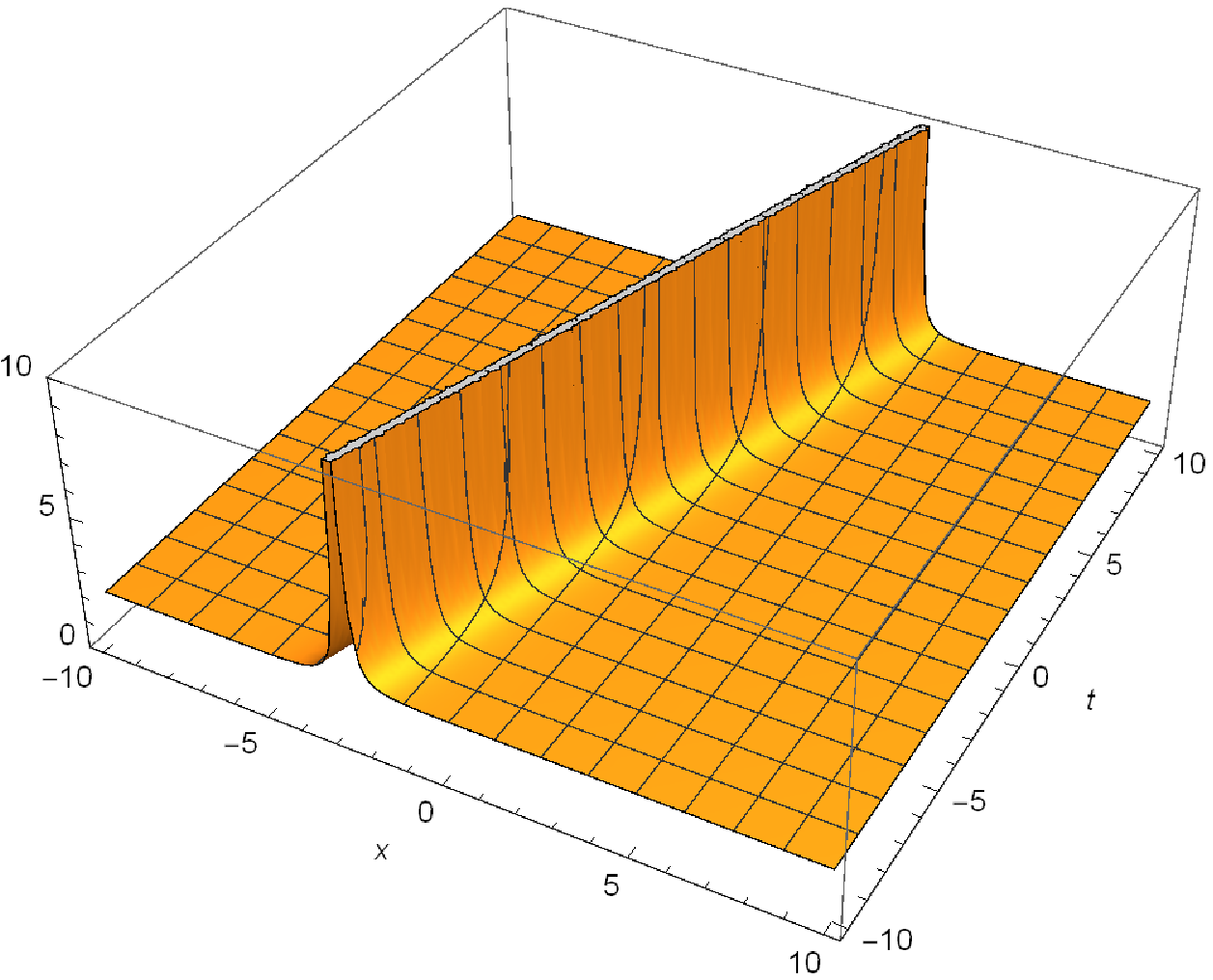}&
\includegraphics[width=0.5\textwidth]{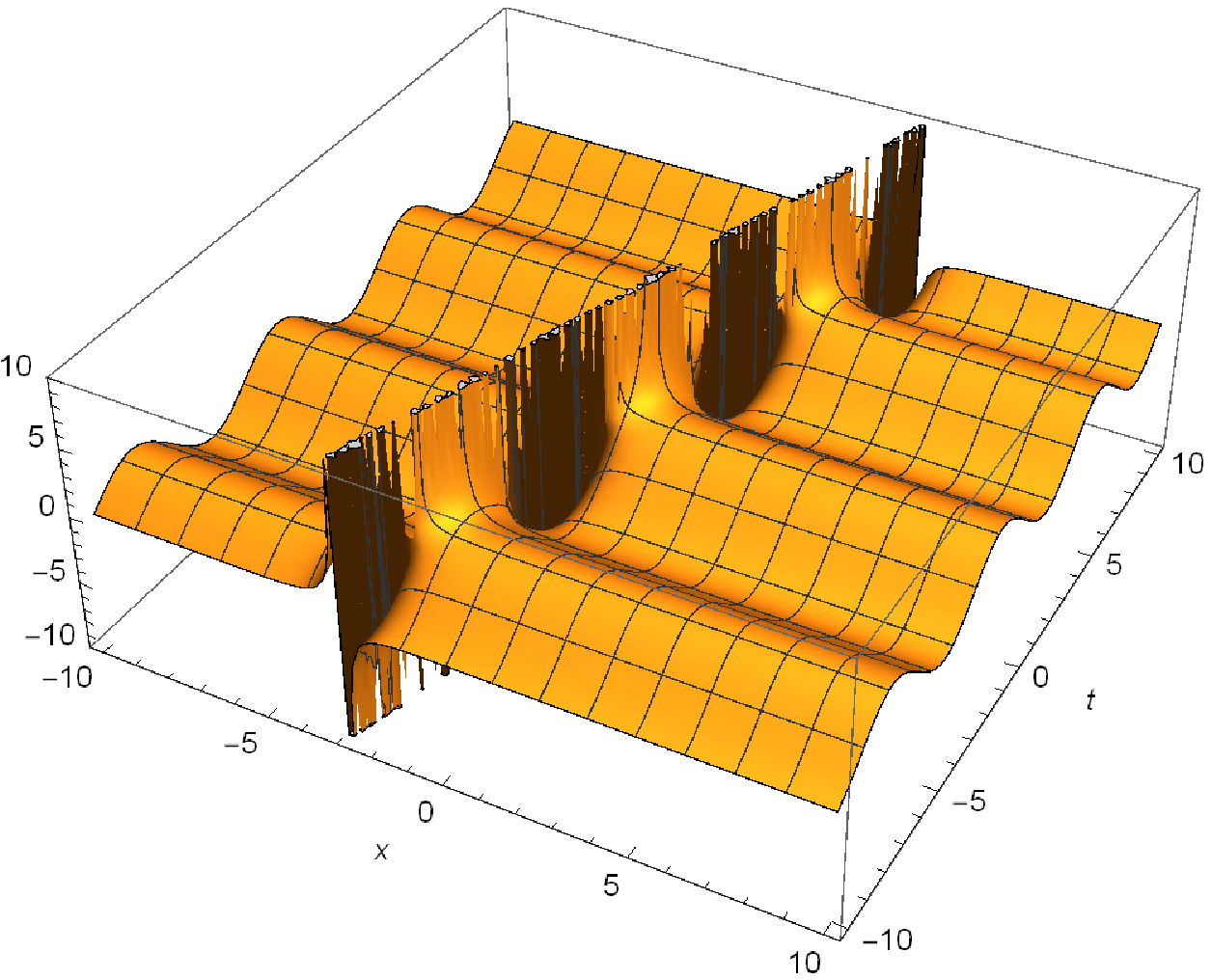}\\
(a) & (b)
\end{tabular}
\caption{(a) The amplitude of $q(x,t)$ for $\delta=1$ with $\theta_{+}=\frac{\pi}{3}$, $\alpha=1$ and $q_{0}=2$.  (b) The real part of $q(x,t)$ for $\delta=1$ with $\theta_{+}=\frac{\pi}{3}$, $\alpha=1$ and $q_{0}=2$.}
\label{Sine 1}
\end{figure}

We find the following "bright" 1-soliton solution with $\delta=1$, which reads
\begin{equation}
q(x,t)=\frac{1}{2}q_{0}e^{i(\alpha t-\theta_{+})}\left[-1+e^{2i\theta_{+}}+(1+e^{2i\theta_{+}})\coth\left(q_{0}x\cos\theta_{+}-\frac{1}{2}\alpha t\cot\theta_{+}\right)\right]
\end{equation}
and
\begin{equation}
s(x,t)=\frac{1}{2}q_{0}\alpha \cos\theta_{+}\cot\theta_{+}\coth^{2}\left(q_{0}x\cos\theta_{+}-\frac{1}{2}\alpha t\cot\theta_{+}\right)
-\frac{1}{2}q_{0}\alpha \cos\theta_{+}\cot\theta_{+}.
\end{equation}
We see that $s(x,t) \rightarrow 0$ as $|x| \rightarrow \infty$.
This solution is singular along the space-time line: $q_{0}x\cos\theta_{+}-\frac{1}{2}\alpha t \cot\theta_{+}=0$.
In Fig. \ref{Sine 1} below we give a typical bright 1-soliton solution.
\subsection{Singular soliton solutions of the nonlocal Sine-Gordon equation with $\theta_{+}+\theta_{-}=0$}
There are 2-soliton solutions with $\delta_{1}\delta_{2}=1$, which are singular along some  lines.
(i) When $\delta_{1}=\delta_{2}=1$, we have
\begin{equation}
\begin{split}
&q(x,t)=\Big(e^{i\alpha t}\Big(-4q_{0}^{5}q_{1}\cdot e^{\frac{2q_{0}^{2}x}{q_{1}}+2q_{1}\left(x+\frac{2iq_{1}\alpha t}{q_{0}^{2}+q_{1}^{2}}\right)}-4q_{0}q_{1}^{5}\cdot e^{2q_{1}x+q_{0}^{2}\left(\frac{2x}{q_{1}}+\frac{4i\alpha t}{q_{0}^{2}+q_{1}^{2}}\right)}\\
&+2q_{0}q_{1}(q_{0}^{2}-q_{1}^{2})^{2}\cdot e^{\frac{2q_{0}^{2}x}{q_{1}}+2q_{1}x+2i\alpha t}+q_{0}q_{1}(q_{0}^{2}+q_{1}^{2})^{2}\cdot e^{\frac{4q_{0}^{2}x}{q_{1}}+2i\alpha t}+q_{0}q_{1}(q_{0}^{2}+q_{1}^{2})^{2}\cdot e^{4q_{1}x+2i\alpha t}\\
&-2q_{0}^{2}(q_{0}^{4}-q_{1}^{4})\cdot e^{\frac{q_{0}^{2}x}{q_{1}}+3q_{1}x+\frac{i\alpha\left(q_{0}^{2}+3q_{1}^{2}\right)t}{q_{0}^{2}+q_{1}^{2}}}
+2q_{0}^{2}(q_{0}^{4}-q_{1}^{4})\cdot e^{\frac{3q_{0}^{2}x}{q_{1}}+i\alpha t+q_{1}\left(x+\frac{2iq_{1}\alpha t}{q_{0}^{2}+q_{1}^{2}}\right)}\\
&+2q_{1}^{2}(-q_{0}^{4}+q_{1}^{4})\cdot e^{\frac{q_{0}^{2}x}{q_{1}}+3q_{1}x+\frac{i\alpha\left(q_{1}^{2}+3q_{0}^{2}\right)t}{q_{0}^{2}+q_{1}^{2}}}
-2q_{1}^{2}(-q_{0}^{4}+q_{1}^{4})\cdot e^{\frac{(3q_{0}^{2}+q_{1}^{2})\left(\left(q_{0}^{2}+q_{1}^{2}\right)x+iq_{1}\alpha t\right)}{q_{1}(q_{0}^{2}+q_{1}^{2})}}\Big)\Big)/\\
&\Big(q_{1}\Big(-4q_{0}^{2}q_{1}^{2}\cdot e^{2q_{1}x+q_{0}^{2}\left(\frac{2x}{q_{1}}+\frac{4i\alpha t}{q_{0}^{2}+q_{1}^{2}}\right)}-4q_{0}^{2}q_{1}^{2}\cdot e^{\frac{2q_{0}^{2}x}{q_{1}}+2q_{1}\left(x+\frac{2iq_{1}\alpha t}{q_{0}^{2}+q_{1}^{2}}\right)}-2(q_{0}^{2}-q_{1}^{2})^{2}\cdot e^{\frac{2q_{0}^{2}x}{q_{1}}+2q_{1}x+2i\alpha t}\\
&+(q_{0}^{2}+q_{1}^{2})^{2}\cdot e^{\frac{4q_{0}^{2}x}{q_{1}}+2i\alpha t}+(q_{0}^{2}+q_{1}^{2})^{2}\cdot e^{4q_{1}x+2i\alpha t}\Big)\Big).
\end{split}
\end{equation}
from (\ref{asympN1c'''}) and (\ref{E:closing system 7}).
(ii) When $\delta_{1}=\delta_{2}=-1$, we have
\begin{equation}
\begin{split}
&q(x,t)=\Big(e^{i\alpha t}\Big(-4q_{0}^{5}q_{1}\cdot e^{\frac{2q_{0}^{2}x}{q_{1}}+2q_{1}\left(x+\frac{2iq_{1}\alpha t}{q_{0}^{2}+q_{1}^{2}}\right)}-4q_{0}q_{1}^{5}\cdot e^{2q_{1}x+q_{0}^{2}\left(\frac{2x}{q_{1}}+\frac{4i\alpha t}{q_{0}^{2}+q_{1}^{2}}\right)}\\
&+2q_{0}q_{1}(q_{0}^{2}-q_{1}^{2})^{2}\cdot e^{\frac{2q_{0}^{2}x}{q_{1}}+2q_{1}x+2i\alpha t}+q_{0}q_{1}(q_{0}^{2}+q_{1}^{2})^{2}\cdot e^{\frac{4q_{0}^{2}x}{q_{1}}+2i\alpha t}+q_{0}q_{1}(q_{0}^{2}+q_{1}^{2})^{2}\cdot e^{4q_{1}x+2i\alpha t}\\
&+2q_{0}^{2}(q_{0}^{4}-q_{1}^{4})\cdot e^{\frac{q_{0}^{2}x}{q_{1}}+3q_{1}x+\frac{i\alpha\left(q_{0}^{2}+3q_{1}^{2}\right)t}{q_{0}^{2}+q_{1}^{2}}}
-2q_{0}^{2}(q_{0}^{4}-q_{1}^{4})\cdot e^{\frac{3q_{0}^{2}x}{q_{1}}+i\alpha t+q_{1}\left(x+\frac{2iq_{1}\alpha t}{q_{0}^{2}+q_{1}^{2}}\right)}\\
&+2q_{1}^{2}(q_{0}^{4}-q_{1}^{4})\cdot e^{\frac{q_{0}^{2}x}{q_{1}}+3q_{1}x+\frac{i\alpha\left(q_{1}^{2}+3q_{0}^{2}\right)t}{q_{0}^{2}+q_{1}^{2}}}
+2q_{1}^{2}(-q_{0}^{4}+q_{1}^{4})\cdot e^{\frac{(3q_{0}^{2}+q_{1}^{2})\left(\left(q_{0}^{2}+q_{1}^{2}\right)x+iq_{1}\alpha t\right)}{q_{1}(q_{0}^{2}+q_{1}^{2})}}\Big)\Big)/\\
&\Big(q_{1}\Big(-4q_{0}^{2}q_{1}^{2}\cdot e^{2q_{1}x+q_{0}^{2}\left(\frac{2x}{q_{1}}+\frac{4i\alpha t}{q_{0}^{2}+q_{1}^{2}}\right)}-4q_{0}^{2}q_{1}^{2}\cdot e^{\frac{2q_{0}^{2}x}{q_{1}}+2q_{1}\left(x+\frac{2iq_{1}\alpha t}{q_{0}^{2}+q_{1}^{2}}\right)}-2(q_{0}^{2}-q_{1}^{2})^{2}\cdot e^{\frac{2q_{0}^{2}x}{q_{1}}+2q_{1}x+2i\alpha t}\\
&+(q_{0}^{2}+q_{1}^{2})^{2}\cdot e^{\frac{4q_{0}^{2}x}{q_{1}}+2i\alpha t}+(q_{0}^{2}+q_{1}^{2})^{2}\cdot e^{4q_{1}x+2i\alpha t}\Big)\Big).
\end{split}
\end{equation}
\subsection{Singular soliton solutions of the nonlocal reverse  NLS equation}
{\bf Case 1.} When $\sigma=1$ and $\theta_{+}+\theta_{-}=0$, we have a singular 1-soliton solution with $\delta=-1$, which is given by
\begin{equation}
\begin{split}
q(x,t)=q_{0}e^{2iq_{0}^{2}t}[\cos\theta_{+}+i\coth(q_{0}(x-2q_{0}t\cos\theta_{+})\sin\theta_{+})\sin\theta_{+}].
\end{split}
\end{equation}
{\bf Case 2.} When $\sigma=1$ and $\theta_{+}+\theta_{-}=\pi$, we have singular 2-soliton solutions with $\delta_{1}\delta_{2}=1$.

When $\delta_{1}=1$ and $\delta_{2}=1$, it yields
\begin{equation}
\begin{split}
q(x,t)=\frac{ie^{-2iq_{0}^{2}t}\left[-q_{0}^{4}e^{i\frac{q_{1}^{4}-q_{0}^{4}}{q_{1}^{2}}t}+q_{1}^{4}e^{i\frac{q_{0}^{4}-q_{1}^{4}}{q_{1}^{2}}t}
+q_{0}q_{1}(q_{0}^{2}+q_{1}^{2})\sinh\left(\frac{q_{0}^{2}-q_{1}^{2}}{q_{1}}x\right)\right]}{q_{1}\left[2iq_{0}q_{1}\sin\left(\frac{(q_{0}^{4}-q_{1}^{4})t}{q_{1}^{2}}\right)+(q_{0}^{2}+q_{1}^{2})\sinh\left(\frac{(q_{0}^{2}-q_{1}^{2})x}{q_{1}}\right)\right]}.
\end{split}
\end{equation}

When $\delta_{1}=-1$ and $\delta_{2}=-1$, it yields
\begin{equation}
\begin{split}
q(x,t)=\frac{e^{-2iq_{0}^{2}t}\left[i(q_{0}^{4}-q_{1}^{4})\cos\left(\frac{q_{0}^{4}-q_{1}^{4}}{q_{1}^{2}}t\right)
+iq_{0}q_{1}(q_{0}^{2}+q_{1}^{2})\sinh\left(\frac{q_{0}^{2}-q_{1}^{2}}{q_{1}}x\right)+(q_{0}^{4}+q_{1}^{4})\sin\left(\frac{q_{0}^{4}-q_{1}^{4}}{q_{1}^{2}}t\right)\right]}{q_{1}\left[-2q_{0}q_{1}\sin\left(\frac{(q_{0}^{4}-q_{1}^{4})t}{q_{1}^{2}}\right)+(q_{0}^{2}+q_{1}^{2})\sinh\left(\frac{(q_{0}^{2}-q_{1}^{2})x}{q_{1}}\right)\right]}.
\end{split}
\end{equation}


{\bf Case 3.} When $\sigma=-1$ and $\theta_{+}+\theta_{-}=\pi$, we have a singular 1-soliton solution with $\delta=1$, which is given by
\begin{equation}
\begin{split}
&q(x,t)=q_{0}e^{2iq_{0}^{2} t}\cdot\\
&\frac{\left(1+e^{4q_{0}\cos\theta_{+}(x+2q_{0}t\sin\theta_{+})}+2e^{2q_{0}\cos\theta_{+}(x+2q_{0}t\sin\theta_{+})}
\right)\cos\theta_{+}+i\sin\theta_{+}(-1+e^{4q_{0}\cos\theta_{+}(x+2q_{0}t\sin\theta_{+})})}{-1+e^{4q_{0}\cos\theta_{+}(x+2q_{0}t\sin\theta_{+})}}.
\end{split}
\end{equation}

\begin{figure}[h]
\begin{tabular}{cc}
\includegraphics[width=0.5\textwidth]{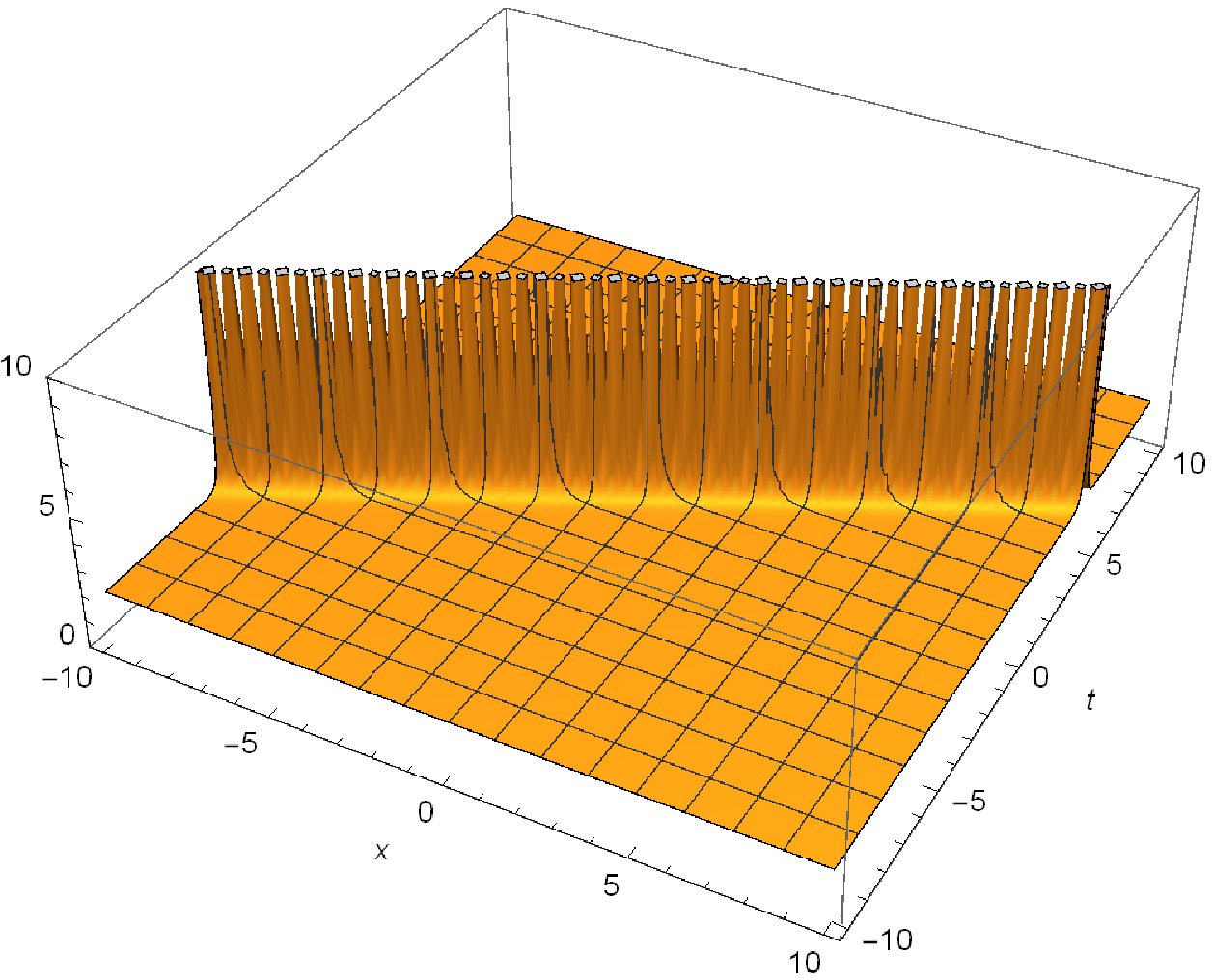}&
\includegraphics[width=0.5\textwidth]{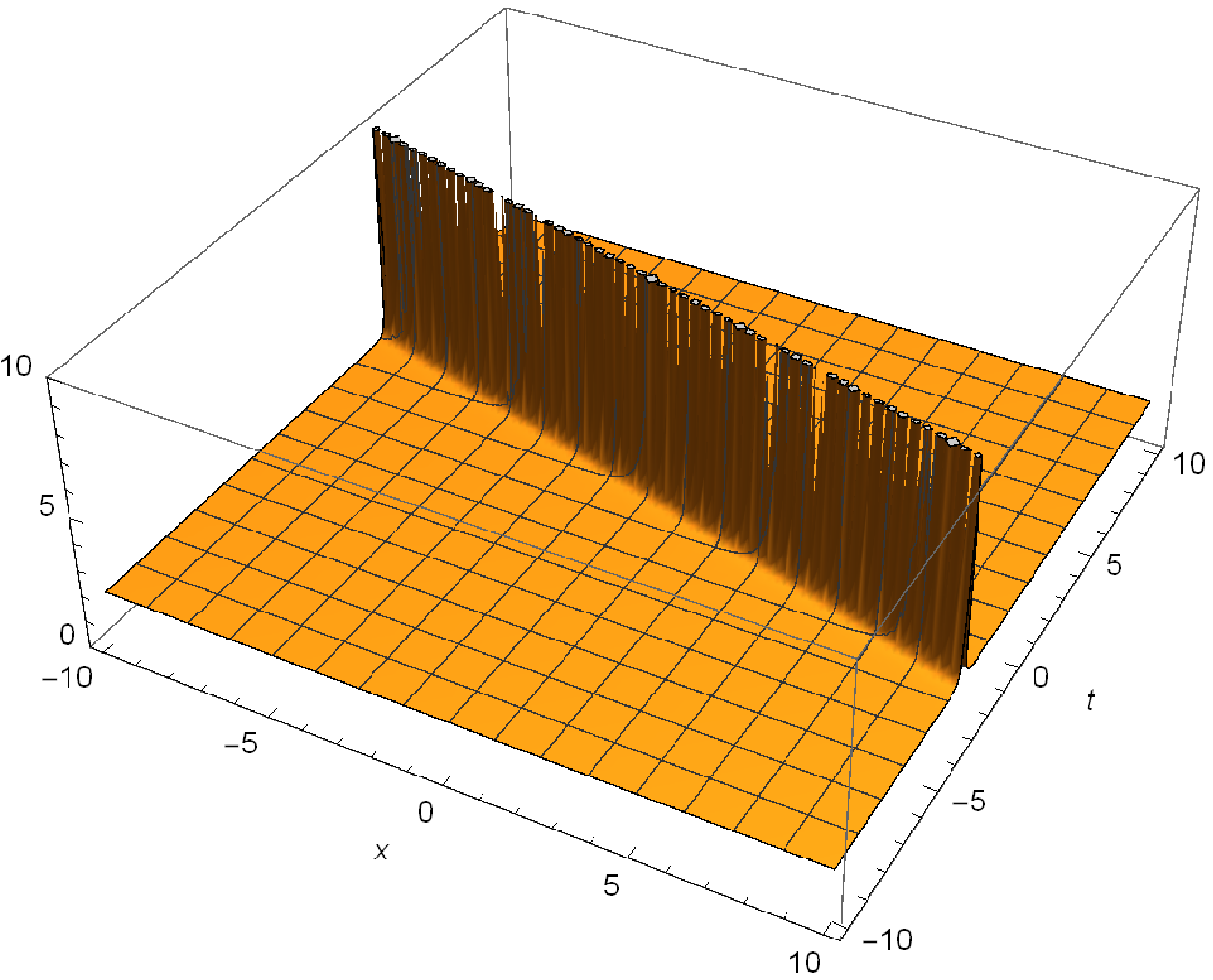}\\
(a) & (b)
\end{tabular}
\caption{(a) The amplitude of $q(x,t)$ $\sigma=1$, $\delta=-1$, $\theta_{+}=\frac{\pi}{3}$ and $q_{0}=2$.  (b) The amplitude of $q(x,t)$ with $\sigma=-1$, $\delta=1$, $\theta_{+}=\frac{\pi}{3}$ and $q_{0}=2$.}
\label{RST 2}
\end{figure}

{\bf Case 4.} When $\sigma=-1$ and $\theta_{+}+\theta_{-}=0$, we have singular 2-soliton solutions with $\delta_{1}\delta_{2}=1$, which are
\begin{equation}
\begin{split}
q(x,t)=\frac{e^{-2iq_{0}^{2}t}\left[q_{0}^{4}e^{i\frac{q_{1}^{4}-q_{0}^{4}}{q_{1}^{2}}t}-q_{1}^{4}e^{i\frac{q_{0}^{4}-q_{1}^{4}}{q_{1}^{2}}t}
+q_{0}q_{1}(q_{0}^{2}+q_{1}^{2})\sinh\left(\frac{q_{0}^{2}-q_{1}^{2}}{q_{1}}x\right)\right]}{q_{1}\left[-2iq_{0}q_{1}\sin\left(\frac{(q_{0}^{4}-q_{1}^{4})t}{q_{1}^{2}}\right)+(q_{0}^{2}+q_{1}^{2})\sinh\left(\frac{(q_{0}^{2}-q_{1}^{2})x}{q_{1}}\right)\right]}.
\end{split}
\end{equation}
for $\delta_{1}=1$ and $\delta_{2}=1$, and
\begin{equation}
\begin{split}
q(x,t)=\frac{e^{-2iq_{0}^{2}t}\left[-q_{0}^{4}e^{i\frac{q_{1}^{4}-q_{0}^{4}}{q_{1}^{2}}t}+q_{1}^{4}e^{i\frac{q_{0}^{4}-q_{1}^{4}}{q_{1}^{2}}t}
+q_{0}q_{1}(q_{0}^{2}+q_{1}^{2})\sinh\left(\frac{q_{0}^{2}-q_{1}^{2}}{q_{1}}x\right)\right]}{q_{1}\left[2iq_{0}q_{1}\sin\left(\frac{(q_{0}^{4}-q_{1}^{4})t}{q_{1}^{2}}\right)+(q_{0}^{2}+q_{1}^{2})\sinh\left(\frac{(q_{0}^{2}-q_{1}^{2})x}{q_{1}}\right)\right]}.
\end{split}
\end{equation}
for $\delta_{1}=-1$ and $\delta_{2}=-1$.
The singular 1-soliton solutions for $\sigma=1, \delta=-1$ and  $\sigma=-1, \delta=1$ are displayed in
Fig. \ref{RST 2} (a) and (b), respectively.

\section{Acknowledgements}
The authors are pleased to acknowledge Justin T. Cole  for assistance in the figures of this paper and in Mathematica respectively. MJA was partially supported by NSF under Grant No. DMS-1310200.


\bibliographystyle{amsplain}

\end{document}